\documentstyle[twoside,ijmpd,overcite,epsf]{article}  
  \input prepictex \input pictex \input postpictex

  \def\mathput#1{\relax \ifmmode \displaystyle #1\else $\displaystyle #1$\fi}
  
  \artsectnumbering

\def\Ref{Ref.~} \def\Refs{Refs.~}

\begin{document}

\runninghead{D.\ Bini, P.\ Carini and R.T.\ Jantzen}{Intrinsic Derivatives and Centrifugal Forces in General Relativity $\ldots$}

\normalsize\textlineskip
\thispagestyle{empty}
\setcounter{page}{1}

\copyrightheading{Vol.\ 6, No.\ 1 (1997) 143--198 [revised 2001]}

\vspace*{0.88truein}

\fpage{1}

  \centerline{\bf
THE INTRINSIC DERIVATIVE AND CENTRIFUGAL FORCES
  }\vspace*{0.035truein}
  \centerline{\bf
IN GENERAL RELATIVITY: II. APPLICATIONS 
  }\vspace*{0.035truein}
  \centerline{\bf
TO CIRCULAR ORBITS IN SOME FAMILIAR 
  }\vspace*{0.035truein}
  \centerline{\bf
STATIONARY AXISYMMETRIC SPACETIMES
  }

  \vspace*{0.37truein}
  \centerline{
DONATO BINI
  }
  \vspace*{0.015truein}
  \centerline{\footnotesize\it 
Istituto per Applicazioni della Matematica C.N.R., 
I--80131 Napoli, Italy 
  and
  }
  \baselineskip=10pt
  \centerline{\footnotesize\it 
International Center for Relativistic Astrophysics, 
University of Rome, I--00185 Roma, Italy
  }
  \vspace*{10pt}
  \centerline{
PAOLO CARINI\footnote{
 Present Address: Physics Department, 
 Amherst College, Amherst, MA 01002, USA.}
  }
  \vspace*{0.015truein}
  \centerline{\footnotesize\it 
GP-B, Hansen Labs, Stanford University, 
Stanford, CA 94305, USA 
and
  }
  \baselineskip=10pt
  \centerline{\footnotesize\it 
International Center for Relativistic Astrophysics,
University of Rome, I--00185 Roma, Italy
  }
  \vspace*{10pt}
  \centerline{
ROBERT T. JANTZEN
  }
  \vspace*{0.015truein}
  \centerline{\footnotesize\it 
Department of Mathematical Sciences, 
Villanova University, Villanova, PA 19085, USA 
and
  }
  \baselineskip=10pt
  \centerline{\footnotesize\it 
International Center for Relativistic Astrophysics, 
University of Rome, I--00185 Roma, Italy
  }

  \vspace*{0.225truein}
  \pub{17 December 1996}

  \vspace*{0.21truein}


\abstracts{
The tools developed in a preceding article for interpreting spacetime
geometry in terms of all possible space-plus-time splitting approaches
are applied to circular orbits in some familiar stationary
axisymmetric spacetimes. This helps give a more intuitive picture of
their rotational features including spin precession effects, and puts
related work of Abramowicz, de Felice, and others on circular orbits
in black hole spacetimes into a more general context. 
}{}{}

\vspace*{10pt}

\textlineskip \typeout{*** correct baselineskip for body.}

\setcounter{footnote}{0}
\renewcommand{\thefootnote}{\alph{footnote}}


\section{Introduction}

``Rotating spacetimes'' have captured people's imaginations ever since
``rigid'' rotations in Minkowski spacetime were considered within the
theory of general relativity. Even this simple example which is the
foundation of the ``fictitious'' centrifugal and Coriolis forces in
classical physics has led to its share of confusion about rotation in
relativity. G\"odel's discovery \cite{god} of the spacetime which
bears his name certainly added fuel to the fire, which was again
stoked by the discovery of the rotating black hole solution of Kerr
\cite{ker} and its generalizations \cite{car,new}. 

The language of gravitoelectromagnetism \cite{mfg}, specialized in the
preceding companion article \cite{idcfi} (to be referred to here as
[BCJ1]) for stationary axially symmetric spacetimes, helps us to
understand the effects of rotation as well as those of acceleration
and spatial curvature in these three classic spacetime examples.
Indeed the lines of force of the various gravitoelectromagnetic vector
fields, especially in the Kerr spacetimes, help give a more tangible
way of interpreting the behavior of test particle motions in the
gravitational field of these spacetimes. 

Here we focus on the simpler case of circular orbits following Killing
trajectories in these spacetimes, confining our attention to the
equatorial plane in the Kerr spacetimes \cite{rinper,korea91}. This
test particle motion is an example of ``purely transverse" relative
acceleration \cite{rok}. By exploring the roles played by the radial
spatial gravitational forces, one obtains a clearer picture of the
action and interrelationships of the various gravitoelectric (GE),
gravitomagnetic (GM), and space curvature (SC) forces that one may
define within each spacetime as well as of the correspondences one may
establish between these different spacetimes. Presenting the details
of these applications also helps make more concrete the somewhat
abstract but powerful language of gravitoelectromagnetism itself,
which can be valuable in interpreting the geometry of other
spacetimes. In particular, previous discussions of circular orbits in
black hole spacetimes by Abramowicz, de Felice, and others
\cite{acl88,a90a,a90b,a90c,ab91,a92,ams93,%
pc90,ip93,a93,anw93,anw95,sem,sem96,nayvis96,pra96,%
def91,defuss91,defuss93,def94,def95,defuss92,defuss96,barboiisr} are
fit into a more general picture which helps to clarify their
particular analyses of the behavior of certain properties of these
orbits. 

Each of these three classes of spacetimes have natural stationary
axisymmetric nonlinear reference frames, i.e., threaded slicings
(hypersurface foliations with transversal congruences of curves) of
the spacetime which are adapted to two Killing vector fields
associated with a 2-dimensional stationary axisymmetry group. The
nonlinear reference frames for the rotating Minkowski and G\"odel
spacetimes have an additional translational symmetry making them
cylindrically symmetric as well. In each case the threading and
slicing families of test observers associated with these nonlinear
reference frames are tied to the geometry of the spacetime and help
elucidate its properties. 

In the case of Kerr, the nonlinear reference frame associated with
Boyer-Lind\-quist coordinates $\{t,r,\theta,\phi\}$ \cite{mtw} has as
its threading observers (following the time coordinate line Killing
trajectories) the distantly nonrotating observers or static observers,
while the slicing observers (moving normal to the time coordinate
hypersurfaces) are the locally nonrotating observers or
zero-angular-momentum observers. Both observer families are
accelerated, the threading observers opposing the dragging along
action of the rotating black hole, while the slicing observers are
dragged along by the hole with respect to spatial infinity. In the
G\"odel spacetime, the nonlinear reference frame of cylindrical
coordinates $\{t,\rho,\phi,z\}$ has the threading observers moving
along the geodesic flow lines of the rotating dust source, while the
accelerated slicing observers oppose the global rotation of the
spacetime. In the rotating Minkowski spacetime, the accelerated
threading observers are uniformly rotating, while the geodesic slicing
observers are a global family of inertial observers associated with
the usual time lines of an orthonormal Cartesian coordinate system.
Each of these three spacetime examples exhibits different
configurations of the various gravitoelectromagnetic fields whose
comparison offers insights about the nature of the spacetimes
themselves. 

The G\"odel spacetime was first studied as an example of a constant
gravitomagnetic field configuration by Wilkins and Jacobs
\cite{wiljac}, based on the analogy between linearized general
relativity and  electromagnetism \cite{for}. Their pioneering work
stimulated the present program of trying to better understand fully
nonlinear general relativity in terms of observer-splittings. 

\section{Gravitoelectromagnetic potentials, fields, and forces}

The ``rotating Minkowski'' spacetime is just Minkowski spacetime
expressed in terms of the nonlinear reference frame associated with a
family of uniformly rotating test observers (threading) and the time
hypersurfaces (slicing) associated with the global inertial Cartesian
coordinates with respect to which this rotation takes place. It is
most easily described in terms of rotating cylindrical coordinates
$\{t,\rho,\phi,z\}$, where $\phi = \phi\,' -\Omega t$ gives the
relationship to the nonrotating angular coordinate $\phi\,'$, as
discussed by Landau and Lifshitz \cite{lanlif75} and in [BCJ1].
Although the coordinate angular velocity $\Omega$ about the $z$-axis
may take any real value (define ${\cal R} = 1/|\Omega|$), here it will
be assumed to be positive in order to discuss corotation and
counter-rotation with respect to the sense defined by the positive
$z$-axis and the righthand rule. 

The G\"odel spacetime is a solution of the Einstein equations with
constant dust energy density $\rho_{(0)}$ and cosmological constant
$\Lambda$ related by $\Lambda= -\Omega^2 = -(2{\cal R}^2)^{-1} =
-4\pi\rho_{(0)}$, where ${\cal R} = |\sqrt{2}\Omega|^{-1}$ is
G\"odel's curvature parameter ``$a$'' and $\Omega$ is the nonzero
constant parametrizing the constant vorticity of the fluid source.
Although it may take any nonzero value, it will be assumed to be
positive here for the same reason as above, so that the threading
observers corotate with respect to the positive $z$-axis in the
cylindrical coordinates used here. These coordinates differ from
others in the literature by a rescaling, chosen so that to linear
order in the vorticity parameter $\Omega$ the G\"odel and rotating
Minkowski metrics are the same, although important differences occur
at second order. In these coordinates the limiting behavior
approaching the axis of symmetry of the associated nonlinear reference
frame is also the same to lowest order in this parameter, and the
limit of both metrics as $\Omega \to0$ is just Minkowski spacetime
expressed in nonrotating cylindrical coordinates. Thus the new
parametrization of the G\"odel metric emphasizes the key feature of
its geometry, namely the global rotation, rather than the related
spatial curvature. 

The Kerr spacetime describes the geometry around a rotating black hole
with mass ${\cal M}$ (directly defining a length scale ${\cal R}$)
and angular momentum per unit mass $a$ as seen
from infinity. The angular velocity parameter $a$ will be assumed
positive as above, so that all three spacetime examples rotate in the
same sense. The limit $a=0$ gives the nonrotating Schwarzschild
spacetime, while $a={\cal M}$ is the ``extreme Kerr" case. The usual
Boyer-Lindquist coordinates $\{t,r,\theta,\phi\}$ will be used here. 

In each spacetime the Killing vector field
$e_0{}^\alpha=\delta^\alpha{}_t$ generates the stationary symmetry
while the spacelike Killing vector field $\delta^\alpha{}_\phi$
generates the axisymmetry, and the spatial coordinates are orthogonal.
For uniformity of discussion, the radial Kerr coordinate symbol $r$
will be used to denote the cylindrical coordinate $\rho$ in the other
two examples as well, while the physical (orthonormal) component along
$-\delta^\alpha{}_\theta$ perpendicular to the equatorial plane in
Kerr will be referred to as along the positive $z$-axis as in the case
of the plane $z=0$ in the other spacetimes and will be indicated by
the index $\hat z$. The latter plane will be referred to as the
equatorial plane in all cases. For graphing purposes it is natural to
introduce the rescaled radial variable $\bar r = r/{\cal R}$ for all
these spacetimes and $\bar a = a/{\cal R} = a/{\cal M}$ for the
additional parameter in the Kerr case. 

For all the three spacetimes under consideration, the spacetime metric
restricted to the world sheet of this plane can be expressed in the
reference (i.e., coordinate), threading, and slicing decompositions
respectively as 
\begin{eqnarray}
       ds^2 &=& {}^{(4)}\kern-1pt  g_{tt} dt^2 
                    + 2{}^{(4)}\kern-1pt  g_{t\phi} dt d\phi
	            +  {}^{(4)}\kern-1pt  g_{\phi\phi} d\phi^2 
                    +  {}^{(4)}\kern-1pt  g_{rr}dr^2
                 \nonumber\\
	&=& -M^2 (dt - M_\phi d\phi)^2 + \gamma_{\phi\phi} d\phi^2
                                     + \gamma_{rr}dr^2 
                  \nonumber\\
    &=& -N^2 dt^2 + g_{\phi\phi}(d\phi + N^\phi dt)^2 + g_{rr}dr^2 
               \ .
\end{eqnarray}
Since the threading and slicing observer-adapted frames associated
with this adapt\-ed coordinate system are obtained by projection of the
coordinate frame, their spatial structure functions vanish. Table
\ref{tab:thdsli} gives the expressions of the various threading and
slicing quantities for the three cases. 

\typeout{*** Table 1. (thdsli)}
\begin{table}[htbp]\footnotesize
\tcaption
{The lapse, and the nonzero observer-adapted components of the shift
and spatial metric in both the threading and slicing points of view
are given for the rotating Minkowski, G\"odel and Kerr spacetimes on
the equatorial plane. Note the G\"odel parameter relation
$\sqrt{2}|\Omega| = 1/{\cal R}$ useful for reinterpreting quantities
in terms of spatial curvature rather than vorticity. 
}
\vbox{%
 \typeout{*** eqnarray struts inserted}
 \def\Strut{\relax\hbox{\vrule width0pt height 10.5pt depth 5.5pt}}
\begin{eqnarray*}
\begin{array}{||l|l|l|l||} \hline \Strut
\rm lapse,\ shift\ and          &\rm Rotating   
   &\hbox{G\"odel}  &\rm Kerr \\ 
\rm spatial\ metric             &\rm  Minkowski &                &  
                     \\ \hline \Strut
N= (-{}^{(4)}\kern-1pt  g^{00})^{-1/2}       
                           & 1  
                           & {\rm c}/\sqrt{1-{\rm s}^2} 
                           & \sqrt{r\Delta/(r^3+a^2r+2a^2{\cal M})}
                      \\ \hline \Strut
N^\phi= N_\phi/g_{\phi\phi}     & \Omega 
                         & \Omega / (1- {\rm s}^2)           
                         & -2a{\cal M}/(r^3 + a^2 r +2a^2{\cal M})
                       \\ \hline \Strut
N_\phi={}^{(4)}\kern-1pt  g_{0\phi}          
                                & \Omega r^2 
                                & 2{\rm s}^2/\Omega 
                                & -2a{\cal M}/r 
                        \\ \hline \Strut
g_{\phi\phi} = {}^{(4)}\kern-1pt  g_{\phi\phi} 
                                & r^2 
                                & 2{\rm s}^2(1-{\rm s}^2)/\Omega^2 
                                & (r^3 + a^2r + 2a^2{\cal M})/r
                        \\ \hline \Strut
g_{rr}=\gamma_{rr} = {}^{(4)}\kern-1pt  g_{rr} 
                                & 1 
                                & 1 
                                & r^2/\Delta 
                         \\ \hline \Strut
M= (-{}^{(4)}\kern-1pt  g_{00})^{1/2}        
                                & \gamma^{-1} 
                                & 1  
                                & \sqrt{(r-2{\cal M})/r} 
                         \\ \hline \Strut
M_\phi = - {}^{(4)}\kern-1pt  g_{0\phi}/{}^{(4)}\kern-1pt  g_{00} 
                                & \Omega r^2 \gamma^2 
                                & 2{\rm s}^2/\Omega 
                                & -2a{\cal M}/(r-2{\cal M}) 
                          \\ \hline \Strut
\gamma_{\phi\phi} = {}^{(4)}\kern-1pt  g_{\phi\phi}
                  -({}^{(4)}\kern-1pt  g_{0\phi})^2
                   /{}^{(4)}\kern-1pt  g_{00}
                                & \gamma^2 r^2 
                                & 2{\rm s}^2{\rm c}^2 /\Omega^2 
                                & r\Delta/(r-2{\cal M})  
                          \\ \hline
\multicolumn{4}{c}{} \\
\multicolumn{4}{l}{%
        \gamma\equiv (1-\Omega^2r^2)^{-1/2}    \, , \ 
	{\rm s}\equiv\sinh(\sqrt{2}\Omega r/2) \, , \
        {\rm c}\equiv\cosh(\sqrt{2}\Omega r/2) \, , \
        {\rm t}\equiv\tanh(\sqrt{2}\Omega r/2) \, ,}\\
\multicolumn{4}{l}{%
        \Strut
        {\rm S}\equiv\sinh(\sqrt{2}\Omega r) \, , \
        {\rm C}\equiv\cosh(\sqrt{2}\Omega r) \, , \
	{\rm T}\equiv\tanh(\sqrt{2}\Omega r) \, , \
        \Delta\equiv r^2 -2{\cal M} r + a^2  \, ,}
\end{array}
\end{eqnarray*}
}
\label{tab:thdsli}
\end{table}
\typeout{*** No visible margin spill although 14pt overfull box reported.}

The general formulas of [BCJ1] for constant speed test particle
circular orbits in stationary axisymmetric spacetimes are easily
evaluated for the present explicit metrics. The test particle moves
along the $\phi$ direction with constant speed. The 4-velocity of a
nonzero rest mass test particle is parametrized by the coordinate
angular velocity $\zeta = \dot\phi = d\phi/dt$ as follows 
\begin{equation}\label{eq:zeta}
   U^\alpha 
    = \Gamma [ \delta^\alpha{}_t + \zeta \delta^\alpha{}_\phi ]
      \ ,
\end{equation}
where $\Gamma = dt/d\tau_U>0$ is defined by
\begin{eqnarray}\label{eq:gammasqd}
  \Gamma^{-2} 
    &=& 
     -[            {}^{(4)}\kern-1pt  g_{tt} 
         + 2 \zeta {}^{(4)}\kern-1pt  g_{t\phi} 
         + \zeta^2 {}^{(4)}\kern-1pt  g_{\phi\phi}]
    =
         - {}^{(4)}\kern-1pt  g_{\phi\phi}
            (\zeta - \zeta_-)(\zeta - \zeta_+)
              \nonumber\\
    &=&  M^2(1- M_\phi \zeta)^2 - \gamma_{\phi\phi} \zeta^2 
    =  N^2 - g_{\phi\phi}(\zeta + N^\phi)^2 
\end{eqnarray}
and $\tau_U$ is a proper time parametrization of the world line. The
timelike condition for the 4-velocity $U^\alpha$ requires $\Gamma^{-2}
> 0$, constraining $\zeta$ to belong to the interval
$[\zeta_-,\zeta_+]$ between the roots of the quadratic equation
$\Gamma^{-2}=0$ in $\zeta$ corresponding to null directions, namely 
\begin{eqnarray}\label{eq:zetapm}
 \zeta_\pm 
   &=& [-{}^{(4)}g_{t\phi} 
           \pm ({}^{(4)}g_{t\phi}{}^2 - g_{\phi\phi}g_{tt})^{1/2}]
                /{}^{(4)}g_{\phi\phi} 
          \nonumber\\
   &=& [-M^2M^\phi \pm M\gamma_{\phi\phi}^{-1/2}]
             / (1-M^2M_\phi M^\phi)
   = -N^\phi \pm N (g_{\phi\phi})^{-1/2} \ .
\end{eqnarray}
The 4-velocity of a zero rest mass particle (for which
$\Gamma^{-2}=0$) has an arbitrary normalization factor
$\Gamma_{\rm(null)}$ in place of $\Gamma$ in equation (\ref{eq:zeta}) 
\begin{equation}
 P_\pm^\alpha 
    = \Gamma_{\rm(null)}
          [ \delta^\alpha{}_t + \zeta_\pm \delta^\alpha{}_\phi ] \ .
\end{equation}
Circular orbits for which $\zeta \ge 0$ or $\zeta<0$ will be referred
to respectively as corotating or counter-rotating (with respect to the
nonlinear reference frame or the threading observers). 

Note that the coordinate angular velocity of the slicing observers is
just the average of the two limiting angular velocities 
\begin{equation}\label{eq:sagnacobservers}
     \zeta_{\rm(sl)}  
      = \zeta_{\rm(nmp)} 
      = (\zeta_- + \zeta_+)/2 
      = - N^\phi\ ,
\end{equation}
which is just equation (33.16) of Misner, Thorne, and Wheeler
\cite{mtw} in the specific context of the Kerr spacetime. Their
exercise (33.3) following the discussion of Bardeen \cite{bar} applies
to the general (orthogonally transitive) stationary axially symmetric
case \cite{greschvis}, so that one may interpret the slicing observers
as the locally nonrotating observers with respect to the Sagnac
effect. They experience no Sagnac effect for the oppositely directed
accelerated photons constrained by mirrors or fiber optical cable to
remain on a given circular orbit, meaning that the alternating meeting
points of these photons lie on the same observer world line. They are
also called the ``zero angular momentum observers" (ZAMO's) since they
are orthogonal to the angular Killing vector and therefore have
vanishing angular momentum. A complementary formula exists for the
angular coordinate component of the threading shift 1-form 
\begin{equation}
   M_\phi = (\zeta_-{}^{-1} + \zeta_+{}^{-1})/2 \ .
\end{equation}
This is related to the Sagnac effect as well, as explained below.

The physical components of the velocities measured by the threading
and slicing observers for such motion are related to the coordinate
angular velocity by linear or fractional linear transformations
\begin{eqnarray}\label{eq:nuUmn}
  \nu(U,m)^{\hat\phi} 
     &=& \gamma_{\phi\phi}{}^{1/2} \zeta / [M(1-M_\phi \zeta)] \ ,
  \quad
  \nu(U,n)^{\hat\phi} 
     = g_{\phi\phi}{}^{1/2} (\zeta + N^{\phi})/N
\end{eqnarray}
and
\begin{eqnarray}\label{eq:zetanuUmn}
   \zeta 
  &=& M \nu(U,m)^{\hat\phi}/ 
      [\gamma_{\phi\phi}{}^{1/2} + M M_\phi \nu(U,m)^{\hat\phi}]
              \nonumber\\
  &=& -N^\phi + N g_{\phi\phi}{}^{-1/2}  \nu(U,n)^{\hat\phi} \ .
\end{eqnarray}
Note that when the shift is nonzero, test particle motions with
angular velocities of equal magnitude but opposite sign lead to
physical velocities which do not have the same magnitude and vice
versa. When $\nu(U,u)^{\hat \phi} = \pm 1$, the latter equation
reduces to Eq.~(\ref{eq:zetapm}). 

The ``coordinate" gamma factor is easily expressed in terms of the
usual Lorentz gamma factor associated with these relative velocities 
\begin{eqnarray}\label{eq:gammagamma}
     \Gamma &=& \gamma(U,m)/ [ M(1-M_\phi \zeta)] \equiv \Gamma(U,m) 
                  \nonumber\\
            &=& \gamma(U,n)/ N \equiv \Gamma(U,n) \ .
\end{eqnarray}
These formulas may be used to express the angular momentum (per unit
mass) 
\begin{equation}\label{eq:angmom}
  p_\phi = U_\phi 
  = \Gamma\nu(U,n)_\phi 
  = g_{\phi\phi} \Gamma(\zeta-\zeta_{\rm(sli)})
\end{equation}
of $U$ defined by the rotational Killing vector $\delta^\alpha{}_\phi$
and its Killing energy (per unit mass) ${\cal E} = -U_t =
M^{-1}\gamma(U,m)$ defined by the Killing vector $\delta^\alpha{}_t$,
both conserved for geodesic motion. These are related to the
coordinate gamma factor by the identity $-1=U_\alpha U^\alpha = \Gamma
(-{\cal E} +\zeta p_\phi)$ in the timelike case and $0 =P_\alpha
P^\alpha = \Gamma_{\rm(null)} (-{\cal E} +\zeta P_\phi)$ in the null
case, where ${\cal E}=-P_t$. Note that the slicing relative velocity
is directly proportional to the angular momentum. 

In the timelike case the ratio
\begin{equation}\label{eq:barU}
  \bar\zeta = \frac{{\cal E}}{p_\phi}
            = -\frac{g_{tt}+\zeta g_{\phi t}}
                    {g_{t\phi}+\zeta g_{\phi\phi}}
\end{equation}
defines the angular velocity of the spacelike circular orbit
orthogonal to $U^\alpha$ with unit tangent $\bar U{}^\alpha
=\bar\Gamma(\delta^\alpha{}_t +\bar\zeta \delta^\alpha{}_\phi)$,
$\bar\Gamma>0$ and having the same sense of rotation. This is the
angular direction of the local rest space of the test particle.

Consider only a nonzero rest mass test particle in what follows. The
various spatial forces acting on such a particle all point along the
radial direction whether expressed in the threading or in the slicing
observer-adapted frame. In the ``spatial equation of motion'' (9.9) of
[BCJ1], the Lie total spatial covariant derivative of the spatial
momentum reduces to minus the space curvature force (see equations
(12.7) and (12.23) of [BCJ1]) and one finds the following simple
results for the equation of motion in the threading, hypersurface, and
slicing points of view respectively 
\begin{eqnarray}\label{eq:FFF}
    -F(U,u)^{\hat{r}}
         &=& F^{\rm(G)}_{\rm(lie)}(U,u)^{\hat{r}}
           + F^{\rm(SC)}(U,u)^{\hat{r}} \ ,
	\qquad u=m,n \ , \nonumber\\
    -F(U,n)^{\hat{r}} 
         &=& F^{\rm(G)}_{\rm(lie)}(U,n,e_0)^{\hat{r}}
           + F^{\rm(SC)}(U,n,e_0)^{\hat{r}}
         \ ,
\end{eqnarray}
namely, minus the relative non-gravitational spatial force must
balance the sum of the Lie spatial gravitational force and the space
curvature force. The spatial gravitational forces in the various
points of view can be separated into their gravitoelectric (GE),
vector gravitomagnetic (GM), and symmetric tensor gravitomagnetic
expansion (EX) components 
\begin{eqnarray}\label{eq:FG}  
  F^{\rm(G)}_{\rm(lie)}(U,m)^{\hat{r}} 
   &=& \gamma(U,m)
         [g(m)^{\hat{r}}+\nu(U,m)^{\hat{\phi}}H(m)^{\hat{z}}]
                         \nonumber\\
   &=& F^{\rm(GE)}(U,m)^{\hat{r}} + F^{\rm(GM)}(U,m)^{\hat{r}} \ ,
                         \nonumber\\
  F^{\rm(G)}_{\rm(lie)}(U,n)^{\hat{r}} 
   &=& \gamma(U,n)[g(n)^{\hat{r}}-
         2\nu(U,n)^{\hat{\phi}}\theta(n)_{\hat{r}\hat{\phi}}]
                         \\
   &=& F^{\rm(GE)}(U,n)^{\hat{r}} + F^{\rm(EX)}(U,n)^{\hat{r}} \ , 
                         \nonumber\\
  F^{\rm(G)}_{\rm(lie)}(U,n,e_0)^{\hat{r}} 
   &=& \gamma(U,n)[g(n)^{\hat{r}}
   + {\textstyle {1\over2}} \nu(U,n)^{\hat{\phi}}H(n,e_0)^{\hat{z}}
   - \nu(U,n)^{\hat{\phi}}\theta(n)_{\hat{r}\hat{\phi}}] 
                          \nonumber\\
   &=& F^{\rm(GE)}(U,n)^{\hat{r}} + F^{\rm(GM)}(U,n,e_0)^{\hat{r}} 
   + F^{\rm(EX)}(U,n,e_0)^{\hat{r}}\ ,\nonumber
\end{eqnarray}
where the sum of terms defines respectively the individually named
force terms. The space curvature forces reduce simply to the
sign-reversal of $\gamma(U,u)$ times the Lie relative centripetal
acceleration 
\begin{eqnarray}\label{eq:FSC}
  F^{\rm(SC)}(U,m)^{\hat{r}}
        &=& - \kappa(\phi,m)^{\hat{r}} 
                 \gamma(U,m)|\nu(U,m)^{\hat{\phi}}|^2
          \ , \nonumber\\
  F^{\rm(SC)}(U,n)^{\hat{r}}
        &=& - \kappa(\phi,n)^{\hat{r}} 
                 \gamma(U,n) |\nu(U,n)^{\hat{\phi}}|^2 
          \ , \\
  F^{\rm(SC)}(U,n,e_0)^{\hat{r}}
        &=& - \kappa(\phi,n)^{\hat{r}} 
                 \gamma(U,n) [\nu(U,n)^{\hat{\phi}}]
      	  [\nu(U,n)^{\hat{\phi}} - \nu(e_0,n)^{\hat{\phi}}]
	\ ,\nonumber 
\end{eqnarray}
where
\begin{eqnarray}\label{eq:kappa}
   \kappa(\phi,m)^{\hat{r}} 
      &=& -(\ln \gamma_{\phi\phi}{}^{1/2})_{,\hat{r}}\ ,
 \qquad
   \kappa(\phi,n)^{\hat{r}} 
      = -(\ln g_{\phi\phi}{}^{1/2})_{,\hat{r}}
\end{eqnarray}
are the signed Lie relative curvatures of the $\phi$
coordinate lines in the threading and hypersurface points of view and 
\begin{equation}
       \nu(e_0,n)^{\hat{\phi}}
         = N^{-1}N^{\hat{\phi}} 
         = (g_{\phi\phi})^{1/2}N^{-1}N^{\phi} 
\end{equation}
is the physical component $\nu(m,n)^{\hat\phi}$ along the positive
$\phi$ direction of the relative velocity of the threading observers
with respect to the slicing observers when $e_0{}^\alpha$ is timelike,
and of the relative velocity of the time coordinate lines in general.
The caret index notation indicates physical (orthonormal) components
along the orthogonal coordinate frame vectors in the slicing point of
view and along their corresponding spatially projected vectors of the
associated orthogonal observer-adapted spatial frame in the threading
point of view. 

Finally, the nonzero components of the gravitoelectric and
gravitomagnetic vector fields together with the expansion can be
calculated from their expressions in the observer adapted frames (see
equations (14.5) and (14.6) of [BCJ1]) 
\begin{eqnarray}
 g(m)_r &=& (\gamma_{rr})^{1/2}g(m)_{\hat{r}} = - (\ln M)_{,r} \ ,
                 \nonumber\\
 g(n)_r &=& (g_{rr})^{1/2}g(n)_{\hat{r}} = - (\ln N)_{,r} \ ,
                 \nonumber\\
 H(m)^z &=& H(m)^{\hat{z}} 
         = M (\gamma_{rr}\gamma_{\phi\phi})^{-1/2} M_{\phi,r} \ ,
                 \nonumber\\
 H(n,e_0)^z &=& H(n,e_0)^{\hat{z}} 
         = N^{-1} (g_{rr}g_{\phi\phi})^{-1/2} N_{\phi,r} \ ,
                 \nonumber\\
 \theta(n)_{r\phi} &=& (g_{\phi\phi}g_{rr})^{1/2}
                             \theta(n)_{\hat{r}\hat{\phi}} 
          = -(1/2)g_{\phi\phi}N^{-1}N^\phi{}_{,r}
            \ .
\end{eqnarray}
Explicit expressions for all of these quantities in the various points
of view are given in Tables \ref{tab:summaryMG} and \ref{tab:summaryK}
for each of the three spacetimes under consideration. 

\typeout{*** Table 2. (summary)}
\begin{table}[ht]\footnotesize 
\tcaption
{Causal restrictions, spatial gravitational force field
quantities, and geodesic and null conditions for circular orbits
in the rotating Minkowski and G\"odel spacetimes.
} 
 \typeout{*** eqnarray struts inserted}
 \def\Strut{\relax\hbox{\vrule width0pt height 10.5pt depth 5.5pt}}
 \def\TopStrut{\relax\hbox{\vrule width0pt height 10.5pt depth 0pt}}
 \def\BotStrut{\relax\hbox{\vrule width0pt height 0pt depth 5.5pt}}
\begin{eqnarray*}
\begin{array}{||l|l|l||} \hline 
 & \hbox{Rotating} \qquad & {\hbox{G\"odel}} 
 \\
 & \hbox{Minkowski} \qquad  &           
 \\ 
\hline\hline \Strut
 \hbox{length scale ${\cal R}$}\qquad & \Omega^{-1} 
                                   & |\sqrt{2}\Omega|^{-1} 
 \\
\hline \TopStrut
 \hbox{threading region}\qquad & r< r_{\rm(h)}=\Omega^{-1} 
                     & {\rm everywhere}   
 \\
 \hbox{of validity} &&
 \\
\hline \TopStrut
 \hbox{slicing region}\qquad   & {\rm everywhere} 
        & r< r_{\rm(h)}= 
                      {\cal R} \,{\mathop{\rm arcsinh}\nolimits}  1
 \\
 \hbox{of validity} && 
 \\
\hline \Strut
 \nu(m,n)^{\hat{\phi}}=-\nu(n,m)^{\hat{\phi}} & \Omega r 
         & \sqrt{2}\,{\rm t} 
 \\
\hline
 \hbox{gravitoelectric}& g(m)^{\hat{r}} =\gamma^2\Omega^2 r >0  
	& g(m)^{\hat{r}}=0 
 \\ 
{\rm field}& g(n)^{\hat{r}}= 0 
   & g(n)^{\hat{r}}= -{\sqrt{2}\Omega\,{\rm t}\over{1-{\rm s}^2}}< 0
 \\
\hline \Strut 
 \hbox{gravitomagnetic} & H(m)^{\hat{z}}=2\Omega\gamma^2 >0 
	& H(m)^{\hat{z}}=2\Omega >0 
 \\ \Strut
 \hbox{field} &H(n,e_0)^{\hat{z}}=2\Omega 
        &H(n,e_0)^{\hat{z}}=2\Omega >0 
 \\
\hline \Strut
 \hbox{expansion} & \theta(n)_{\hat{r}\hat{\phi}} = 0
   &\theta(n)_{\hat{r}\hat{\phi}}
    = -\frac{\Omega{\rm s}^2}{1-{\rm s}^2} <0
 \\
\hline \Strut
 \hbox{signed relative}
    & \kappa(\phi,m)^{\hat{r}}= -\frac{\gamma^2}{r} <0 
    &\kappa(\phi,m)^{\hat{r}}= - \sqrt{2}\Omega\,{\rm T}^{-1}
 \\ \Strut
 \hbox{curvature}& \kappa(\phi,n)^{\hat{r}}= -\frac{1}{r} <0
    & \kappa(\phi,n)^{\hat{r}}
  =
    - \frac{\sqrt{2}\Omega (1-2{\rm s}^2)}{2\,{\rm t}(1-{\rm s}^2)} 
 \\
\hline 
 \hbox{counter-rotating}& \dot{\phi}_{-} = -\Omega 
    & \dot{\phi}_\pm = -\frac{2\Omega}{1-2{\rm s}^2}\ , 0
 \\
 \hbox{and co-rotating}& \nu(U_-,m)^{\hat{\phi}} = -\Omega r 
    &\nu(U_\pm,m)^{\hat{\phi}} = -\sqrt{2}\,{\rm T}\ , 0
 \\
 \hbox{timelike geodesics}&\nu(U_-,n)^{\hat{\phi}} = 0
  &\nu(U_\pm,n)^{\hat{\phi}} 
         = -\frac{\sqrt{2}\,{\rm t}}{1-2{\rm s}^2} \ ,
		\sqrt{2}\,{\rm t}  
 \\
\hline \Strut
 \hbox{null orbits}& \zeta_\pm = -\Omega \pm 1/r 
    & \zeta_\pm =
	\frac{\Omega}{1-{\rm s}^2}[ -1 \pm t^{-1}/\sqrt{2}] 
 \\
\hline
\end{array}
\end{eqnarray*}	
\label{tab:summaryMG}
\end{table}

\typeout{*** Table 3.}

\begin{table}[htb]\footnotesize 
\tcaption
{Causal restrictions, spatial gravitational force field
quantities, and geodesic and null conditions for circular orbits
in the Kerr spacetime.
} 
 \typeout{*** eqnarray struts inserted}
 \def\Strut{\relax\hbox{\vrule width0pt height 10.5pt depth 5.5pt}}
 \def\TopStrut{\relax\hbox{\vrule width0pt height 10.5pt depth 0pt}}
 \def\BotStrut{\relax\hbox{\vrule width0pt height 0pt depth 5.5pt}}
\begin{eqnarray*}
\begin{array}{||l|l||} \hline 
 & {\rm Kerr} \\
\hline\hline \Strut
 \hbox{length scale ${\cal R}$}\qquad 
  &{\cal M}\\
\hline \TopStrut
 \hbox{threading region}\qquad 
  &r> r_{\rm(erg)}=2{\cal M}\\
 \hbox{of validity} &\\ 
\hline \TopStrut
 \hbox{slicing region}\qquad   
  & r> r_{\rm(h)}={\cal M} +\sqrt{{\cal M}^2 - a^2} \\ \BotStrut
 \hbox{of validity} &\\ 
\hline \Strut
 \nu(m,n)^{\hat{\phi}}=-\nu(n,m)^{\hat{\phi}} 
	 & - \frac{2a{\cal M}}{r\sqrt{\Delta}} \\ 
\hline \Strut
 \hbox{gravitoelectric} 
 &
 g(m)^{\hat{r}}= -{{\cal M}\sqrt{\Delta}\over{r^2(r-2{\cal M})}} <0
        \\ \Strut
{\rm field} 
   & g(n)^{\hat{r}}= -\frac{{\cal M} [(r^2+a^2)^2-4a^2{\cal M}
	r]}{r^2\sqrt{\Delta}(r^3+a^2r+2a^2{\cal M} )} < 0 \\ 
\hline \Strut
 \hbox{gravitomagnetic} 
	& H(m)^{\hat{z}}=\frac{2a{\cal M}}{r^2(r-2{\cal M})} >0
        \\ \Strut
 \hbox{field} 
	&H(n,e_0)^{\hat{z}}=\frac{2a{\cal M}}{r^3} >0 \\ 
\hline \Strut
 \hbox{expansion} 
   &\theta(n)_{\hat{r}\hat{\phi}}
    = -\frac{a{\cal M} (3r^2+a^2)}{r^2(r^3+a^2r+2a^2{\cal M} )}<0 
	\\ 
\hline \Strut
 \hbox{signed relative}
    &\kappa(\phi,m)^{\hat{r}}= - \frac{r(r-2{\cal M} )^2-{\cal M}
		a^2}{r^2\sqrt{\Delta}(r-2{\cal M} )}
       \\ \Strut
 \hbox{curvature} 
    & \kappa(\phi,n)^{\hat{r}} 
           = - \frac{\sqrt{\Delta}(r^3-a^2{\cal M})}
                                  {r^2(r^3+a^2r+2a^2{\cal M} )} \\
\hline \Strut
 \hbox{counter-rotating} 
    &\dot{\phi}_\pm = \frac{\pm\sqrt{{\cal M}/r^3}}
                                  {1\pm a\sqrt{{\cal M}/r^3}}
       \\ \Strut
 \hbox{and co-rotating} 
    &\nu(U_\pm,m)^{\hat{\phi}} 
      = \frac{\sqrt{\Delta}}{a\pm (r-2{\cal M})\sqrt{r/{\cal M}}}
  	\\ \Strut
 \hbox{timelike geodesics} 
  &\nu(U_\pm,n)^{\hat{\phi}} = \frac{a^2 \mp 2a\sqrt{{\cal M} r} 
         + r^2}{\sqrt{\Delta}(a\pm r\sqrt{r/{\cal M}})}  \\
\hline \Strut
 \hbox{null orbits} 
    &\zeta_\pm = \frac{2a{\cal M} \pm r\sqrt{\Delta}}
                               {r^3 + a^2 r + 2a^2 {\cal M}} \\
\hline
\end{array}
\end{eqnarray*}	
\label{tab:summaryK}
\end{table}

The threading point of view is valid for those points of spacetime
where the threading is timelike, i.e., where ${}^{(4)}\kern-1pt 
g_{tt} < 0$ (or $M>0$), while the hypersurface/slicing points of view
are valid where the slicing is spacelike, or equivalently where the
normal to the slicing is timelike, i.e., where ${}^{(4)}\kern-1pt 
g^{tt}<0$ (or $N>0$). If the threading observers become spacelike in a
region where the slicing observers are still timelike, then the
threading point of view is still valid where the magnitude of the
relative velocity of the threading observers with respect to the
slicing observers is $|\nu(m,n)^{\hat{\phi}}|=|N^{-1}N^{\hat{\phi}}| <
1$. Vice versa, if the slicing observers become spacelike in a region
where the threading observers are still timelike, then the slicing
point of view is still valid in the region where
$|\nu(n,m)^{\hat{\phi}}| < 1$. The regions of validity of the
threading and slicing  points of view are given in Tables
\ref{tab:summaryMG} and \ref{tab:summaryK}.

For both the rotating Minkowski and G\"odel spacetimes, the magnitude
of the relative velocity is an increasing function of $r$ which  leads
to the existence of an outer light cylinder at $r=r_{\rm(h)}$ (\lq\lq
h" for observer \lq\lq horizon") where the worldlines of one of the
families of test observers become null and the corresponding point of
view is no longer valid. In the rotating Minkowski case the light
cylinder exists only in the threading point of view and it occurs at
the radius for which $|\nu(m,n)^{\hat{\phi}}|=1$, while in the G\"odel
case it exists only in the slicing point of view where it occurs when
$|\nu(n,m)^{\hat{\phi}}|=1$. However, beyond that horizon
$g_{\phi\phi}$ becomes negative, leading to the famous closed timelike
$\phi$ coordinate lines. In the Kerr case where discussion is confined
to the equatorial plane, the situation is reversed and the relative
velocity is a decreasing function of $r$ leading to an inner light
radius. The slicing point of view is valid outside the event horizon
which occurs at the value of $r$ for which $\Delta=0$, while the
threading point of view is valid for $r> r_{\rm(erg)}$ outside the
ergosphere (which in turn surrounds the event horizon), where
$r_{\rm(erg)}$ is determined by the condition
$|\nu(m,n)^{\hat{\phi}}|=1$. The G\"odel slicing observers attempt to
resist the global rotation of the spacetime, but are forced to
corotate at the outer observer horizon; similarly the threading and
slicing observers in Kerr are forced to corotate at their respective
inner observer horizons, namely the ergosphere and the event horizon. 

The gravitoelectric field is the sign-reversed acceleration of the
observer congruence, thus revealing the accelerations which
characterize the threading and slicing observers. For example, in the
rotating Minkowski spacetime the threading observers are accelerated
radially inward while the counter-rotating slicing observers (at rest
in a global inertial frame) have zero acceleration. In the G\"odel
spacetime the situation is reversed and the threading observers are
not accelerated while the counter-rotating slicing observers are
accelerated radially outward. In both cases an outward acceleration
must be added to resist the global rotation of spacetime (more
precisely, of the nonlinear reference frame) by counter-rotating. The
equatorial Kerr slicing and threading observers are both accelerated
radially outward to oppose the attraction of the central mass, leading
to inward gravitoelectric fields which allow circular orbits even in
the nonrotating Schwarzschild limit.  In the Kerr case the threading
observers counter-rotate with respect to the slicing observers and one
finds that their acceleration is larger then the acceleration of the
slicing observers. 
 
In each case considered here the gravitomagnetic fields are along the
positive $z$-direction. In both the rotating Minkowski and G\"odel
spacetime this is due to the fact that the shift 1-forms are along the
positive $\phi$ direction and their physical components along that
direction are increasing functions of $r$, while in Kerr spacetime the
same 1-form physical component is a decreasing function of $r$ but its
sign is reversed. Thus in each case the sign of the radial
gravitomagnetic force component depends only on the sign of the
relative velocity of the test particle along the angular direction,
namely a positive (outward) force for co-rotating orbits and a
negative (inward) force for counter-rotating orbits, where here the
terms corotating and counter-rotating are with respect to the given
observer family. Note also that in the G\"odel spacetime the threading
and slicing gravitomagnetic fields are the same and uniform (spatially
covariant constant). That they are equal can also be seen directly
from the transformation law for the gravitomagnetic vector field given
in equation (11.6) of \Ref\citen{mfg} and using the fact that the 
threading
gravitoelectric field and the Lie derivative along $e_0{}^\alpha$ of
the shift are zero and that the relative projection $P(m,n)^{-1}$
reduces to the identity along the radial direction orthogonal to the
plane of the relative motion of the two observers. As discussed in
[BCJ1], the expansion tensor, zero in the threading point of view, has
one possibly nonvanishing $r$-$\phi$ component which is zero (rotating
Minkowski) or negative (G\"odel and Kerr) in the hypersurface and
slicing points of view. Like the radial gravitomagnetic vector force,
the radial expansion force is also positive (outward) for co-rotating
orbits and negative (inward) for counter-rotating orbits, again with
the sense of rotation referred here to the observer family. The sum of
the gravitomagnetic vector force and the expansion force equals the
total gravitomagnetic tensor force as discussed in [BCJ1]. 

\begin{figure}[htbp]\footnotesize
\centerline{\epsfbox{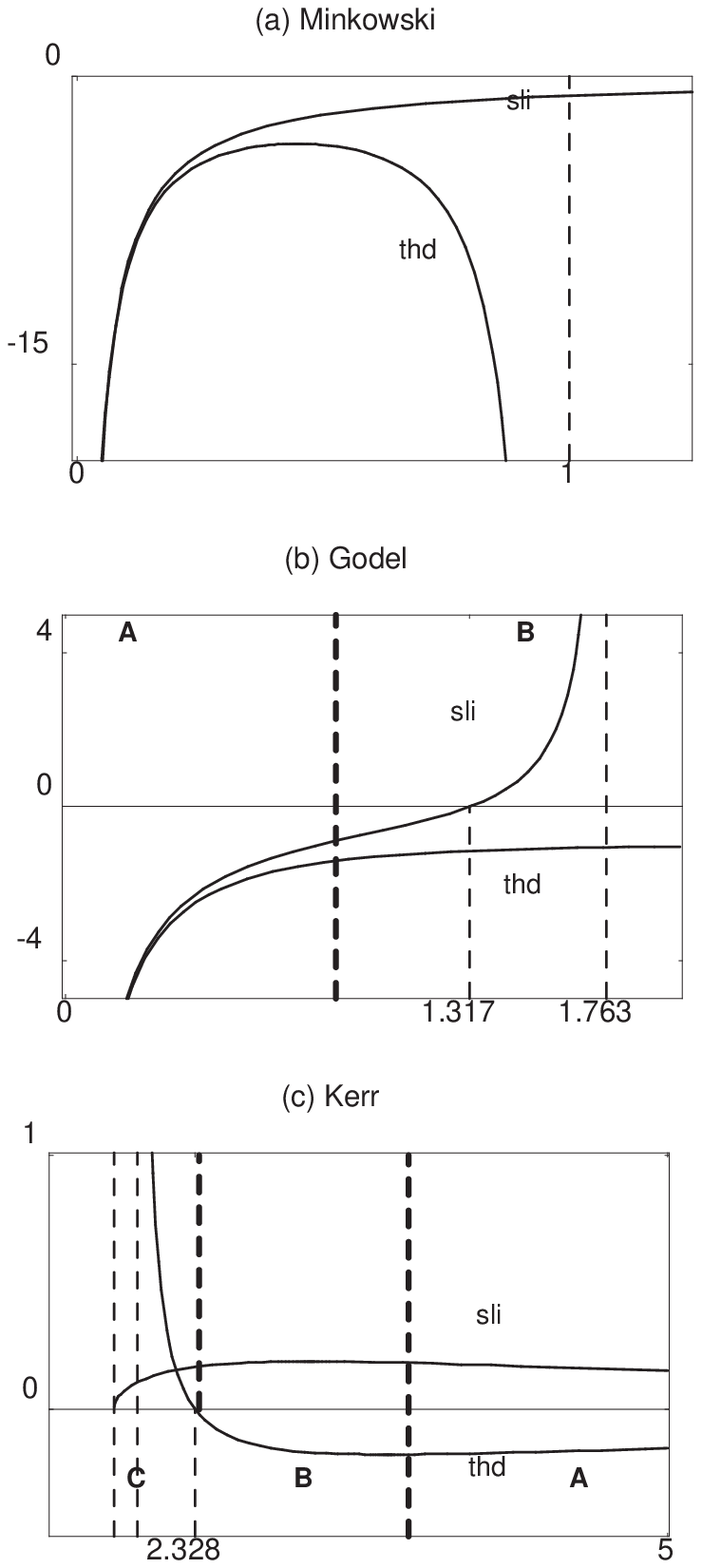}}
\fcaption
{$\kappa$ versus $\bar r$: plots of the signed Lie relative curvature
$\kappa(\phi,u)^{\hat r}$ of the circular orbits in the equatorial
plane of the rotating Minkowski, G\"odel, and $a/{\cal M}=1/2$ Kerr
spacetimes in the threading ($u=m$) and hypersurface ($u=n$) points of
view as functions of $\bar r = r/{\cal R}$ (Minkowski, G\"odel) or
$\bar r = r/{\cal M}$ (Kerr). In this particular Kerr case the
relatively straight circle occurs just inside the region C (which
begins at $\bar r \approx 2.35$) described below. 
}
\label{fig:mgk3}
\end{figure}

On the other hand, the signed relative curvatures can change sign
within the range of validity of one point of view yielding radial
centripetal accelerations (sign-reversal of the space curvature
forces) that can be either inward ($\kappa(\phi,u)^{\hat r}<0$, the
usual case) or outward ($\kappa(\phi,u)^{\hat r}>0$). Figure
\ref{fig:mgk3} shows $\kappa(\phi,u)^{\hat r}$ for the various cases
in the threading and hypersurface points of view. The local extrema of
$g_{\phi\phi}$ and $\gamma_{\phi\phi}$ are the points where the
relative Lie centripetal acceleration changes sign and the relative
curvature of the spatial trajectory vanishes, yielding the Lie
relatively straight trajectories which are also spatial geodesics in
each point of view. 

In Figure \ref{fig:mgk3} and successive figures, the G\"odel and Kerr
diagrams are divided into regions A, B, and C by thick vertical dashed
lines according to whether 2, 1, or 0 of the two oppositely directed
circular geodesics are timelike. The boundaries of these regions, as
discussed in detail in section 4, occur at the radii of the two null
circular geodesics. Thin vertical dashed lines also mark the horizon
and outer ergosphere radii in the Kerr case. 

\typeout{*** Figure 1. (mgk3)}

\typeout{*** (Full page figure on page 12.)}

The behavior of the Lie relative curvature depends on the intrinsic
geometry of the Riemannian metric on the quotient space appropriate to
each point of view. In the threading case this metric is just the
natural projection of the spatial metric to the observer-quotient
space, while in the slicing point of view, it is instead the pullback
of the spatial metric to the quotient by the threading congruence. For
the case of planar orbits, 2-dimensional embedding diagrams in either
3-dimensional Euclidean space $E_3$ or 3-dimensional Minkowski
spacetime $M_3$ as appropriate are useful to interpret the effects of
this spatial geometry on the orbits in this plane, as well as on the
precession of the spin of a gyroscope following such an orbit. Thorne
\cite{tho81} has given an neat illustration of this latter effect in
terms of the conical defect of the tangent cone in the embedding
space. The details of the embedding are explained in the appendix,
together with the actual diagrams of the cross-sectional curves of
these surfaces of revolution. 

\section{Circular geodesics} \label{sec:cirgeo}

The timelike circular geodesics are the circular orbits along which
the relative spatial force $F(U,u)^{\hat{r}}$ vanishes. The relative
velocities corresponding to these geodesics can then be found by
setting to zero the sum of the Lie spatial gravitational force and the
space curvature force,  leading to a quadratic equation in those
velocities or in the coordinate angular velocities (linear when
$\kappa(\phi,u)^{\hat r}=0$). The regions where the geodesics are
timelike are those regions where the corresponding spatial relative
speeds are smaller than 1 (provided that the observers are also
timelike). Let $U_\pm^\alpha$, $\nu(U_\pm,u)^\alpha$, and
$\dot\phi_\pm$ be the 4-velocity, relative velocity, and coordinate
angular velocity of the corotating ($+$) and counter-rotating ($-$)
geodesics in each spacetime, when they exist. 

For example, in the rotating Minkowski case there are only
counter-rotating circular geodesics, which correspond to the points
fixed in the global inertial frame with respect to which the rotating
nonlinear reference frame rotates, namely the orbits of the slicing
observers ($\nu(U_-,n)^{\hat{\phi}} = 0$ in 
Tables \ref{tab:summaryMG} and \ref{tab:summaryK}.
These worldlines are clearly timelike everywhere. Their spatial
velocity relative to the threading observers becomes larger than the
velocity of light when $\Omega r > 1$ only because the threading
observer congruence becomes spacelike there. In the G\"odel spacetime
the co-rotating geodesics are instead the orbits of the threading
observers ($\nu(U_+,m)^{\hat{\phi}}=0$). These worldlines are then
timelike in the region of validity of the threading point of view,
i.e.,  everywhere. On the other hand the counter-rotating geodesics
are timelike as long as $|\nu(U_-,m)^{\hat{\phi}}| < 1$ or
$|\nu(U_-,n)^{\hat{\phi}}| < 1$ which gives $r< {\textstyle {1\over2}}
r_{\rm(h)}$. In the Kerr spacetime the equations
$|\nu(U_\mp,n)^{\hat{\phi}}| =1$ and $|\nu(U_\mp,m)^{\hat{\phi}}| =1$
have roots (valid only within the region of validity of each point of
view) at $r(r-3{\cal M})=\pm 2 a {\cal M}\sqrt{r/{\cal M}}$.  In the
extreme Kerr case $a={\cal M}$, these roots are $r={\cal M}$ for the
co-rotating geodesics and $r=4{\cal M}$ for the counter-rotating
geodesics. 

Tables \ref{tab:summaryMG} and \ref{tab:summaryK}
show that for the G\"odel and Kerr spacetimes which
have two oppositely directed circular geodesics, both the magnitude of the
coordinate angular velocity $\zeta$ and of the angular component of the
velocity $\nu(U,u)^{\hat\phi}$ are larger for geodesics in the counter-rotating
direction than in the corotating direction in both the threading and slicing
points of view. This immediately implies that the geodesic gamma factor
$\gamma(U_\pm,u)$ and with some additional reasoning the coordinate gamma
factor $\Gamma(U_\pm,u)$ are also both larger in the counter-rotating direction
than in the corotating direction. The latter follows from the former using the
slicing representation of the coordinate gamma factor given in
Eq.~(\ref{eq:gammagamma}). This general counter-rotation effect manifested in
this asymmetry between the co-rotating and counter-rotating directions is the
direct consequence of an upward gravitomagnetic field which adds an inward
radial force to the total sum for the counter-rotating case increasing the
speed, and an outward radial force to the total sum for the corotating case
decreasing the speed. 

The asymmetry in the coordinate angular velocity between the
corotating and the counter-rotating circular geodesics is the origin
of the precession of their alternating meeting points (after a full
revolution, not half a revolution) in the counter-rotating direction.
Analogous to the zero Sagnac effect (slicing) observers which follow
Killing trajectories containing the meeting points of oppositely
directed circular null paths, one can introduce ``geodesic meeting
point observers" (only in the equatorial plane in Kerr \cite{mit})
containing the meeting points of the oppositely directed circular
geodesics. Their angular velocity is analogously the average (see
Tables \ref{tab:summaryMG} and \ref{tab:summaryK}) 
\begin{equation}
         \zeta_{\rm(gmp)} = (\dot\phi_- + \dot\phi_+)/2 \ .
\end{equation}
For Kerr this has the negative value $\zeta_{\rm(gmp)} = - a{\cal M}
r^{-3}/(1-a^2{\cal M} r^{-3})$, and for G\"odel it is also negative
$\zeta_{\rm(gmp)}= \dot\phi_{-}/2$ since $\dot\phi_{+}=0$. Using
Eqs~(\ref{eq:nuUmn}), (\ref{eq:zetanuUmn}), one can show that the
slicing velocities are also related by averaging 
\begin{equation}
    \nu(U_{\rm(gmp)},n)^{\hat\phi}
      =  [\nu(U_-,n)^{\hat\phi} + \nu(U_+,n)^{\hat\phi}]/2 \ ,
\end{equation}
as is trivially the case for the slicing velocities of the null
meeting point observers and the oppositely directed null paths. 

\begin{figure}[htbp]\footnotesize
\centerline{\epsfbox{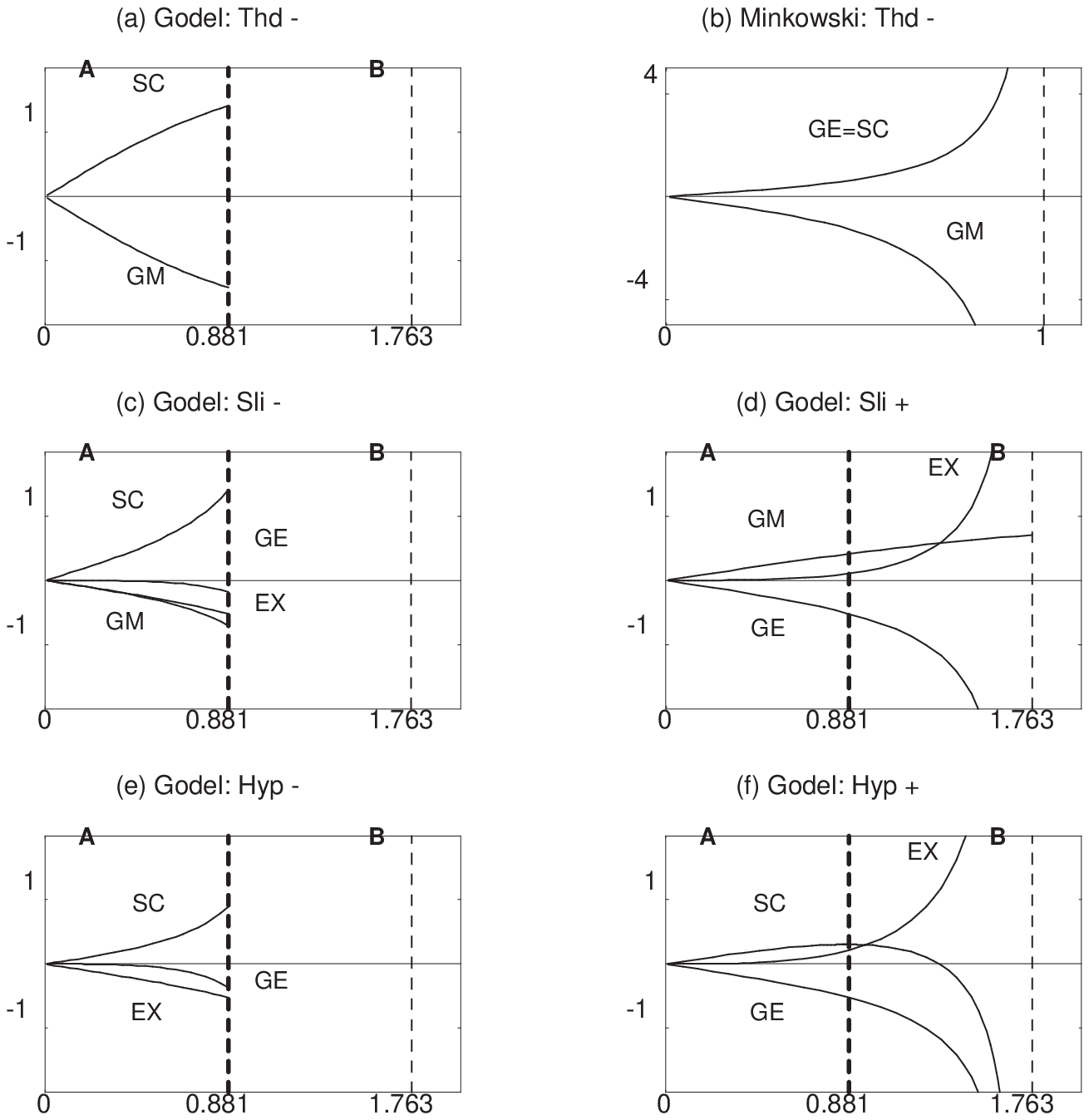}}
\fcaption
{$F$ versus $\bar r$: plots of the space curvature (SC),
gravitomagnetic (GM), expansion (EX), and gravitoelectric (GE) spatial
radial forces (divided by a common $\gamma(U,u)$ factor and multiplied
by ${\cal R}$) along the corotating $(+)$ and counter-rotating $(-)$
circular geodesics in the rotating Minkowski and G\"odel spacetimes in
the threading (thd), hypersurface (hyp) and slicing (sli) points of
view, plotted versus the rescaled radial coordinate $\bar{r}$. The
horizontal ranges for the G\"odel counter-rotating/corotating cases
are respectively $r\in[0,r_{\rm(h)}/2]$  and $r\in[0,r_{\rm(h)}]$. The
thick dashed vertical line separates the region A near the origin
where both circular geodesics are timelike from the region B where
only the corotating geodesic is timelike. The observer horizons are
marked by a thin dashed vertical line. 
}
\label{fig:mgk1}
\end{figure}

\begin{figure}[htbp]\footnotesize
\centerline{\epsfbox{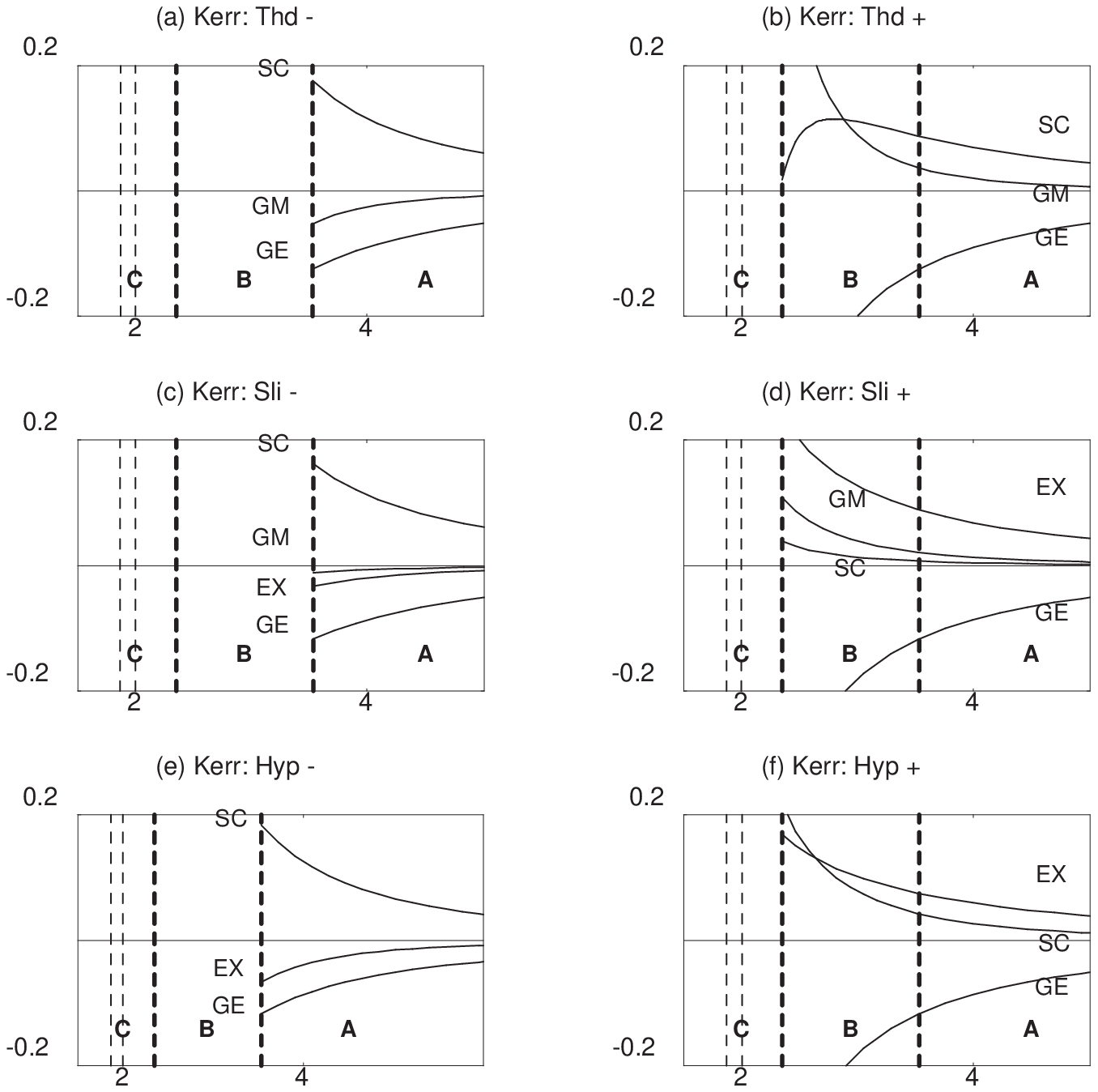}}
\fcaption
{$F$ versus $\bar r$: plots of the space curvature (SC),
gravitomagnetic (GM), expansion (EX) and gravitoelectric (GE) spatial
radial forces (multiplied by ${\cal M}/\gamma(U,u)$) along the 
corotating $(+)$ and counter-rotating $(-)$ equatorial circular
geodesics around a  Kerr black hole with $a/{\cal M}=0.5$, in the
threading (thd), hypersurface (hyp), and slicing (sli) points of view
as a function of $\bar r = r/{\cal M}$. The thick dashed vertical
lines at $\bar r \approx 2.35, 3.53$ separate the three regions A
which extends out to infinity, C which extends inward to the horizon
($\bar r_{\rm(hor)} \approx 1.87$) and B which is sandwiched in
between them and  which contains the ergosphere boundary ($\bar
r_{\rm(erg)} = 2$). The observer horizons are marked by thin dashed
vertical lines. 
}
\label{fig:mgk2}
\end{figure}

The spatial equation of motion evaluated along the circular geodesics
with zero total spatial force describes how the various radial spatial
forces balance. In the threading point of view the expansion is zero
because of the stationary symmetry so only the gravitoelectric, the
vector gravitomagnetic, and the space curvature forces are present. 
In the hypersurface point of view the gravitomagnetic field vanishes
and one is left with only the vector gravitoelectric, expansion, and
space curvature forces. In the slicing point of view all the forces
contribute to the radial force balance equation.  Figures
\ref{fig:mgk1} and \ref{fig:mgk2} show the plots of the various radial
forces (divided by the common factor $\gamma(U,u)$) evaluated along
the circular geodesics as functions of the radial coordinate $r$ in
the several points of view and for the various cases. To better
visualize the results, some suggestive 3-dimensional diagrams showing
the balance of the various forces in space may be found in Carini,
Bini and Jantzen \cite{korea91}, with relative magnitudes of the
various forces given in the limit of small rotational speeds, i.e.,
near the axis of symmetry in the rotating Minkowski and G\"odel cases,
and far from the black hole in the Kerr case. We now discuss the force
balance for each of the three spacetimes under consideration. 

\typeout{*** Figure 2. (mgk1)}

\typeout{*** (Full page figure on page 15.)}

\typeout{*** Figure 3. (mgk2)}

\typeout{*** (Full page figure on page 16.)}

\subsection{Rotating Minkowski spacetime}

In the rotating Minkowski spacetime the radial force equation reduces
respectively in the threading, hypersurface, and slicing points of
view to 
\begin{eqnarray} 
   -F(U,m)^{\hat{r}}
       &=& F^{\rm(SC)}(U,m)^{\hat{r}} + F^{\rm(GM)}(U,m)^{\hat{r}} 
                                + F^{\rm(GE)}(U,m)^{\hat{r}}
                \nonumber\\
       &=& \gamma(U,m)\biggl\{\biggl[
                \gamma^2 \frac{|\nu(U,m)^{\hat{\phi}}|^2}{r}\biggr] 
	  +\biggl[2\gamma^2\nu(U,m)^{\hat{\phi}}\Omega\biggr] 
	  +\biggl[\gamma^2 \Omega^2 r\biggr]\biggr\}
                \nonumber\\
	&=& \gamma(U,m) \frac{\gamma^2}{r}(\nu(U,m)^{\hat{\phi}} 
                                              + \Omega r)^2 \ ,
                \nonumber\\
  -F(U,n)^{\hat{r}}
    &=& F^{\rm(SC)}(U,n)^{\hat{r}} 
        =\gamma(U,n)\frac{|\nu(U,n)^{\hat{\phi}}|^2}{r} \ ,
                \nonumber\\
  -F(U,n)^{\hat{r}}
      &=& F^{\rm(SC)}(U,n,e_0)^{\hat{r}} 
            + F^{\rm(GM)}(U,n,e_0)^{\hat{r}}
                \nonumber\\
      &=& \gamma(U,n)\biggl\{\biggl[
              \frac{|\nu(U,n)^{\hat{\phi}}|^2}{r} 
               - \Omega\nu(U,n)^{\hat{\phi}}\biggr] 
        + \biggl[\Omega\nu(U,n)^{\hat{\phi}}\biggr]\biggr\} \ . 
\end{eqnarray}
Table \ref{tab:minkgeo} shows the explicit expressions of the various
spatial forces evaluated along the circular geodesics where they
balance. 

\typeout{*** Table 4. (minkgeo)}
\begin{table}[htbp]\footnotesize
\tcaption
{The various spatial radial forces divided by a common factor
$\gamma(U,u)$ evaluated along the equatorial counter-rotating circular
geodesics are given as functions of $r$ in the various points of
view.
} 
 \typeout{*** eqnarray struts inserted}
 \def\Strut{\relax\hbox{\vrule width0pt height 10.5pt depth 5.5pt}}
\begin{eqnarray*}
\begin{array}{||l|l|l|l|l||} \hline
\multicolumn{5}{|c|}{\hbox{Rotating Minkowski Spacetime}} \\ 
\hline\hline \Strut
    & \gamma^{-1}F^{\rm(SC)}{}^{\hat{r}} 
    & \gamma^{-1} F^{\rm(GM)}{}^{\hat{r}}
    & \gamma^{-1}F^{\rm(EX)}{}^{\hat{r}} 
    & \gamma^{-1} F^{\rm(GE)}{}^{\hat{r}} \\ \hline \Strut
{\rm Thd_{-}}      & \gamma^2\Omega^2 r 
    & -2\gamma^2\Omega^2 r & 0 
		& \gamma^2\Omega^2 r  \\ \hline \Strut
{\rm Hyp_{-}}   & 0 & 0 & 0 & 0 \\ \hline \Strut
{\rm Sli_{-}}      & 0 & 0 & 0 & 0 \\ \hline
\end{array}
\end{eqnarray*}
\label{tab:minkgeo}
\end{table}

The geodesics are the orbits of particles at rest in the associated
global inertial frame with respect to which the rotation takes place.
In the threading point of view these orbits must counter-rotate in
order to compensate for the rotation of the threading observers and
the outward gravitoelectric force and space curvature force along them
are equal and add together to balance the inward gravitomagnetic
force. On the other hand the slicing observers are simply the fixed
observers in the nonrotating Minkowski spacetime and so they
experience no spatial gravitational field in the hypersurface point of
view. However, in cylindrical coordinates they measure a radially
outward space curvature force, which vanishes only along the circular
geodesics and it is due to the curvature of the circular $\phi$
coordinates rather than to a real curvature of the space which is
actually flat. 

In the slicing point of view there is a gravitomagnetic force and a
space curvature force but again both vanish along the circular
geodesics. The relation between the threading and the hypersurface
points of view in this case is given simply by the addition of
velocity formula 
\begin{eqnarray}
  \nu(U,n)^{\hat{\phi}}
    &=& 
  \frac{\nu(U,m)^{\hat{\phi}}+\Omega r}
       {1+\nu(U,m)^{\hat{\phi}}\Omega r} 
            \ , \nonumber\\
  \gamma(U,n) &=& \gamma \gamma(U,m)
                     (1+\nu(U,m)^{\hat{\phi}}\Omega r) \ ,
\end{eqnarray}
which just shows that the two points of view are related by a boost.

\subsection{G\"odel spacetime}

The radial force equations for the circular orbits in the G\"odel
spacetime in the threading, hypersurface, and slicing points of view
respectively are 
\begin{eqnarray}
  -F(U,m)^{\hat{r}}
    &=& F^{\rm(SC)}(U,m)^{\hat{r}} + F^{\rm(GM)}(U,m)^{\hat{r}} 
                  \nonumber\\
    &=& \gamma(U,m)\biggl\{\biggl[
       \sqrt{2}\Omega\,{\rm T}^{-1}|\nu(U,m)^{\hat{\phi}}|^2\biggr]
	 + \biggl[2\Omega\nu(U,m)^{\hat{\phi}}\biggr]\biggr\} \ ,
                  \nonumber\\
 -F(U,n)^{\hat{r}}
    &=& F^{\rm(SC)}(U,n)^{\hat{r}} + F^{\rm(EX)}(U,n)^{\hat{r}} +
		F^{\rm(GE)}(U,n)^{\hat{r}}
                  \nonumber\\
    &=& \gamma(U,n)\biggl\{\biggl[\frac{\sqrt{2}\Omega(1-2s^2)}
		{2\,{\rm t}(1-s^2)}|\nu(U,n)^{\hat{\phi}}|^2\biggr] 
                  \nonumber\\ \quad & &
     + \biggl[\frac{2\Omega s^2}{1-s^2}\nu(U,n)^{\hat{\phi}}\biggr]
     - \biggl[\frac{\sqrt{2}\Omega\,{\rm t}}{1-s^2}\biggr]\biggr\} 
            \ ,
                  \nonumber\\
 -F(U,n)^{\hat{r}}
   &=& F^{\rm(SC)}(U,n,e_0)^{\hat{r}} 
                + F^{\rm(GM)}(U,n,e_0)^{\hat{r}}
		+ F^{\rm(EX)}(U,n,e_0)^{\hat{r}}
                  \nonumber\\ && 
                + F^{\rm(GE)}(U,n)^{\hat{r}}
                  \nonumber\\
   &=& \gamma(U,n)\biggl\{\biggl[\frac{\sqrt{2}\Omega(1-2s^2)}
         {2\,{\rm t}(1-s^2)}|\nu(U,n)^{\hat{\phi}}|^2 
	 -\frac{\Omega(1-2s^2)}{1-s^2}\nu(U,n)^{\hat{\phi}}\biggr]
                  \nonumber\\
      & & + \biggl[\Omega\nu(U,n)^{\hat{\phi}}\biggr]
      +\biggl[\frac{\Omega s^2}{1-s^2}\nu(U,n)^{\hat{\phi}}\biggr]
      -\biggl[\frac{\sqrt{2}\Omega\,{\rm t}}{1-s^2}\biggr]\biggr\}
         \ .
\end{eqnarray}
Table \ref{tab:godgeo} shows the values of the various spatial forces
evaluated along the two sets of geodesics where the forces balance. 

\typeout{*** Table 5. (godgeo)}
\begin{table}[htbp]\footnotesize
\tcaption
{The various spatial radial forces divided by a common factor
$\gamma(U,u)$ evaluated along the counter-rotating (subscript $-$) and
corotating (subscript $+$) circular geodesics are given as functions
of $r$ in various points of view.
} 
 \typeout{*** eqnarray struts inserted}
 \def\Strut{\relax\hbox{\vrule width0pt height 10.5pt depth 5.5pt}}
\begin{eqnarray*}
\begin{array}{||l|l|l|l|l||} \hline
\multicolumn{5}{|c|}{\hbox{G\"odel spacetime}} \\ \hline\hline \Strut
        & \gamma^{-1}F^{\rm(SC)}{}^{\hat{r}} 
        &\gamma^{-1} F^{\rm(GM)}{}^{\hat{r}}
	& \gamma^{-1}F^{\rm(EX)}{}^{\hat{r}} 
        &\gamma^{-1} F^{\rm(GE)}{}^{\hat{r}} \\ \hline \Strut
{\rm Thd_{-}} & 2\sqrt{2}\Omega {\rm T} 
                    & - 2\sqrt{2}\Omega {\rm T} 
                & 0 & 0  \\ \hline \Strut
{\rm Hyp_{-}}   
         & \frac{\sqrt{2}\Omega\,{\rm t}}{(1-s^2)(1-2s^2)} & 0 
	 & - \frac{2\sqrt{2}\Omega\,{\rm t}s^2}{(1-s^2)(1-2s^2)} 
	 & - \frac{\sqrt{2}\Omega\,{\rm t}}{1-s^2} \\ \hline \Strut
{\rm Sli_{-}}  & \frac{2\sqrt{2}\Omega\,{\rm t}}{1-2s^2} 
	 & -\frac{\sqrt{2}\Omega\,{\rm t}}{1-2s^2} 
       	 & - \frac{\sqrt{2}\Omega\,{\rm t}s^2}{(1-s^2)(1-2s^2)} 	
	 & - \frac{\sqrt{2}\Omega\,{\rm t}}{1-s^2} \\ \hline \Strut
{\rm Thd_{+}}      & 0 & 0 & 0 & 0  \\ \hline \Strut
{\rm Hyp_{+}}   
         & \frac{\sqrt{2}\Omega\,{\rm t}(1-2s^2)}{1-s^2} 
	 & 0 & \frac{2\sqrt{2}\Omega\,{\rm t}s^2}{1-s^2} 
	 & - \frac{\sqrt{2}\Omega\,{\rm t}}{1-s^2} \\ \hline \Strut
{\rm Sli_{+}} & 0 & \sqrt{2}\Omega\,{\rm t}
	 & \frac{\sqrt{2}\Omega\,{\rm t}s^2}{1-s^2} 
	 & - \frac{\sqrt{2}\Omega\,{\rm t}}{1-s^2} \\ \hline
\end{array}
\end{eqnarray*}
\label{tab:godgeo}
\end{table}

In this case the threading observers move along (corotating) geodesics
which are the same trajectories as the dust particles, while a second
family of geodesics counter-rotates as in the rotating Minkowski case.
In the threading point of view the outward space curvature force
balances the inward gravitomagnetic force along these counter-rotating
geodesics, while all forces vanish for the corotating geodesics. In
the slicing point of view, the space curvature force for the
corotating geodesics vanishes due to the factor $[\nu(U,n)^\phi -
N^{-1} N^\phi]\nu(U,n)^\phi = N^{-1}\dot{\phi}\,\nu(U,n)^\phi$ in
equation (\ref{eq:FSC}) which vanishes for $\dot{\phi}=0$. Both the
hypersurface and slicing space curvature forces for the
counter-rotating geodesics and the hypersurface space curvature force
become negative, corresponding to a Lie centripetal acceleration
directed radially outward, for $r > r_{\rm(rs)} > r_{\rm(h)}/2$, where
$\bar r_{\rm(rs)} = 2 {\mathop{\rm arcsinh}\nolimits}(2^{-1/2})
\approx 1.317$ defines the Lie relatively straight trajectory. This
effect is closely related to the result discussed at length by
Abramowicz et al in the case of circular orbits in the Schwarzschild
spacetime and just corresponds to the turning back of the embedded
2-surface of revolution towards its axis as shown in the appendix.
However, the counter-rotating geodesic is only timelike for $r <
r_{\rm(h)}/2$. 

\subsection{Kerr spacetime}

Here one finds the following spatial forces respectively in the
threading, hypersurface, and slicing points of view 
\begin{eqnarray}
  -F(U,m)^{\hat{r}}
     &=& F^{\rm(SC)}(U,m)^{\hat{r}} + F^{\rm(GM)}(U,m)^{\hat{r}} 
                                 + F^{\rm(GE)}(U,m)^{\hat{r}} 
                   \nonumber\\
     &=& \gamma(U,m)\biggl\{\biggl[
         \frac{r(r-2{\cal M} )^2-{\cal M} a^2}
              {r^2\sqrt{\Delta}(r-2{\cal M} )}
              |\nu(U,m)^{\hat{\phi}}|^2\biggr]
                    \nonumber\\
     && + \biggl[\frac{2a{\cal M} }{r^2(r-2{\cal M} )}
                           \nu(U,m)^{\hat{\phi}}\biggr]
    - \biggl[\frac{{\cal M} \sqrt{\Delta}}{r^2(r-2{\cal M} )}
                                                 \biggr]\biggr\} 
                         \ , 
                   \nonumber\\
 -F(U,n)^{\hat{r}} 
        &=& F^{\rm(SC)}(U,n)^{\hat{r}} 
          + F^{\rm(EX)}(U,n)^{\hat{r}} 
	  + F^{\rm(GE)}(U,n)^{\hat{r}}
                   \nonumber\\
	&=& \gamma(U,n)\biggl\{\biggl[
           \frac{\sqrt{\Delta} (r^3-a^2{\cal M} )}
                  {r^2(r^3+a^2r+2a^2{\cal M} )}
                      |\nu(U,n)^{\hat{\phi}}|^2\biggr]
                   \nonumber\\
	& & + \biggl[ \frac{2a{\cal M} (3r^2+a^2)}
                         {r^2(r^3+a^2r+2a^2{\cal M} )}
                                  \nu(U,n)^{\hat{\phi}}\biggr]
                   \nonumber\\
        & &  -\biggl[\frac{{\cal M} [(r^2+a^2)^2-4a^2{\cal M} r]}
          {r^2\sqrt{\Delta}(r^3+a^2r+2a^2{\cal M} )}
                                       \biggr]\biggr\} \ ,
                   \nonumber\\
 -F(U,n)^{\hat{r}}
        &=& F^{\rm(SC)}(U,n,e_0)^{\hat{r}} 
          + F^{\rm(GM)}(U,n,e_0)^{\hat{r}}
	  + F^{\rm(EX)}(U,n,e_0)^{\hat{r}} 
                    \nonumber\\ & &
          + F^{\rm(GE)}(U,n)^{\hat{r}} 
                    \nonumber\\
	&=& \gamma(U,n)\biggl\{\biggl[
           \frac{\sqrt{\Delta} (r^3-a^2{\cal M} )}
                        {r^2(r^3+a^2r+2a^2{\cal M} )}
                                |\nu(U,n)^{\hat{\phi}}|^2
                    \nonumber\\ & &
	 - \frac{2a{\cal M} (a^2{\cal M}  - r^3)}
                     {r^3(r^3 +a^2 r + 2a^2 {\cal M} )}
         \nu(U,n)^{\hat{\phi}}\biggr] 
                     \nonumber\\
	& &
	 +\biggl[\frac{a{\cal M} }{r^3}
                       \nu(U,n)^{\hat{\phi}}\biggr] 
	 +\biggl[\frac{a{\cal M} (3r^2 + a^2)}
                    {r^2(r^3+a^2r +2a^2 {\cal M} )}
                          \nu(U,n)^{\hat{\phi}}\biggr]
                     \nonumber\\
        & &
	 -\biggl[\frac{{\cal M} [(r^2+a^2)^2-4a^2{\cal M} r]}
         {r^2\sqrt{\Delta}(r^3+a^2r+2a^2{\cal M} )}
                                          \biggr]\biggr\}
                 \ .
\end{eqnarray}

In both the threading and slicing points of view the addition of an
upward ($+z$ direction) gravitomagnetic field, in comparison with the
corresponding Schwarzschild case, adds a radially inward
gravitomagnetic force for the counter-rotating circular geodesics and
a radially outward one for the corotating geodesics. The coordinate
angular velocity $\dot{\phi}_{-}$ of the counter-rotating geodesics
thus increases in magnitude with respect to the Schwarzschild case
while the coordinate angular velocity $\dot{\phi}_{+}$ of the
corotating geodesics decreases in magnitude. 

\begin{figure}[htbp]\footnotesize
\centerline{\epsfbox{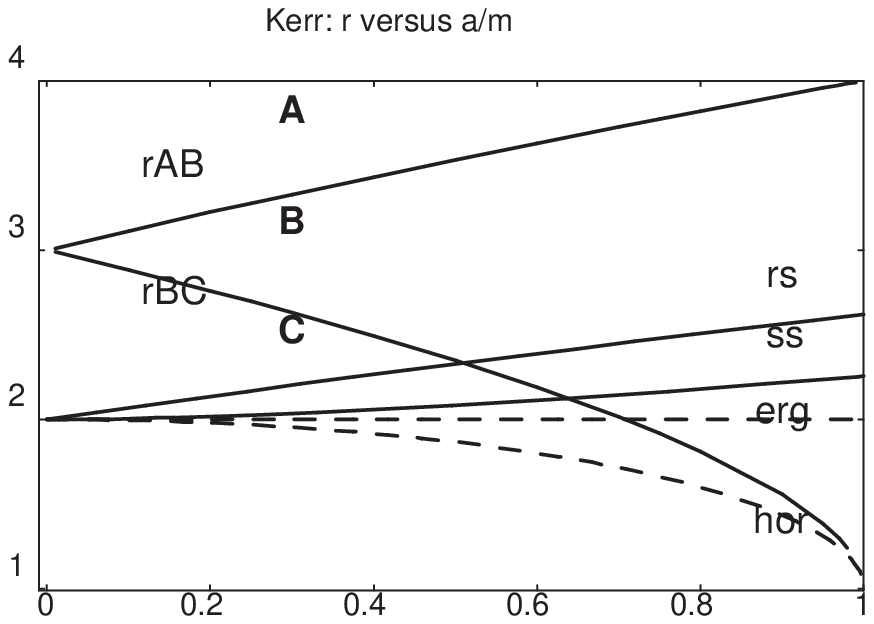}} 
\fcaption
{$\bar r$ versus $\bar a$: a plot of the Kerr Boyer-Lindquist radial
coordinate $\bar r$ versus $\bar a = a/{\cal M}\in [0,1]$ for the
horizon (hor), outer ergosphere boundary (erg), the two boundaries
between the regions A, B, and C, and the threading relatively straight
circles (rs) and embedding signature-switching circles (ss) for the 
equatorial plane.
}
\label{fig:rversusa}
\end{figure}

In the threading point of
view, the space curvature force changes sign at the single real root
$r_{\rm(rs)}$ of the equation $r(r-2{\cal M})^2 -{\cal M} a^2 = 0$
which occurs outside the ergosphere in regions C and B,
as shown in Figure~\ref{fig:rversusa},
thus only affecting the corotating geodesics in region B.
This corresponds to a Lie relatively straight
trajectory which in the extreme case $\bar a=1$ occurs at $\bar
r_{\rm(rs)} = (3+\sqrt{5})/2  \approx 2.618$,
while for the example $\bar a = 0.5$ one has
$\bar r_{\rm(rs)} \approx 2.328$.
Note that the Lie centripetal acceleration
points radially outward for $r < r_{\rm(rs)}$ while in G\"odel it
points outward for $r > r_{\rm(rs)}$, i.e., approaching the observer
horizon in both cases. 

\typeout{*** Figure 4. (rversusa)}

\typeout{*** (Partial page figure on top of page 20.)}

\section{Accelerated circular orbits}

The physical non-gravitational force acting on a test particle that
moves with 4-velocity $U$ is just the 4-force $f(U)^\alpha$ which is
related to the rescaled apparent 3-force by the equation (9.9) of
[BCJ1], namely 
\begin{equation}
  F(U,u)^\alpha =\gamma(U,u)^{-1}P(u)^\alpha{}_\beta f(U)^\beta \ .
\end{equation}
In the purely transverse relative acceleration applications that are
considered here, the  4-force is along the radial direction where the
projection reduces to the identity. The equation of motion may then be
simply expressed in terms of the physical force by equation (14.7) of
[BCJ1], namely 
\begin{equation}\label{eq:fFFFa}
 f(U)^{\hat r} = \gamma(U,u)F(U,u)^{\hat r}
     = \gamma(U,u) [ - F^{\rm(SC)}(U,u)^{\hat r}
                         - F^{\rm(G)}_{\rm(lie)}(U,u)^{\hat r} ]
         = a(U)^{\hat r} \ .
\end{equation} 
The middle equality is just the first of equations (\ref{eq:FFF}).
This relationship enables one to express the single
observer-independent quantity in terms of all the various points of
view using equations (\ref{eq:FG}) and (\ref{eq:FSC}). However, it is
important to remember that for relative motion not of this kind, the
situation is much more complicated. 

In the present case the expression for the physical force (namely, the
4-accelera\-tion by the equation of motion) is a quadratic expression in
the relative velocity physical component $\nu(U,u)^{\hat\phi}$
multiplied by the square of the associated gamma factor, leading to a
quotient of two quadratic expressions multiplied by a space curvature
factor. The zeros of the denominator occur when the velocity
approaches $\pm 1$, i.e., corresponding to photon orbits, values which
bound the domain $(-1,1)$ of allowed values of the velocity. The zeros
of the numerator are just the velocities which correspond to circular
geodesic motion and may occur inside (subluminal) or outside
(superluminal) the physical domain $(-1,1)$ of the function of the
velocity. 

As noticed by Barrabes, Boisseau and Israel \cite{barboiisr} in their
discussion of the hypersurface point of view for Kerr and
Schwarzschild, it is very useful to express the numerator of the
physical radial force expression in the explicitly factorized form 
\begin{eqnarray}\label{eq:ffactorized}
     f(U)^{\hat{r}}
         &=& \kappa(\phi,u)^{\hat r} \gamma(U,u)^2
            [\nu(U,u)^{\hat{\phi}} - \nu(U_-,u)^{\hat{\phi}}]
            [\nu(U,u)^{\hat{\phi}} - \nu(U_+,u)^{\hat{\phi}}]
          \nonumber\\
         &=& - \kappa(\phi,n)^{\hat r} 
             \frac{(\zeta-\dot\phi_-)(\zeta-\dot\phi_+)}
                  {(\zeta-\zeta_-)(\zeta-\zeta_+)}
\end{eqnarray}
valid when $\kappa(\phi,u)^{\hat r} \neq0$, where the roots of these
factors in the first case are the geodesic velocities
($\nu(U_-,u)^{\hat{\phi}} \leq \nu(U_+,u)^{\hat{\phi}}$), at least one
of which blows up when $\kappa(\phi,u)^{\hat r}=0$ so that the force
remains finite when other terms besides the centripetal acceleration
term are present. This also applies to the Minkowski spacetime with
$\nu(U_-,u)^{\hat\phi} = \nu(U_+,u)^{\hat\phi}$. Although the right
hand side of this equation contains observer-dependent quantities, the
expression itself is observer-independent since it represents the
physical force acting on the test particle. Thus in general, even when
the spatial curvature force changes sign due to a change in sign of
$\kappa(\phi,u)^{\hat r}$, the total force will not. The analogous
expression for the zero rest mass particle is 
\begin{eqnarray}\label{eq:ffactorizednull}
     f(P_\pm)^{\hat{r}}
         &=& \kappa(\phi,u)^{\hat r} E(P_\pm,u)^2
            [\pm 1 - \nu(U_-,u)^{\hat{\phi}}]
            [\pm 1 - \nu(U_+,u)^{\hat{\phi}}]
          \nonumber\\
         &=& 4 E(P_\pm,n)^2 \kappa(\phi,n)^{\hat r} 
             (\zeta_\pm -\dot\phi_-)(\zeta_\pm -\dot\phi_+) /
                  {(\zeta_- - \zeta_+)^2} \ .
\end{eqnarray}

de Felice \cite{def91,defuss91} first used such a factorization
(\ref{eq:ffactorized}) of the physical radial force for Kerr expressed
in terms of the coordinate angular velocities rather than the physical
component of the relative velocities along the orbit. The new angular
velocity parameter $y = \zeta/(1-a \zeta)$ (a fractional linear
tranformation) he uses in place of the parameter $\zeta$ (his 
$\Omega$) of equation (\ref{eq:zeta}) for constant angular velocity
world lines corresponds locally but not globally to a new choice of
slicing by a local time function  $t' = t- a \phi$ as discussed in the
appendix of Greene, Schucking, and Vishveshwara \cite{greschvis}.
Although a global such function does not exist, one can introduce it
as a well-defined change of parametrization on the world lines of test
particles. The advantage of this is that the new expressions for the
coordinate angular velocity $d\phi/dt'$ of timelike circular geodesics
in the equatorial plane of Kerr are reduced to the simpler
Schwarzschild form, while retaining the feature that the physical
force is still a quotient of quadratic expressions in this new
parameter, as in the case of the original angular velocity parameter
$\zeta$ and also for the physical component of the relative velocity
in each point of view. Note that the physical force is a fractional
quadratic function of the relative velocity, as it is also of any new
variable differing from the relative velocity by a fractional linear
transformation, like $\zeta$ or $y$, or the change in relative
velocity from changing the observer. 

Comparing the full expression for the physical force expressed in
terms of the relative acceleration, gravitoelectric, and
gravitomagnetic terms in the threading point of view 
\begin{equation}
     f(U)^{\hat{r}}
         = \gamma(U,m)^2 [
             \kappa(\phi,m)^{\hat r} \nu(U,m)^{\hat{\phi}}{}^2
             - \nu(U,m)^{\hat{\phi}} H(m)^{\hat z}
             - g(m)^{\hat r} ]
\end{equation}
leads to the identification
\begin{eqnarray}\label{eq:kgH}
 g(m)^{\hat r} &=& - \kappa(\phi,m)^{\hat r} 
                 \nu(U_-,m)^{\hat\phi} \nu(U_+,m)^{\hat\phi} \ ,
                 \nonumber\\
 H(m)^{\hat z} &=&   \kappa(\phi,m)^{\hat r} 
             [ \nu(U_-,m)^{\hat\phi} + \nu(U_+,m)^{\hat\phi} ] \ .
\end{eqnarray}
Similar relations hold for the hypersurface point of view with
$H(m)^{\hat z}$ replaced by $-2\theta(n)_{r\phi} = H(n,e_0)^{\hat z} +
2 \kappa(\phi,n)^{\hat r} \nu(e_0,n)^{\hat \phi}$. In the usual case
in which the signed Lie curvature is negative, an upward
gravitomagnetic vector field corresponds to a negative sum for the two
geodesic velocities, as will be assumed in discussion below unless
explicitly specified. 

Consider the case of a circular orbit for which the Lie signed
relative curvature $\kappa(\phi,u)^{\hat r} < 0$, i.e., the familiar
case in which the appropriate derivative of the Lie relative tangent
vector is inward along the radial direction. Then the physical force
is positive (outward) if $\nu(U_-,u)^{\hat{\phi}} <
\nu(U,u)^{\hat{\phi}} < \nu(U_+,u)^{\hat{\phi}}$ and it is negative
(inward) if $\nu(U,u)^{\hat{\phi}} > \nu(U_+,u)^{\hat{\phi}}$ or
$\nu(U,u)^{\hat{\phi}} < \nu(U_-,u)^{\hat{\phi}}$. This result is
exactly what one would expect in Newtonian gravity: namely, an outward
push is necessary to remain on a circular orbit if the speed is less
than the geodesic speed (Keplerian speed in the Newtonian context),
while an inward push is necessary if the speed is larger than the
geodesic speed. 

However, again as in Newtonian gravity, if the speed is larger than
the geodesic speed and one increases it, the inward force necessary to
maintain the circular orbit must also increase in magnitude and vice
versa. This is not always true in the general relativistic case, not
only in strong gravitational fields but also in certain weak
gravitational fields. To investigate this property, one needs not only
the sign of $f(U)^{\hat{r}}$ but also the sign of its derivative with
respect to the velocity of the test particle $\nu(U,u)^{\hat{\phi}}$
for a given point of view. 

This analysis is governed by the 3-parameter family of functions
\begin{equation}\label{eq:calF}
   {\cal F}(\nu;\kappa,\nu_-,\nu_+) 
        = \kappa[\nu-\nu_-][\nu-\nu_+]/[1-\nu^2]
\end{equation}
of the real variable $\nu$ and by its derivative with respect to that
variable. A specific spacetime and choice of observers collapses this
family of functions to a 1-parameter family in which
$\kappa,\nu_-,\nu_+$ all depend on the radial coordinate. The way in
which this collapse takes place leads to many possible behaviors. One
must also be careful taking limits of this expression in which $\nu$
and another velocity parameter go to $\pm1$, since the order in which
the limits are taken matters. 

The qualitative behavior of family of force versus velocity graphs
(for different radii) is characterized first by the number of distinct
real roots of the force function for a given radius, and second by the
number of such roots which lie in the physical interval $(-1,1)$ of
``subluminal" values for nonzero rest mass test particle motion. One
can distinguish three different regions of the spacetime (which may or
may not be present) containing those radii for which one of the
following conditions holds (assuming $\nu_- + \nu_+ \le0$)
\begin{description}
\item[region A:] $\nu_-,\nu_+$ both subluminal,
\item[region B:] $\nu_+$ subluminal, $\nu_-$ superluminal,
\item[region C:] $\nu_-,\nu_+$ both superluminal.
\end{description}
When they exist, the interfaces between regions A and B and regions B
and C are at the radii $r_{\rm(AB)}$ and $r_{\rm(BC)}$
at which the circular null geodesics occur. 

Within each of these three regions, the qualitative properties of the
force function are similar and follow from the expressions for its
first and second derivatives 
\begin{eqnarray}\label{eq:Fdiff}
 && d{\cal F}(\nu;\kappa,\nu_-,\nu_+)/d\nu
     \nonumber\\
 &&\quad
     = - \kappa 
        [ ( \nu_- + \nu_+ ) (\nu^2 + 1) - 2\nu(1+\nu_-\nu_+) ]
                                                    /[1 - \nu^2]^2
     \nonumber\\
 &&\quad
     = - \kappa ( \nu_- + \nu_+ ) 
        [ \nu^2 - 2\nu/\nu_{\rm(rel)} + 1 ]/[1 - \nu^2]^2
     \nonumber\\
 &&\quad
     = - \kappa ( \nu_- + \nu_+ ) 
        [ \nu - \nu_{\rm(crit)-}][\nu - \nu_{\rm(crit)+}]
                                                    /[1 - \nu^2]^2
     \nonumber\ ,\\
 &&
  d^2{\cal F}(\nu;\kappa,\nu_-,\nu_+)/d\nu^2
     \nonumber\\
 &&\quad
     = - 2\kappa
        [( \nu_- + \nu_+ ) (\nu^2 + 3)\nu
             - (3\nu^2 + 1)(1+\nu_-\nu_+) ]/[1 - \nu^2]^3
      \nonumber\\
 &&\quad
     = - 2\kappa ( \nu_- + \nu_+ ) 
      [ (\nu^2 + 3)\nu - (3\nu^2 + 1)/\nu_{\rm(rel)} ]/[1 - \nu^2]^3
     \ ,
\end{eqnarray}
where 
\begin{equation}\label{eq:nurel}
    \nu_{\rm(rel)} = (\nu_- + \nu_+)/(1 + \nu_-\nu_+) 
\end{equation}
up to sign is the relativistic difference of the two velocities and
vanishes when $\nu_- + \nu_+ = 0$ as occurs in the Schwarzschild case.
This is a subluminal velocity only when $\nu_-$ and $\nu_+$ are either
both subluminal or both superluminal (regions A and C). 

The extrema of the force function are found by examining its critical
points which only exist in regions A and C where $ \nu_{\rm(rel)} $ is
subluminal. The critical points occur at the roots (complex in region
B) of the quadratic factor in its numerator 
\begin{equation}\label{eq:nucrit}
  \nu_{\rm(crit)\pm} 
   = {\cal V}_\pm (\nu_{\rm(rel)})
   = \left[ 1 \pm \sqrt{1-\nu_{\rm(rel)}{}^2} \,\right]
                                          / \nu_{\rm(rel)} \ ,
\end{equation}
which satisfy 
\begin{equation}\label{eq:recip}
    \nu_{\rm(crit)-} \nu_{\rm(crit)+} = 1 \ .
\end{equation}
Under the assumption that $\nu_- + \nu_+ < 0$, they also satisfy
$\nu_{\rm(crit)+} < -1 < \nu_{\rm(crit)-}  < 0$ in region A where
$-1<\nu_{\rm(rel)}<0$, and $0 < \nu_{\rm(crit)-} < 1 <
\nu_{\rm(crit)+} $ in region C where $0<\nu_{\rm(rel)}<1$. Thus the
only critical point in the physical region occurs at the minus root 
\begin{eqnarray}
    \nu_{\rm(ext)} 
       &=&  \nu_{\rm(crit)-} 
          = \gamma_{\rm(rel)}\nu_{\rm(rel)} / (1+\gamma_{\rm(rel)})
                   \nonumber\\
        &\to & 
            {\textstyle {1\over2}} \nu_{\rm(rel)} 
            \to (\nu_- + \nu_+)/2
             \hbox{\ as\ } \nu_{\rm(rel)} \to 0\ .
\end{eqnarray}

In region A this leads to a maximum of the force function when $\kappa
< 0$ (minimum when $\kappa > 0$) and a minimum in region C when
$\kappa < 0$ (maximum when $\kappa > 0$), with the extreme value
\begin{equation}\label{eq:Fext}    
   {\cal F}(\nu_{\rm(ext)};\kappa,\nu_-,\nu_+)
       = \kappa [1 + (1 +\gamma_{\rm(rel)}) \nu_-\nu_+] 
                      / \gamma_{\rm(rel)}\ . 
\end{equation}
In region B there are no critical points and no extrema, but one point
of inflection. $\nu_{\rm(rel)}$ has a vertical asymptote at $\nu_-
\nu_+ = -1$, dividing region B into two parts: on the region A side
($\nu_- \nu_+ > -1$) it is negative as in region A, while on the
region C side ($\nu_- \nu_+ < -1$) it is positive as in region C. 

In region A where both $\nu_-$ and $\nu_+$ are subluminal, the value
$\nu_{\rm(ext)}$ has a simple and elegant interpretation in terms of
1-dimensional motion in special relativity. Consider three relative
velocities $\nu_\pm, \nu_{\rm(ext)}$ of three observers with respect
to a fixed fourth observer. Determine the velocity $\nu_{\rm(ext)}$ so
that in its own rest frame, the other two observers (corresponding to
$\nu_\pm$) have relative velocities which only differ in sign. Using
the relativistic addition formula for velocities along a fixed
direction, this condition is just 
\begin{equation}
       (\nu_+ -\nu_{\rm(ext)})/(1 - \nu_+ \nu_{\rm(ext)})
    = - (\nu_- -\nu_{\rm(ext)})/(1 - \nu_- \nu_{\rm(ext)}) \ .
\end{equation}
Cross-multiplying this equation leads to the same quadratic equation
in $\nu_{\rm(ext)}$ as follows for $\nu$ by setting the derivative
(\ref{eq:Fdiff}) to zero.  Thus the extremal force observer sees the 
two oppositely rotating geodesics with the same relative speed.

It follows that the 4-velocity of the extremal velocity is just the
renormalized average of the 4-velocities of the two oppositely
rotating circular geodesics. This immediately yields the result 
\begin{equation}
   \nu_{\rm(ext)} = \nu_{\rm(crit)-} 
      = [\gamma_- \nu_- + \gamma_+ \nu_+]/[\gamma_- + \gamma_+ ] \ .
\end{equation}
From this and the reciprocal relation (\ref{eq:recip}) one easily
finds 
\begin{equation}
   \nu_{\rm(crit)+} 
      = [\gamma_- \nu_- - \gamma_+ \nu_+]/[\gamma_- - \gamma_+ ] \ ,
\end{equation}
from which it follows that the 4-velocity of the second critical
velocity is just the renormalized difference of the 4-velocities of
the two oppositely rotating circular geodesics. 

The geometry of the relative observer plane of the circular motion
(the $t$-$\phi$ subspace of the tangent space) is very useful in
visualizing the various velocities which arise here and in the spin
precession analysis, and in extending this discussion to region C. For
a given observer with 4-velocity $u$ in this relative observer plane,
denote the map which reflects across $u$ by a tilde 
\begin{equation}
   U = \gamma(u + \nu \hat e) \to \tilde U = \gamma(u - \nu \hat e)
\end{equation}
and denote by a bar the commuting map which reflects (relative to $U$)
across the nearest forward null direction to the orthogonal direction 
\begin{equation}
   U = \gamma(u + \nu \hat e) 
        \to \bar U 
                 = \gamma \nu (u + 1/\nu \hat e) \ ,
\end{equation}
where $\hat e$ is a unit vector orthogonal to $u$ in the positive
$\phi$ direction in the relative observer plane of the circular
motion. The pair $(U,\bar U)$ arises from either $(u,\hat e)$ or
$(u,-\hat e)$ by a boost, for a timelike future-pointing $U$. For the
threading observers, this corresponds to the bar map introduced in
Eq.~(\ref{eq:barU}). Also let $U_\pm = \gamma_\pm(u + \nu_\pm \hat e)$
be the geodesic (unit) 4-velocities whenever they are not null,
whether timelike or spacelike, and the same for the critical
4-velocities $U_{\rm(crit)\pm} = \gamma_{\rm(crit)\pm}(u +
\nu_{\rm(crit)\pm} \hat e)$. In regions A and C the latter pair are
related to each other by the bar map because of Eq.~(\ref{eq:recip}). 

Figure \ref{fig:rok} illustrates the geometry of the various
velocities in the relative observer plane of the circular motion for
each of the three regions. In region A where the geodesic 4-velocities
$U_\pm$ are both timelike, the relative difference velocity
$-\nu_{\rm(rel)} > 0$ is the (scalar) relative velocity of $\tilde
U_-$ with respect to $U_+$. The critical velocities correspond to the
4-velocities which are the normalized sum/difference of the geodesic
4-velocities 
\begin{eqnarray}
      \nu_{\rm(crit)\pm} 
         &=& {\cal V}_\pm (\nu_{\rm(rel)})\ ,
            \nonumber\\
      U_{\rm(crit)\pm}^\alpha 
         &=& (U_-^\alpha \mp U_+^\alpha)/||U_+ \mp U_-|| \ .
\end{eqnarray}

\typeout{*** Figure 5. (rok)}

\typeout{*** (Full page figure on page 26
         with caption on top of page 27 to right.)}

\begin{figure}[htbp]\footnotesize
\typeout{*** *** pictex: loading
         idcf5.tex:  a,b,c pictex relative observer plane diagrams}
\typeout{*** *** pictex  fig a}
$$ \vbox{
\beginpicture
\typeout{*** *** Units set at 2.25cm tickmarks.}
  \setcoordinatesystem units <2.25cm,2.25cm> point at 0 0 


\setdashes
\setlinear
  \plot  0 0   1.2 1.2   /       %
  \plot -1.6 1.6   0 0   /       %
\setsolid


\setdashes 
\setquadratic
\plot
0 1    -.2 1.0198    -.4 1.07793    -.6 1.16619   -.8 1.28062 
-1 1.41421   -1.2 1.56205  -1.4 1.72047 -1.6 1.8868 /

\plot
0 1    .2 1.0198    .4 1.07793  .6 1.16619  .8 1.28062  
1 1.41421   1.2 1.56205  /

\plot 
-1 0  -1.0198 .2  -1.07793 .4  -1.16619 .6   -1.28062 .8  -1.41421 1
-1.56205 1.2  /

\plot
1 0  1.0198 .2  1.07793 .4  1.16619 .6   1.28062 .8  /

\setlinear

\setdashes <2pt>

\plot -1 1.4 -0.5 2.5  /
\plot -0.2 1 -0.5 2.5 /
\plot 0.5 1.1 -0.5 2.5 /
\plot -1 1.4 0.5 1.1 /

\plot -1.01 0.2 -1.5 0.3 /
\plot -1 1.4  -1.5 0.3 /

\setlinear
\setsolid


\plot 0 0  .679 .95 / 

\put {\mathput{a) \hbox{\ Region A}}}      [t]   at  0 -0.2

\put {\mathput{u}}                       [lb]   at   0.1 0.8
\put {\mathput{\hat e}}                  [rb]   at  0.9 0.1

\put {\mathput{U_-}}                     [lb]   at  -.9 1.4
\put {\mathput{U_+}}                     [b]   at  0.55 1.25 
\put {\mathput{U_{{\rm(c)}-}}}        [rb]   at  -0.2 0.85
\put {\mathput{U_{{\rm(c)}+}}}        [lb]   at  -.95 0.25 
\put {\mathput{U_- - U_+}}            [t]   at  -1.5 0.2
\put {\mathput{U_- + U_+}}            [lt]   at  -0.4 2.5

\put {\mathput{-\nu_{\rm (rel)}}}          [lt]   at   0.55 1.1

\put {\mathput{\tilde U_-}}              [rb]   at  1 1.5

\arrow <.3cm> [.1,.3]    from  0 0 to  0 1         
\arrow <.3cm> [.1,.3]    from  0 0 to  1 0         
\arrow <.3cm> [.1,.3]    from  0 0 to -1 0         

\arrow <.3cm> [.1,.3]    from  0 0  to  0.5 1.1   
\arrow <.3cm> [.1,.3]    from  0 0  to   -1 1.4   
\arrow <.3cm> [.1,.3]    from  0.82 1.15 to 1 1.4 

\arrow <.3cm> [.1,.3]    from 0 0  to  -0.2 1.01  
\arrow <.3cm> [.1,.3]    from 0 0  to   -1.01 0.2 

\arrow <.3cm> [.1,.3]    from  0.5 1.1 to  .914 1.28 

\endpicture}$$

\typeout{*** *** pictex fig b}
$$ \vbox{
\beginpicture
\typeout{*** *** Units set at 2.25cm tickmarks.}
  \setcoordinatesystem units <2.25cm,2.25cm> point at 0 0 


\setdashes
\setlinear
  \plot  0 0   .85 .85   /       %
  \plot  1.1 1.1   1.4 1.4   /       %

  \plot -1.6 1.6   0 0   /       %
\setsolid


\setdashes 
\setquadratic
\plot
0 1    -.2 1.0198    -.4 1.07793    -.6 1.16619   -.8 1.28062 
-1 1.41421   -1.2 1.56205  -1.4 1.72047 -1.6 1.8868 /

\plot
0 1    .2 1.0198    .4 1.07793  .6 1.16619  .8 1.28062  
1 1.41421   1.2 1.56205  /

\plot 
-1 0  -1.0198 .2  -1.07793 .4  -1.16619 .6   -1.28062 .8  -1.41421 1
-1.56205 1.2  /

\plot
1 0  1.0198 .2  1.07793 .4  1.16619 .6   1.28062 .8  1.41421 1
  1.56205 1.2  /

\setlinear

\setdashes <2pt>

\plot -1 1.4 -0.5 2.5  /
\plot -0.2 1 -0.5 2.5 /
\plot 0.5 1.1 -0.5 2.5 /
\plot -1 1.4 0.5 1.1 /

\plot -1.01 0.2 -1.5 0.3 /
\plot -1 1.4  -1.5 0.3 /

\setlinear
\setsolid


\plot 0 0  .679 .95 / 

\put {\mathput{b) 
       \hbox{\ Region B ($-1<\nu_- \nu_+ < 0$)}}} [t]   at  0 -0.2

\put {\mathput{u}}                       [lb]   at   0.1 0.8
\put {\mathput{\hat e}}                  [rb]   at  0.9 0.1

\put {\mathput{\bar U_-}}                     [lb]   at  -.9 1.4
\put {\mathput{U_+}}                     [b]   at  0.55 1.25 
\put {\mathput{U_{{\rm(sc)}-}}}        [rb]   at  -0.2 0.85
\put {\mathput{U_{{\rm(sc)}+}}}        [lb]   at  -.95 0.25 
\put {\mathput{\bar U_- - U_+}}            [t]   at  -1.5 0.2
\put {\mathput{\bar U_- + U_+}}            [lt]   at  -0.4 2.5

\put {\mathput{\tilde U_-}}                      [lt]   at   1.4 .9
\put {\mathput{U_-}}                      [rt]   at   -1.4 .9

\put {\mathput{-1/\nu_{\rm (rel)}}}          [lt]   at   0.525 1.1

\put {\mathput{\tilde{\bar U}_-}}              [rb]   at  1 1.5

  
\arrow <.3cm> [.1,.3]    from  0 0 to  0 1         
\arrow <.3cm> [.1,.3]    from  0 0 to  1 0         
\arrow <.3cm> [.1,.3]    from  0 0 to -1 0         

\arrow <.3cm> [.1,.3]    from  0 0  to  0.5 1.1   
\arrow <.3cm> [.1,.3]    from  0 0  to   -1 1.4   
\arrow <.3cm> [.1,.3]    from  0.82 1.15 to 1 1.4 

\arrow <.3cm> [.1,.3]    from  0 0  to  1.4 1     
\arrow <.3cm> [.1,.3]    from  0 0  to -1.4 1     

\arrow <.3cm> [.1,.3]    from 0 0  to  -0.2 1.01  
\arrow <.3cm> [.1,.3]    from 0 0  to   -1.01 0.2 

\arrow <.3cm> [.1,.3]    from  0.5 1.1 to  .914 1.28 

\endpicture}$$

\typeout{*** *** pictex  figc}
$$ \vbox{
\beginpicture
\typeout{*** *** Units set at 2.25cm tickmarks.}
  \setcoordinatesystem units <2.25cm,2.25cm> point at 0 0 



\setdashes
\setlinear
  \plot  0 0   1.6 1.6   /       %
  \plot -1.6 1.6   0 0   /       %


\setdashes 
\setquadratic

\plot
0 1 -.2 1.0198 -.4 1.07793  -.6 1.16619  -.8 1.28062 -1 1.41421
 -1.2 1.56205  -1.4 1.72047 -1.6 1.8868 /

\plot
0 1 .2 1.0198 .4 1.07793  .6 1.16619  .8 1.28062 1 1.41421
 1.2 1.56205 1.4 1.72047 1.6 1.8868 /

\plot 
1 0  1.0198 .2  1.07793 .4  1.16619 .6   1.28062 .8  1.41421 1
  1.56205 1.2  /

\plot 
-1 0  -1.0198 .2  -1.07793 .4  -1.16619 .6   -1.28062 .8  -1.41421 1
  -1.56205 1.2 /

\setlinear

\setdashes <2pt>

\plot -0.5 1.1   1 1.4 /
\plot -0.5 1.1 0.5 2.5 /
\plot   0.2 1.01    0.5 2.5 /
\plot 0 0 1.5 0.3 /
\plot 0.5 2.5  1.5 0.3 /

\setlinear
\setsolid

\put {\mathput{\hat e}}                  [rt]   at  0.9 -0.1
\put {\mathput{u}}                       [rb]   at   -0.075 0.8

\put {\mathput{U_{{\rm (c)}+}}}          [rb]   at  1 0.25
\put {\mathput{U_{{\rm (c)}-}}}          [rb]   at  0.2 1.25
\put {\mathput{-\nu_{\rm (rel)}}}         [lb]   at   0.5 0.85

\put {\mathput{\tilde{\bar U}_-}}         [rb]   at  0.6 1.35
\put {\mathput{\bar U_-}}                 [lb]   at  -0.65 1.2
\put {\mathput{\bar U_+}}                 [b]    at  1.05  1.55
\put {\mathput{U_+}}                      [lt]   at   1.4 .9
\put {\mathput{U_-}}                      [rt]   at   -1.15 0.55

\put {\mathput{\bar U_+ - \bar U_-}}            [lt]   at  1.2 0.15
\put {\mathput{\bar U_+ + \bar U_-}}            [lt]   at  0.6 2.5

\put {\mathput{c) 
   \hbox{\ Region C}}}                   [t] at 0 -0.2

\arrow <.3cm> [.1,.3]    from  0 0 to  0 1            
\arrow <.3cm> [.1,.3]    from  0 0 to  1 0            
\arrow <.3cm> [.1,.3]    from  0 0 to -1 0            

\arrow <.3cm> [.1,.3]    from  0 0 to  0.5 1.1   

\arrow <.3cm> [.1,.3]    from  0 0  to 1.4 1     
\arrow <.3cm> [.1,.3]    from  0.714 1  to  1 1.4    

\plot 0 0 .571 0.8 /

\arrow <.3cm> [.1,.3]    from  0 0  to  -0.5 1.1 
\arrow <.3cm> [.1,.3]    from  0 0  to  -1.1 0.5 

\arrow <.3cm> [.1,.3]    from  0 0 to  0.2 1.01  
\arrow <.3cm> [.1,.3]    from  0 0 to   1 0.2    

\arrow <.3cm> [.1,.3]    from 1 1.4 to  0.4 0.9  

\endpicture}$$

\end{figure}

\begin{figure}[t]\footnotesize
\vspace{0cm}
\fcaption
{An illustration of the relative observer plane geometry of the
various velocities which arise in the analysis of the physical radial
force function for circular motion,  for each of the three regions A,
B, and C. In regions A and C, these determine the critical velocities
(abbreviated by $U_{\rm(c)\pm}$) of the force function directly, while
in region B where the critical velocities are complex, the geometry
shown instead determines the spin-critical velocities (abbreviated by
$U_{\rm(s,c)\pm}$) for the spin precession function of section
\ref{sec:spin}. The case $\nu_- \nu_+ < -1$ of region B is obtained
from the case $-1 < \nu_- \nu_+ < 0$ shown in $b)$ by interchanging
$U_+$ and $\tilde{\bar U}_-$, reversing the sign of the relative
velocity. The case $\nu_- \nu_+ = -1$ separating these two has instead
$U_+ = \tilde{\bar U}_-$. 
}
\label{fig:rok}
\end{figure}


In region C both $U_\pm$ are spacelike. Here one may rewrite the
relative velocity $\nu_{\rm(rel)} > 0$ in the following form 
\begin{equation}
    \nu_{\rm(rel)} = (\nu_-{}^{-1} + \nu_+{}^{-1})
                       /(1 + \nu_-{}^{-1} \nu_+{}^{-1}) \ ,
\end{equation}
revealing that its sign-reversal may be interpreted as the relative
velocity of $\tilde{\bar U}_-$ with respect to $\bar U_+$, both
timelike in region C. The critical velocities now correspond to the
4-velocities which are the normalized sum/difference of the timelike
barred geodesic 4-velocities or equivalently of the tachyonic
four-velocities themselves 
\begin{eqnarray}
      U_{\rm(crit)\pm}^\alpha 
         &=& (\bar U_+^\alpha \mp \bar U_-^\alpha)
               / ||\bar U_+ \mp \bar U_-|| 
               \nonumber\\
         &=& (U_+^\alpha \mp U_-^\alpha)
               / ||U_+ \mp  U_-|| \ .
\end{eqnarray}
Thus in region C the extremal observer sees the two tachyonic geodesic
velocities with equal but opposite velocities as in region A. 

In region B where $U_+$ is still timelike but $U_-$ is spacelike, the
spin-critical velocities introduced below in section \ref{sec:spin}
are determined by a similar relationship but with $\nu_{\rm(rel)}$
replaced by its reciprocal (i.e., $U_-$ replaced by $\bar U_-$) 
\begin{equation}
      \nu_{\rm(crit,spin)\pm} 
         = {\cal V}_\pm (1/\nu_{\rm(rel)})
             = \left[1 \pm \sqrt{1-1/\nu_{\rm(rel)}{}^2}\,\right]
                     \nu_{\rm(rel)} \ .
\end{equation}
But $-\nu_{\rm(rel)}$ is the relative velocity of $\tilde U_-$ with
respect to $U_+$ and $-1/\nu_{\rm(rel)}$ is the relative velocity of
$\tilde{\bar U}_-$ with respect to $U_+$. Therefore the corresponding
spin-critical 4-velocities are related  to the critical ones in the
same way that the critical ones are related to the geodesic ones 
\begin{equation}
      U_{\rm(crit,spin)\pm}^\alpha 
         = (U_{\rm(crit)+}^\alpha \mp U_{\rm(crit)-}^\alpha)
                      /||U_{\rm(crit)+} \mp U_{\rm(crit)-}|| \ .
\end{equation}
Thus if one reinterprets $U_\pm$ in Figure~\ref{fig:rok}$(b)$ as the
critical velocities $U_{\rm(crit)}$, then the four-velocities
$U_{\rm(sc)}$ of the figure become the spin-critical velocities
$U_{\rm(crit,spin)}$. A similar diagram for the region B case
$\nu_-\nu_+ <-1$ has $U_+$ and $\tilde{\bar U}_-$ interchanged
relative to $u$. 

In regions A and C one can introduce a family of ``extremal force"
observers with 4-velocity $U_{\rm(ext)}^\alpha$ and its associated
relative velocity $\nu_{\rm(ext)}(u)^{\hat\phi} =
\nu_{\rm(ext)}(U_{\rm(ext)},u)^{\hat\phi}$ for both the Kerr and
G\"odel spacetimes, both of which have a pair of oppositely directed
circular geodesics. The expression (\ref{eq:Fext}) with appropriate
arguments then defines the acceleration of the extremal force
observers. In region A their angular velocity 
\begin{equation}
  \zeta_{\rm(ext)} 
     = (\Gamma_- \dot\phi_- + \Gamma_+ \dot\phi_+) 
             / (\Gamma_- + \Gamma_+)
     < 0
\end{equation}
is easily evaluated by averaging the geodesic 4-velocities in the form
(\ref{eq:zeta}), and is negative since $\Gamma_- > \Gamma_+$ as
explained at the beginning of section \ref{sec:cirgeo}. For low speeds
($\Gamma_\pm\to1$), this reduces to the average angular velocity
$\zeta_{\rm(gmp)}$. In region C their angular velocity is given by the
analytic continuation of above formula ($\Gamma_\pm$ are purely
imaginary so $\zeta_{\rm(ext)}$ remains real) but is instead positive.

Thus the extremal force condition picks out the family of observers
which see two oppositely moving geodesic test particles symmetrically,
i.e., with the same relative speed. Expressing the physical force
function with respect to these observers leads to an expression which
is an even function of $\nu$ since ${\cal F}(\nu;\kappa,-\nu_+,\nu_+)
= \kappa[\nu^2-\nu_+{}^2]/[1-\nu^2]$. However, a pair of oppositely
directed geodesic test particles which start from a given such
observer do not return simultaneously to the same observer after each
full revolution of the orbit because of the same asymmetry between
corotation and counter-rotation which appears in the relationship
(\ref{eq:nuUmn}) between the physical velocity component and the
angular velocity. This is described below as a geodesic analog of the
Sagnac effect. 

The difference
\begin{equation}
     \zeta_{\rm(ext)} - \zeta_{\rm(gmp)} 
         = {\textstyle\frac12}
           [(\Gamma_- -\Gamma_+)/(\Gamma_- + \Gamma_+)] 
           [\dot\phi_- - \dot\phi_+]    < 0 
\end{equation}
is negative since $\Gamma_- > \Gamma_+$ and 
$\dot\phi_- <0, \dot\phi_+>0$, so
the extremal force observers counter-rotate with respect to the
geodesic meeting point observers. The negative values of the angular
velocities of both families of observers leading to counter-rotation
with respect to the threading observers is due to the asymmetry in
their equation of motion introduced by the gravitomagnetic vector
force. In both cases these counter-rotate with respect to the slicing
observers (which are determined by the null circular orbit meeting
points) since the asymmetry introduced by the gravitomagnetic force,
being proportional to the speed, is smaller for timelike velocities
compared to the speed of light, and so the corotating ``dragging"
effect is smaller. 

It seems  reasonable that at least in the limit of small angular
speeds, the physical force maintaining the orbits of the extremal
force observers is maximum since in some sense they have the least
relative angular motion with respect to the geometry, and hence
require the most force to keep from falling towards the center of
symmetry. This is intuitively clear in Kerr where the attraction
towards the center is provided by the gravitoelectric field of the
mass giving rise to the gravitational field. In the G\"odel spacetime
in the slicing point of view, the gravitoelectric field is also
inward, so a test particle initially at rest would also begin to fall
towards the axis of symmetry as it follows a circular orbit with a
different center. 

\typeout{*** Table 6. (fcritsign)}
\begin{table}[htbp]\footnotesize
\tcaption
{Signs of the physical component of the physical radial force
and its velocity derivatives for circular orbits for the usual case
$\kappa(\phi,u)^{\hat r}<0$. The abbreviations 
$\nu_\pm=\nu(U_\pm,u)^{\hat\phi}$ and
$\nu=\nu(U,u)^{\hat\phi}$ are used.
}
 \typeout{*** eqnarray struts inserted}
 \def\Strut{\relax\hbox{\vrule width0pt height 10.5pt depth 5.5pt}}
\begin{eqnarray*}
\begin{array}{|l|l|l|l|} \hline
 & {\rm Region\ A} \qquad & {\rm Region\ B} \qquad 
                              & {\rm Region\ C} \\ 
\hline
 &\nu_- > -1 & \nu_- < -1 & \nu_- < -1\\
 &\nu_+ <  1 & \nu_+ <  1 & \nu_+ >  1\\ 
\hline 
f(U)^{\hat{r}} & 
 \begin{array}{l}
  {\rm positive \ if:} \\
   \nu_- < \nu < \nu_+ \\
  {\rm negative \ if:} \\
   \nu   > \nu_+ \ ; \\
   \nu   < \nu_-
   \end{array} 
 & \begin{array}{l} 
    {\rm positive \ if:} \\
	\nu < \nu_+ \\
    {\rm negative \ if:} \\
        \nu > \nu_+ 
  \end{array}
 & {\rm positive}
\\ \hline \Strut
 \frac{df(U)^{\hat{r}}}{d\nu} & {\rm any\ sign}
         & {\rm negative}  & {\rm any\ sign} 
\\ \hline \Strut
 \frac{d^2f(U)^{\hat{r}}}{d\nu{}^2} & {\rm negative}  
         & {\rm any\ sign} & {\rm positive} 
\\ \hline
\end{array}
\end{eqnarray*}	
\label{tab:fcritsign}
\end{table}

Table \ref{tab:fcritsign} gives the sign of the physical force and of
its first and second derivatives for each of these regions. For the
G\"odel spacetime, there is always at least one timelike geodesic at
every radius (the threading observers) so region C does not exist,
while region C disappears from the Kerr spacetime in the extreme case
$a={\cal M}$. Similarly the region B, whose existence is due to the
asymmetry in the angular motion introduced by the rotation of the
reference frame, disappears in the Schwarzschild limit  $a=0$ of Kerr.
In the rotating Minkowski spacetime, the two real roots of the force
function coincide, so there is only one circular geodesic (the
counter-rotating slicing observer) and it is timelike at each radius
where the threading point of view is valid, so only region A exists in
both points of view. Figure \ref{fig:rversusa} shows the dependence of
the radius $r$ as a function of $a$ for Kerr for the two boundaries
$r_{\rm(AB)}$, $r_{\rm(BC)}$ between the three regions A, B, and C,
together with the radii of the relatively straight circles and
embedding-signature-switching circles for the threading point of view
discussed in the appendix. 

For each value of the radial variable, one can plot the physical force
$f(U)^{\hat{r}}$ versus the physical component $\nu(U,u)^{\hat\phi}$
of the relative velocity in the angular direction for both the
threading and slicing points of view. Figure~\ref{fig:phyfor} shows a
representative sample of these curves for the rotating Minkowski
spacetime, the G\"odel spacetime, the Kerr spacetime with $a/{\cal
M}=0.5$, and the Schwarzschild spacetime $a=0$. Shown also is the
curve whose intersection with each force graph occurs at the velocity
of the other family of observers. In the G\"odel slicing case, for
example, this is just the positive horizontal axis since the theading
observers are geodesic and corotating. The curves of relative extrema
of the force graphs are also plotted. 

\typeout{*** Figure 6.  (phyfor): moved back to get even-odd page placement}

Note that in the slicing point of view in Minkowski spacetime, only
the space curvature force due to the circular orbit is nonzero. This
is just the usual inward centripetal acceleration required to maintain
a circular orbit under the influence of no real forces, scaled by a
squared gamma factor. It has a maximum at the relative velocity of the
slicing observers, which are the extremal force observers in this
case. Increasing the radius flattens the force graph due to its
inverse dependence on $r$. In the threading point of view, the force
graph also migrates to the left as it follows the velocity of the
single counter-rotating circular geodesic followed by the slicing
observers at each radius, going infinite at the threading observer
horizon radius. 

\typeout{*** Figure 6.  (phyfor)}

\typeout{*** (Full page figure on page 30 (rotating by 90 degrees)
         with caption on top of adjacent page 31 to right.)}

\begin{figure}[p]\footnotesize
\vspace{4cm}
\centerline{\epsfbox{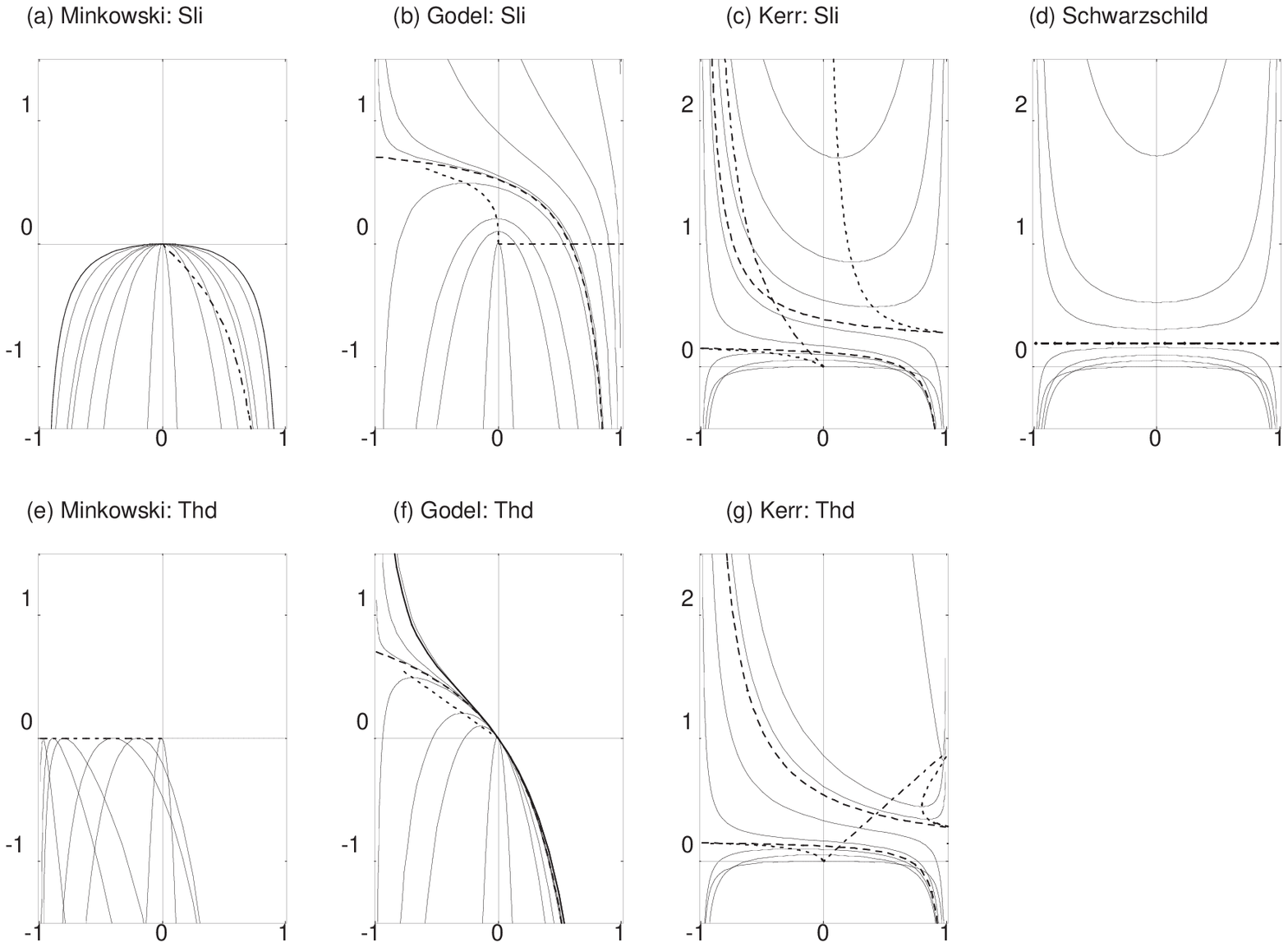}} 
\end{figure}

\begin{figure}[t]\footnotesize
\fcaption
{$f $ versus $\nu$:  plots of the radial physical component of the
force acting on a nonzero rest mass test particle moving on an
equatorial circular orbit as a function of the physical component of
the relative velocity of the particle in the angular direction for the
rotating Minkowski, G\"odel, Kerr ($a/{\cal M}=0.5$), and
Schwarzschild ($a=0$)  spacetimes. The solid curves are the graphs of
the force plotted versus the relative velocity for selected values of
the radius. For the spacetimes other than rotating Minkowski, regions
A, B, and C are the regions which are separated by the long dashed
lines, from bottom to top. Region C is missing in G\"odel and region B
in Schwarzschild. The short dashed lines connect the local extrema of
the force graphs. The long-short dashed line marks the point on a
given such graph which occurs at the velocity of the other family of
observers. 

For Kerr the force curves move higher along the vertical axis with
decreasing radius from region A extending out to infinity through
region B into region C containing the ergosphere and finally
approaching the observer horizon. This is also true for Schwarzschild
except that region B disappears between the interface of regions A and
C at $r=3m$. For G\"odel they move higher along the vertical axis
(slicing diagram) or the dashed-dotted line (threading diagram) with
increasing radius from region A near $r=0$ into region B approaching
the observer horizons. For the rotating Minkowski slicing diagram,
they move outward from the vertical axis with increasing radius, and
in the threading diagram they move to the left with increasing radius
toward the observer horizon. The thickened force curve represents the
radius of the ergosphere (threading observer horizon) in the Kerr
slicing case and the corresponding slicing observer horizon radius in
the G\"odel threading case. Almost no change in the force curve occurs
beyond that radius in the G\"odel threading case. 
}
\label{fig:phyfor}
\end{figure}


For a rotating spacetime the physical force $f(U)^{\hat{r}}$ is not a
symmetric function of the relative velocity $\nu(U,u)^{\hat{\phi}}$
since the gravitomagnetic vector force introduces an asymmetry between
the corotating and counter-rotating orbits. For the Kerr and G\"odel
spacetimes the maxima and minima of the force graphs in regions A and
C do not occur at zero velocity. At these extrema the derivative of
the force with respect to the velocity changes sign at a nonzero
relative velocity, leading to effects which run counter to our
ordinary Newtonian intuition about the radial force for circular
orbits in a force field. 

For those spacetimes with a region A (Kerr and G\"odel), the sign of
the radial force in that region behaves according to our Newtonian
intuition about circular orbits. A positive (outward) radial force is
required to support a circular orbit with speed less than the
Keplerian speed, and a negative (inward) force is required to hold in
a circular orbit with more than that speed. In the Schwarzschild case,
the extrema of the force graphs occur at zero velocity and are local
maxima in this region, so increasing the speed decreases the outward
radial force needed to maintain the orbit. However, for the Kerr and
G\"odel spacetimes, the local maxima occur at negative (slicing or
threading) velocities, so for the counter-rotating orbits between zero
velocity and the one at the local maximum, increasing the speed
increases the outward radial force necessary to maintain the orbit,
which is the opposite of what happens in Newtonian gravity. We refer
to this simply as ``counter-intuitive radial force behavior." Thus in
region A, extending to infinite radius in Kerr but from zero radius
out to the radius at which the counter-rotating geodesic becomes
spacelike in the G\"odel spacetime, those counter-rotating orbits with
speeds less than the speed at which the maximum of the force occurs
experience this behavior. 

For those spacetimes with a region B (Kerr and G\"odel) where only the
counter-rotating geodesic is spacelike (superluminal velocity), there
is no local extremum on each force graph in this region. A particle on
a counter-rotating circular orbit is forced to move with a velocity
smaller than the (unphysical) geodesic speed and so it must be pushed
outward (positive force) to maintain the orbit. This is as expected.
However, if the speed of a counter-rotating orbit increases then the
positive force increases as well, which is again counterintuitive. 

In the region C in the Kerr spacetime where both the corotating and
the counter-rotating circular geodesics are spacelike (superluminal
velocities), both corotating and counter-rotating test particles are
forced to move with a speed less than the geodesic speed thus always
yielding a positive (outward) radial force. As their speed approaches
1, the Lorentz gamma factor goes to infinity causing the physical
force to increase to infinity as well. Since a local minimum occurs at
a corotating (positive) velocity, increasing the speed increases the
outward force for velocities outside the interval between zero
velocity and the positive velocity where the minimum occurs. Thus some
corotating and all counter-rotating orbits exhibit counter-intuitive
radial force behavior. 

This happens for all circular orbits in region C in the Schwarzschild
spacetime, where the interval between zero velocity and the velocity
of this minimum shrinks to zero width. The corotating and
counter-rotating geodesic velocities take the relativistically
corrected Keplerian values 
\begin{equation}
  \nu(U_\mp,u)^{\hat{\phi}}
     =   \mp\sqrt{{\cal M}/r}/\sqrt{1-2{\cal M}/r}
\end{equation} 
and the physical force reduces to
\begin{equation}
  f(U)^{\hat{r}}
    = \kappa(\phi,u)^{\hat r} \gamma(U,u)^2 
          [(\nu(U,u)^{\hat{\phi}})^2
                    - (\nu(U_\mp,u)^{\hat{\phi}})^2] \ ,
\end{equation}
where the signed relative curvature is 
\begin{equation}
   \kappa(\phi,u)^{\hat r} = -\sqrt{1-2{\cal M}/r}/r \ .
\end{equation}
The two circular geodesics become null at $r=3{\cal M}$ where
$|\nu(U_\mp,u)^{\hat\phi}| = 1$, and then spacelike for smaller radii.
At this coordinate radius the physical force reduces to
$f(U)^{\hat{r}}= -\kappa(\phi,u)^{\hat r} > 0$ which is independent of
the velocity of the test particle, leading to the dashed line force
graph separating the two regions A and C in
Figure~\ref{fig:phyfor}$(d)$. For $2{\cal M} < r < 3{\cal M}$ (region
C), the circular geodesics are both spacelike and the physical force
on the test particle must be positive. Moreover $f(U)$ is an
increasing function of $|\nu(U,u)^{\hat{\phi}}|$ in that region and so
increasing the speed of the test particle increases the outward
physical force, contrary to Newtonian intuition. 

This counter-intuitive radial force behavior was first discovered by
Abramo\-wicz \cite{a90a,a90b} 
in the Schwarzschild case inside $r=3{\cal
M}$, the outer boundary in the equatorial plane of his ``rotosphere''
\cite{a90b}, and he explains it as due to a reversal of the ``optical
centrifugal force" in a decomposition of the physical force into a sum
of terms. de Felice \cite{def91,defuss91} has noticed this behavior in
the Kerr spacetime using a factorization of the physical force in
terms of angular velocity, which is closely connected to the threading
point of view relative velocity, and he uses the term ``prehorizon
regime" to characterize it, since initially he interpreted this
behavior as a way of signaling the approach to a horizon. Later he
discovered that it can occur for counter-rotating orbits even a great
distances from a rotating object \cite{def95}. Barrab\`es, Boisseau,
and Israel \cite{barboiisr} have used the hypersurface point of view
to describe the same phenomenon for both Schwarzschild and Kerr, but
using a factorization of the physical force in terms of the slicing
relative velocities. 

Of course all of these effects depend on the zero point used for the
relative speed, and are therefore observer-dependent. The choice of
observer used to describe the effect depends on exactly what one wants
to measure or explain. The threading point of view is relevant to
behavior as seen from infinity, while the hypersurface point of view
might be appropriate for local considerations like accretion disks.
For the extremal force observers in region A introduced above, the
extrema of the physical force occur at zero relative velocity. A
similar family exists in region C. For example, in the region C of the
Kerr diagram, one sees that the corotating family of observers for
which the minimum occurs approaches the slicing observers at the
horizon but increasingly corotates faster than those observers until
it experiences its own outer observer horizon as one moves out to the
radius of the corotating photon orbit. In this region C the
counter-intuitive radial force behavior occurs for all relative
velocities with respect to these observers just as in the
Schwarzschild case. For a certain interval of larger radii (region B),
no extremum occurs. Then in the region A outside the radius at which
the counter-rotating photon orbit occurs, one has a counter-rotating
family for which the maxima occur. This second family has an inner
horizon at that radius and counter-rotates with respect to the
threading observers but approaches them at infinite radius. For this
second family the counter-intuitive radial force behavior does not
occur. Figure \ref{fig:kerrobs}$(a)$ shows the coordinate angular
velocity of each of the geometrically defined observer families for
the Kerr spacetime as a function of the radius (see Fig.~2 of de
Felice and Usseglio-Tomasset \cite{defuss91}), including the Carter
family associated with the usual orthogonal frame in which the Kerr
metric is stated in Boyer-Lindquist coordinates (Eq.~(33.2) of Misner,
Thorne, and Wheeler \cite{mtw}), an observer family which has a
coordinate angular velocity $\zeta_{\rm(car)} = a/(r^2+a^2)$ and
simultaneously diagonalizes the electric and magnetic parts of the
curvature tensor. Figure \ref{fig:kerrobs}$(b)$ shows the
corresponding plot of the observers' physical relative velocities in
the slicing point of view. 

\typeout{*** Figure 7.   (kerrobs): moved back for even/odd page placement}

Finally consider the less usual case in which the Lie signed relative
curvature is positive, $\kappa(\phi,u)^{\hat r} > 0$, which would seem
to reverse the sign of the physical force as one crosses the Lie
relatively straight trajectory from the side for which
$\kappa(\phi,u)^{\hat r}<0$. However, the superluminal velocity
$\nu(U_-,u)$ blows up at this radius, switching signs so that the
physical force remains finite and does not change sign, even though
the relative centripetal acceleration itself does change sign
(``reversal of the centrifugal force" in the language of Abramowicz et
al \cite{a90a}, but in the true relative geometry rather than the
optical one). For example, in the extreme Kerr case in the threading
point of view, the relative curvature becomes positive for
$r<r_{\rm(rs)}$ and the space curvature force along the corotating
circular orbit becomes negative while $\nu(U_-,m)^{\hat r}$ blows up
there. In the G\"odel spacetime this instead occurs in the slicing
point of view where the relative curvature becomes positive for
$r>r_{\rm(rs)}$. 

\typeout{*** Figure 7.   (kerrobs)}
\typeout{*** (Full page figure on page 34
          with caption on top of page 35 to right.)}

\begin{figure}[htbp]\footnotesize
\centerline{\epsfysize=\textheight
\epsfbox{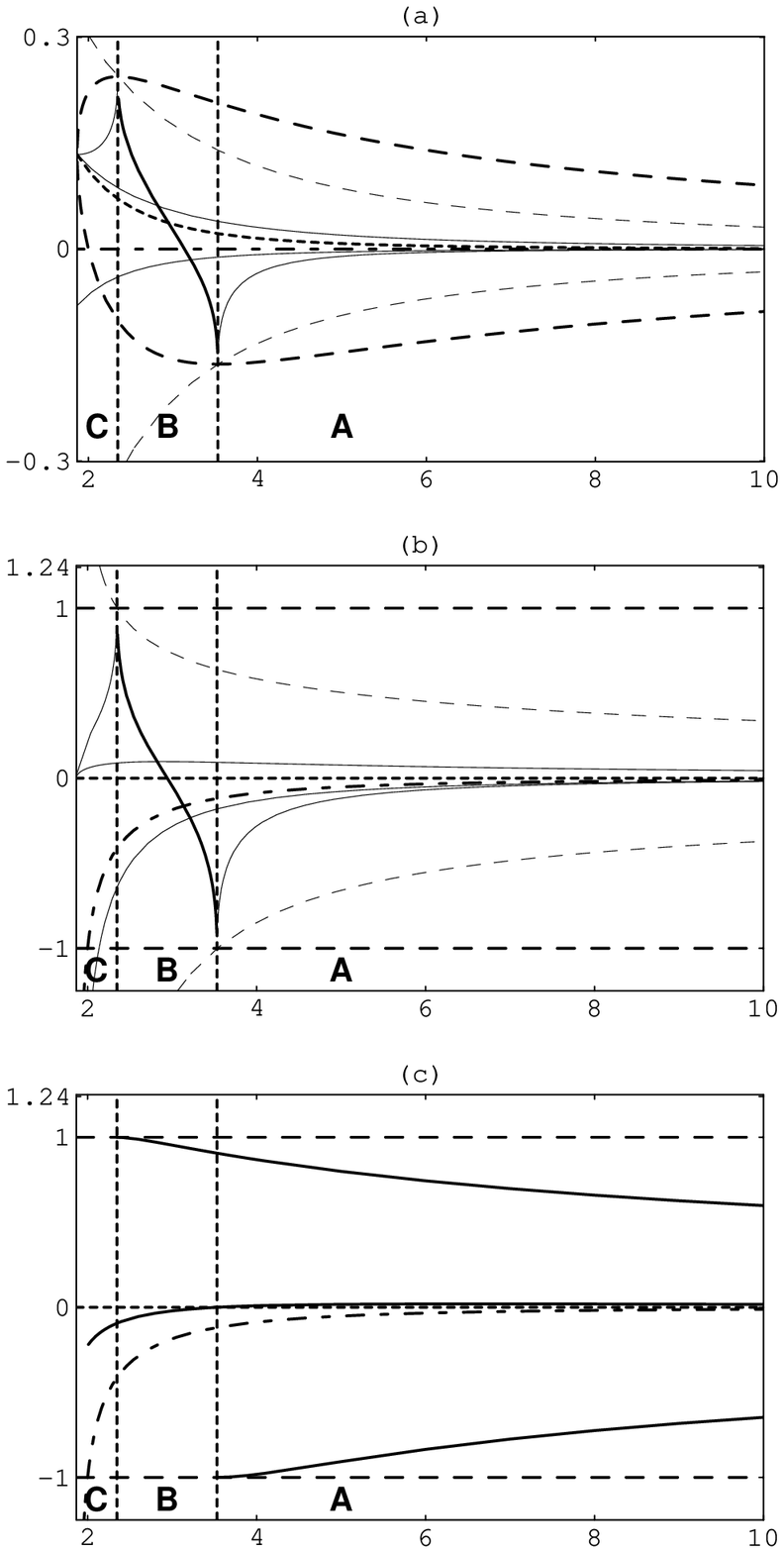}} 
\end{figure}

\begin{figure}[t]\footnotesize
\fcaption
{$(a)$. $\zeta$ versus $ \bar r$: plots of the coordinate angular
velocities of the preferred observer families in the equatorial plane
of the Kerr spacetime as a function of the radial coordinate $\bar r$
for $\bar a = 0.5$. The thick outer long dashed curves ($\zeta_+$
above, $\zeta_-$ below) correspond to the null orbits which enclose
the physical region of timelike orbits, a region divided in two by the
thick short dashed curve $\zeta_{\rm(sli)}$; these three curves meet
at the horizon at the left edge of the graph. The thick long-short
dashed horizontal axis corresponds to $\zeta_{\rm(thd)}=0$. The two
solid curves immediately above and below $\zeta_{\rm(sli)}$ and
$\zeta_{\rm(thd)}=0$ respectively are $\zeta_{\rm(car)}$ and
$\zeta_{\rm(gmp)}$. The thin dashed curves are $\zeta_{\rm(geo)\pm} =
\dot\phi_\pm$, which intersect the null orbit curves at the interface
between regions A, B, and C by definition. The solid curves in regions
A and C terminating in these same intersection points correspond to
($\zeta_{\rm(ext)}$).
The curve decreasing from the upper geodesic photon point to the lower
geodesic photon point in region B is 
$\zeta_{\rm(crit,spin)}$ where the threading precession is extremal. 
\\
$(b)$. $\nu$ versus $\bar r$:
the plot of the corresponding physical relative velocities in the
slicing point of view as a function of the radius. 
\\
$(c)$. $\nu$ versus $\bar r$:
the same plot as in $(b)$ showing the three physical relative
velocities at which the corrected spin precession $\zeta(U,m,E)^{\hat
z}$ vanishes. 
}
\label{fig:kerrobs}
\end{figure}

\section{Optical geometry and inertial forces in the static case}

The preceding companion article [BCJ1] shows how the optical gauge
conformal transformation of the spatial metric changes the relative
centripetal acceleration to the optical relative centripetal
acceleration and reshuffles the various spatial forces in the spatial
equation of motion. As discussed there, the optical geometry is only
natural in the static case, which in the present context applies to
the Schwarzschild limit of the Kerr spacetime. 

Those results are easily specialized to this latter case, so that one
can see how the various radial spatial forces for circular orbits are
transformed under the conformal transformation. Agreeing that
orthonormal components of optical quantities are normalized in the
optical geometry, the transformation of the signed Lie relative
curvature (equation (14.3) of [BCJ1]) is 
\begin{equation}
   \tilde\kappa(\phi,u)^{\hat r} 
        = \sigma^{-1} [ \kappa(\phi,u)^{\hat r} 
                          - (\ln\sigma)_{,\hat r} ]  \ ,
\end{equation}
while the curvatures themselves are
\begin{equation}
   \tilde\kappa(\phi,u)^{\hat r} = -(1-3{\cal M}/r)/r\ ,\quad
         \kappa(\phi,u)^{\hat r} = -(1-2{\cal M}/r)^{1/2}/r\ .
\end{equation}
With the static relation $g(u)^{\hat r} = (\ln\sigma)_{,\hat r}$ that
follows from the optical choice $\sigma = M^{-1}$ or $\sigma = N^{-1}$
respectively, this leads to equation (A.23) of [BCJ1] for the
transformation of the centripetal acceleration, here specialized to
orthonormal components 
\begin{equation}
     \sigma^{-1}
     \tilde a_{(\rm tem)}^{(\bot)}(U,u)^{\hat{r}}
         =  a_{(\rm tem)}^{(\bot)}(U,u)^{\hat{r}}
                - |\nu(U,u)^{\hat\phi}|^2 g(u)^{\hat{r}} \ .
\end{equation}
For comparison with the literature, note that the conformal factor is
not present in the covariant (unnormalized) coordinate component form
of this equation. 

Using this result in equation (\ref{eq:fFFFa}), one finds the
following expression for the physical force necessary to maintain the
circular orbit in terms of the Lie optical relative acceleration 
\begin{eqnarray}
  f(U)^{\hat{r}}
    &=& 
      \sigma^{-1}
      \gamma(U,u)^2 \tilde{a}_{\rm(lie)}^{(\bot)}(U,u)^{\hat{r}} 
         - \gamma(U,u) F^{\rm(G)}_{\rm(lie)}(U,u)^{\hat{r}}
         + \gamma(U,u)^2 |\nu(U,u)^{\hat\phi}|^2 g(u)^{\hat{r}}
                     \nonumber\\
    &=&    \sigma^{-1}
       \gamma(U,u)^2 \tilde{a}_{\rm(lie)}^{(\bot)}(U,u)^{\hat{r}} 
          - g(u)^{\hat r} \ ,
\end{eqnarray}
where the extra term $\gamma(U,u)^2 |\nu(U,u)^{\hat\phi}|^2
g(u)^{\hat{r}}$ which comes from the conformal transformation of the
centripetal acceleration combines with the term 
$- \gamma(U,u)^2 g(u)^{\hat{r}}$\typeout{*** margin problems: 3.3pts;
  enough to worry??}
in the spatial gravitational force to form a
velocity-independent force, namely minus the gravitoelectric field
$-g(u)^{\hat{r}}$. Abramowicz refers to this as the gravitational
force \cite{a93}, but with respect to the local rest space of the test
particle itself, not the static observers fixed in the geometry of the
spacetime. 

The remaining term in the physical force, namely the sign reversal of
the Lie optical relative centripetal acceleration multiplied by the
proper time correction factor $\gamma(U,u)^2$ and the conformal
transformation factor for relative acceleration, is what Abramowicz
identifies as his centrifugal force in the static case. The Lie
optical relative centripetal acceleration is (see [BCJ1]) is
explicitly 
\begin{equation}
     \tilde{a}{}_{\rm(lie)}^{(\bot)}(U,u)^{\hat{r}}
       = -  |\tilde \nu(U,u)^{\hat\phi}|^2
            [\ln \tilde{g}(u)_{\phi\phi}{}^{1/2}]_{,\hat{r}} \ ,
\end{equation}
where $\tilde\nu(U,u)^{\hat r} = \sigma \nu(U,u)^{\hat r}$. In their
attempts to extend Abramowicz's work to the Kerr spacetime, Iyer and
Prasanna \cite{ip93,nayvis96,pra96} evaluate this result for the
slicing decomposition, while Prasanna and Chakrabarti \cite{pc90}
evaluate it for the threading decomposition, modulo conformal factors.

The step which goes from the spatial equation of motion expressed in
terms of the sum of the relative centripetal acceleration and the
spatial gravitoelectric force to the same equation expressed in terms
of the sum of the optical relative centripetal acceleration and the
Abramowicz gravitational force in the Schwarzschild case is explicitly
\begin{eqnarray}
  f(U)^{\hat r} &= &
      \left[ -\gamma(U,u)^2|\nu(U,u)^{\hat\phi}|^2 
                \frac{r-2{\cal M}}{r^2} 
                 + \gamma(U,u)^2 \frac{{\cal M}}{r^2} \right]
         (1-2M/r)^{-1/2}
                \nonumber\\
      &=& \left[- \gamma(U,u)^2 |\nu(U,u)^{\hat\phi}|^2 
                  \frac{r-3{\cal M}}{r^2} 
                         + \frac{{\cal M}}{r^2}\right]
          (1-2M/r)^{-1/2} 
            \ .
\end{eqnarray}
This clearly shows the result that at $r=3{\cal M}$, the physical
force acting on the test particle is independent of its velocity. 
This last property of the physical force is lost in the corresponding
stationary case because of the presence of the velocity-dependent
gravitomagnetic force. 

\section{Spin precession} \label{sec:spin}

A test gyroscope following a test particle world line in spacetime
undergoes Fermi-Walker transport along that worldline. Precession of
the spin direction is a relative effect, depending on both the family
of test observers and the single test observer following the test
particle world line. Spin precession for geodesic motion will be
analyzed first, and then followed by the case of accelerated motion.
Papapetrou \cite{pap} has shown that a spinning particle does not
follow a geodesic in spacetime but is subject to Riemannian tensor
forces. These forces and any non-gravitational forces acting on a
gyroscope are represented by the force $F(U,u)^\alpha$, which leads to
the Thomas precession term given below for accelerated motion. 

The precession angular velocity of the spin of a gyroscope (with
4-velocity $U^\alpha$) as seen in its own local rest space is a
relative effect, parametrizing the derivative of the relative boost
between the local rest space of an observer congruence and the local
rest space of the gyroscope along its world line. Given an
observer-adapted orthonormal spatial frame $\{E_a{}^\alpha\}$ on
spacetime which is tied to the observer congruence by spatial
co-rotating Fermi-Walker transport, one can boost the frame to the
local rest space of the gyroscope, and the components of its spin with
respect to that boosted frame will undergo a time-dependent rotation
with an associated proper time angular velocity having components 
\begin{equation}\label{eq:spinpre}
  \zeta(U,u,E)^a 
   = \gamma(U,u) [ \zeta_{\rm(cfw)}(U,u)^a 
                   + \zeta_{\rm(sc)}(U,u,E)^a ] \ ,
\end{equation}
where
\begin{eqnarray}\label{eq:zetacfw}
   \zeta_{\rm(cfw)}(U,u)^a 
     &=& -{\textstyle {1\over2}}  H(u)^a
               \nonumber\\ &&
     - \gamma(U,u)   [ \gamma(U,u) + 1]^{-1} 
             [\nu(U,u) \times_u F(U,u)]^a
               \nonumber\\ &&
       + [\gamma(U,u)   + 1 ]^{-1} \, 
         [\nu(U,u) \times_u F^{\rm(G)}_{\rm(fw)}(U,u)]^a
               \nonumber\\
   &=& \zeta_{\rm(gm)}(u)^a + \zeta_{\rm(thom)}(U,u)^a 
                            + \zeta_{\rm(geo)}(U,u)^a
\end{eqnarray}
is the precession angular velocity relative to a spatial (with respect
to $u^\alpha$) co-rotating Fermi-Walker frame transported along
$U^\alpha$ (and consisting of gravitomagnetic, Thomas, and geodesic
terms), while the space curvature term 
\begin{equation}
   \zeta_{\rm(sc)}(U,u,E)^a
      = {\textstyle {1\over2}} 
          \eta(u)^{abc} \Gamma(u)_{[b|d|c]}\nu(U,u)^d
\end{equation}
is the relative angular velocity of the latter frame with respect to
the observer-adapted frame $\{E_a{}^\alpha\}$, and the factor
$\gamma(U,u)$ in Eq.~(\ref{eq:spinpre}) is necessary to convert the
observer proper time derivative to the gyroscope proper time
derivative. Conversely, removing the gamma factor in
Eq.~(\ref{eq:spinpre}) one obtains the precession of the spin as seen
by the sequence of observers along its path, removed of the
directional distortions caused by the Lorentz boost between the two
local rest spaces. 

This description is valid for an arbitrary spacetime \cite{mfg}. For
the threading observers in the nonlinear reference frame associated
with post-Newtonian coordinates in that approximation to general
relativity, one obtains the Schiff formula for the precession with
respect to the ``distant stars.'' One can use those same general
results in the threading point of view for any of the three spacetimes
under consideration to obtain an exact precession formula for a
gyroscope in an equatorial plane circular orbit. The natural
observer-adapted (right-handed) orthonormal spatial frame
$E_a{}^\alpha$ which is tied to the threading observer congruence by
corotating Fermi-Walker transport (in the equatorial plane) consists
of unit vectors along the coordinate radial $r$ direction, the
observer local rest space $\phi$ direction, and the $-\theta$ or
positive $z$ direction, respectively. The relative observer boost of
this spatial frame to the local rest space of the gyro, completed to a
spacetime frame by the gyro 4-velocity, is called a phase-locked frame
by de Felice and Usseglio-Tomasset \cite{defuss92,defuss96} who
investigate who investigate the Kerr case, and it is also the
spacetime Serret-Frenet frame for the circular orbit discussed in
detail by Iyer and Vishveshwara \cite{iyevis93}, who treat the general
stationary axisymmetric case. 

For Kerr, the resulting spin precession formula has the same
interpretation as the Schiff formula, namely the precession of the
spin with respect to a local frame locked onto the distant stars. For
the G\"odel spacetime, one obtains the precession with respect to a
local frame locked into the perfect fluid source of the gravitational
field, which reduces to the gravitomagnetic term alone in the
corotating geodesic case where the relative velocity is zero. For the
rotating Minkowski spacetime, the circular geodesics correspond to
points fixed in the inertial coordinates of the slicing observers, so
the precession question is not interesting without considering
accelerated circular orbits, where the Thomas precession is obtained
in the slicing point of view \cite{mfg}. In all cases the only
nonvanishing component of the spin precession is 
\begin{eqnarray}\label{eq:spinpresph}
     \zeta(U,m,E)^{\hat z}
   &=& \gamma(U,m) 
       [ \zeta_{\rm(cfw)}(U,m)^{\hat z} 
             + \zeta_{\rm(sc)}(U,m,E)^{\hat z} ] \nonumber\\
   &=& \gamma(U,m) 
        \zeta_{\rm(sph)}(U,m,E)^{\hat z} \ ,
\end{eqnarray}
where $\zeta_{\rm(sph)}(U,m,E)^{\hat z}$ is the precession of the spin
with respect to the static spherical frame as seen by the sequence of
observers along the gyro world line. 

First consider the case of geodesic motion so that the Thomas
precession is zero. For an equatorial circular geodesic in an
axisymmetric stationary spacetime, the angular velocity terms in the
threading point of view take the form 
\begin{equation}
  \zeta_{\rm(cfw)}(U,m)^{\hat z} 
      = - {\textstyle {1\over2}} H(m)^{\hat z}
        - \frac{\nu(U,m)^{\hat{\phi}}}{\gamma(U,m)+1} 
                     F^{\rm(G)}_{\rm(fw)}(U,m)^{\hat{r}} \ ,
\end{equation}
where
\begin{equation}
  F^{\rm(G)}_{\rm(fw)}(U,m)^{\hat{r}} 
   = \gamma(U,m)[g(m)^{\hat{r}} 
       + {\textstyle {1\over2}} 
            \nu(U,m)^{\hat{\phi}} H(m)^{\hat z}] \ ,
\end{equation}
and
\begin{eqnarray}
  \zeta_{\rm(sc)}(U,m,E)^{\hat z}
   &=& - {\textstyle {1\over2}} \nu(U,m)^{\hat{\phi}} 
                 (\ln\gamma_{\phi\phi})_{,\hat{r}}
    = \nu(U,m)^{\hat{\phi}}\kappa(\phi,m)^{\hat{r}} 
                    \nonumber\\
   &=& \nu(U,m)^\phi 
         \mathop{\rm sgn}(\kappa(\phi,m)^{\hat{r}}) \,
                             R(m)/\rho(\phi,m)\ , 
\end{eqnarray}
where $R(m) = \gamma_{\phi\phi}{}^{1/2}$.

For the Kerr spacetime the relative velocities and gamma factors of
the circular geodesics are 
\begin{eqnarray}
  \nu(U_\pm,m)^{\hat{\phi}} 
    &=& \sqrt{\Delta}/[a\pm\sqrt{r/{\cal M} }(r-2{\cal M} )]
              \ ,\\ 
  \gamma(U_\pm,m) 
    &=& (1-\frac{2{\cal M}}{r} \pm a \sqrt{\frac{{\cal M}}{r^3}})
       \left[(1-\frac{2{\cal M}}{r})(1-\frac{3{\cal M}}{r}\pm 2a  
              \sqrt{\frac{{\cal M}}{r^3}}) \right]^{-1/2}\ ,
       \nonumber
\end{eqnarray}
and the gravitomagnetic, geodesic, and space curvature angular
velocities are 
\begin{eqnarray}
  \zeta_{\rm(gm)}(m)^{\hat z} 
         &=& -\frac{a{\cal M}}{r^2(r-2{\cal M} )} \ ,
                \nonumber\\
  \zeta_{\rm(geo)}(U_\pm,m)^{\hat z} 
    &=& -\frac{{\cal M} \gamma(U_\pm,m)\nu(U_\pm,m)^{\hat{\phi}}
        (-\sqrt{\Delta} + a\nu(U_\pm,m)^{\hat{\phi}})}
                       {(\gamma(U_\pm,m)+1)r^2(r-2{\cal M} )} 
                \nonumber\\
   &=&\pm \frac{\gamma(U_\pm, m) -1}{\gamma (U_\pm,m )}
                                 \sqrt{\frac{{\cal M}}{r^3}}\ ,
                \nonumber\\
  \zeta_{\rm(sc)}(U_\pm,m,E)^{\hat z} 
         &=& -\frac{\nu(U_\pm,m)^{\hat{\phi}}[r(r-2{\cal M} )^2
	-{\cal M} a^2]}{\sqrt{\Delta}r^2(r-2{\cal M} )}
                \nonumber\\
   &=& \frac{a{\cal M} }{r^2(r-2{\cal M} )} 
             \mp \sqrt{\frac{{\cal M} }{r^3}}\ ,
\end{eqnarray}
The sum of these three terms
\begin{eqnarray}\label{eq:zetasph}
   \zeta_{\rm(sph)}(U_\pm,m,E)^{\hat z} 
   &=&  \zeta_{\rm(gm)}(m)^{\hat z} 
      + \zeta_{\rm(geo)}(U_\pm,m)^{\hat z}
      + \zeta_{\rm(sc)}(U_\pm,m,E)^{\hat z}
                \nonumber\\
   &=&  \mp\gamma(U_\pm,m)^{-1}\sqrt{\frac{{\cal M}}{r^3}}
\end{eqnarray}
gives the total precession angular velocity of the spin with respect
to the spherical static frame in terms of the threading observer
proper time. It vanishes in the limit $\gamma(U_\pm,m)^{-1} \to 0$ as
$|\nu(U_\pm,m)| \to 1$ which occurs for the two freefall photon
orbits, corresponding to a locking of the spin to that frame. This is
consistent with the picture of a photon as a massless spinning
particle with spin along the direction of motion, i.e., locked to the
$\phi$-direction in the case of a circular orbit. On the other hand,
the precession from the point of view of the gyro has the same simple
form $\zeta(U_\pm,m,E)^{\hat z} = \mp \sqrt{{\cal M}/r^3}$ as in the
Schwarzschild limit. 

However, since the static spherical frame is locked to the radial
direction, it undergoes a rotation of $2\pi \mathop{\rm
sgn}\nu(U,n)^{\hat\phi}$ during each revolution with respect to
Cartesian-like frames along the gyro worldline and so the spin rotates
an additional $-2\pi \mathop{\rm sgn}\nu(U,n)^{\hat\phi}$ with respect
to the Cartesian-like frames. This corresponds to a precession angular
velocity of the spherical frame components of the spin of the
gyroscope equal to the orbital coordinate angular velocity which must
therefore be added to the space curvature precession to correct it,
once suitably scaled to account for the difference between coordinate
and observer proper times 
\begin{equation}\label{eq:dphidtau}
  d\phi_\pm / d\tau_{(U_\pm,m)}
   =\nu(U_\pm,m)^\phi 
   = \pm \sqrt{\frac{{\cal M} }{r^3}} 
                \sqrt{1-\frac{2{\cal M} }{r}} /
       \left[1-\frac{2{\cal M}}{r} 
               \pm a \sqrt{\frac{{\cal M} }{r^3}}\right] \ .
\end{equation}
The corrected space curvature angular velocity is then
\begin{eqnarray}\label{eq:Delta}
  \zeta_{\rm(sc,cor)}(U_\pm,m,E)^{\hat z} 
     &=& \zeta_{\rm(sc)}(U_\pm,m,E)^{\hat z}   
         + \nu(U_\pm,m)^\phi 
\\
     &=&   [1 + \mathop{\rm sgn}(\kappa(\phi,m)^{\hat{r}}) 
                      R(m)/\rho(U_\pm,m) ] \nu(U_\pm,m)^\phi 
                  \nonumber\\
     &=& \frac{a{\cal M} }{r^2(r-2{\cal M} )}
       \mp \sqrt{\frac{{\cal M} }{r^3}}
                \left(1-\frac{\sqrt{1-\frac{2{\cal M} }{r}}}
          {1-\frac{2{\cal M} }{r}
            \pm a\sqrt{\frac{{\cal M} }{r^3}}}\right) \ .
     \nonumber
\end{eqnarray}
For $r>r_{\rm(rs)}$ when $\mathop{\rm sgn}(\kappa(\phi,m)^{\hat{r}}) <
0$, this reduces to $[\Delta(m) / 2\pi] \nu(U_\pm,m)^\phi $, where the
quantity the quantity $\Delta(m)$ is the deficit angle of the tangent
cone to the embedding of the $r$-$\phi$ coordinate surface (with the
threading geometry) in $E_3$, as explained in the appendix. Thus
Thorne's conical deficit argument using the threading embedding space
\cite{tho81} holds exactly in this region. 

\begin{figure}[p!]\footnotesize
\centerline{\epsfbox{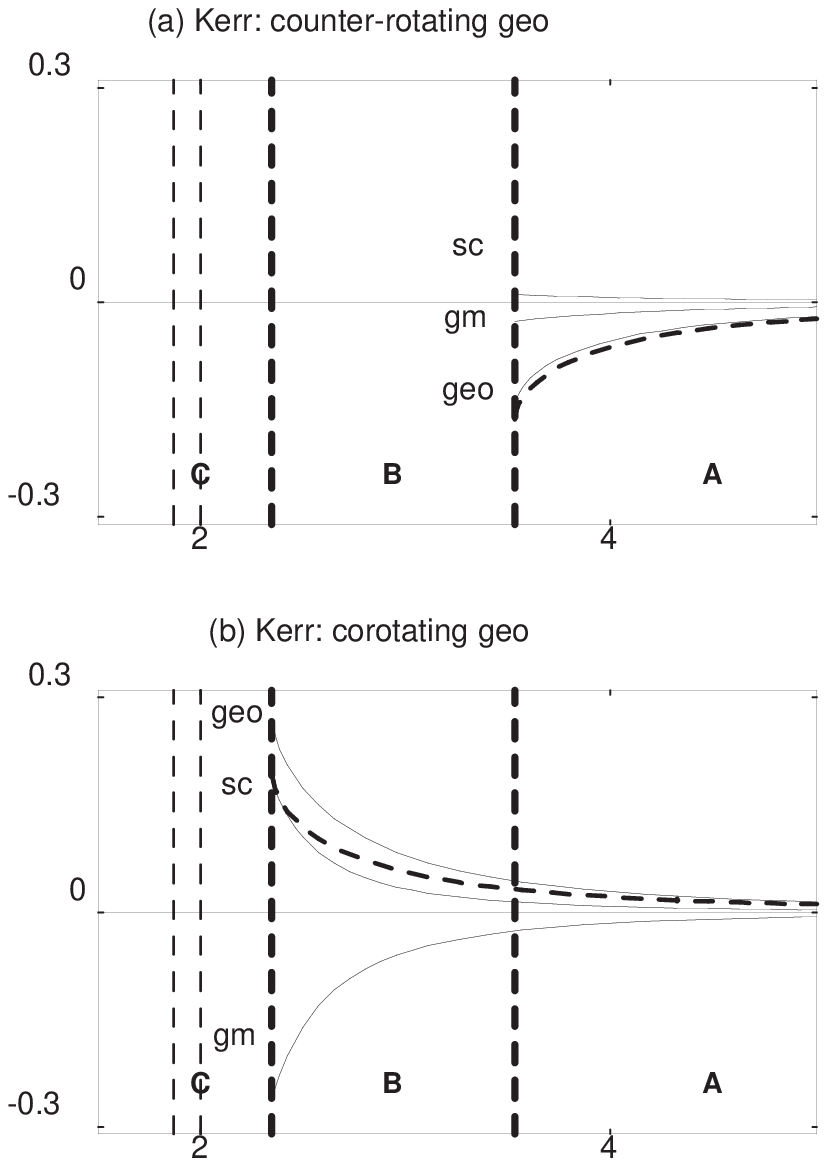}}
\fcaption
{$\zeta$ versus $\bar r$: plots of the geodesic, space curvature
(corrected), and gravitomagnetic precession angular velocity of the
spin of a gyroscope moving on $(a)$ counter-rotating and $(b)$
corotating equatorial circular geodesics of a Kerr black hole with
$\bar a=0.5$ as functions of the radius $\bar r$. Their sum gives the
corrected precession $\zeta_{\rm(cor)}(U,m,E)^{\hat z} / \gamma(U,m)$
with respect to the threading observers, shown as the long dashed
curve. Since these are all finite, the corrected precession terms
themselves all diverge at the ergosphere boundary.
}
\label{fig:gyropre}
\end{figure}

The total precession angular velocity of the gyroscope with respect to
the distant stars can then be written 
\begin{eqnarray}
  \zeta_{\rm(cor)}(U_\pm,m,E)^{\hat z}
      &=& \gamma(U_\pm,m)[\zeta_{\rm(gm)}(m)^{\hat z} 
             + \zeta_{\rm(geo)}(U_\pm,m)^{\hat z} 
             + \zeta_{\rm(sc,cor)}(U_\pm,m,E)^{\hat z}]
             \nonumber\\
  &=& \mp\sqrt{\frac{{\cal M}}{r^3}}[1-(1-\frac{3{\cal M}}{r}\pm 2a 
          \sqrt{\frac{{\cal M}}{r^3}})^{-1/2}] \ .
\end{eqnarray}
Figure \ref{fig:gyropre} shows the gravitomagnetic, geodesic, and
corrected space curvature angular velocity for extreme Kerr as
functions of $r/{\cal M}$. 

\typeout{*** Figure 8.  (gyropre)}

\typeout{*** (Partial page figure eating whole page, then delayed till 
              end of compuscript.)}
\typeout{*** (Made into full page figure on page 40.)}

\typeout{*** Table 7. (spinprec)}
\begin{table}[htbp]\footnotesize
\tcaption
{The spin precession $\zeta(U,m)^{\hat z}$ terms
for geodesic circular orbits, including
low speed limits indicated by arrows. 
The abbreviation $\gamma_-=\gamma (U_-,m)$ is used.
Note that $\gamma_- =C(1-S^2)^{-1/2}$ in the G\"odel case.
} 
 \typeout{*** eqnarray struts inserted}
 \def\Strut{\relax\hbox{\vrule width0pt height 10.5pt depth 5.5pt}}
\begin{eqnarray*}
\begin{array}{|l|l|l|l|} \hline
 &\rm Rotating        \hfill&\hbox{G\"odel:} \hfill&\rm Kerr: \hfill\\ 
 &\rm  Minkowski: U_- \hfill& U_-            \hfill&  U_\pm   \hfill
                      \\ \hline \Strut
 \zeta_{\rm(gm)}^{\hat z} 
    & -\Omega \gamma^2  \to -\Omega 
    & -\Omega  
    & \to  -\frac{a{\cal M}}{r^3} 
                      \\ \hline \Strut
 \zeta_{\rm(geo)}^{\hat z}
    & 0
    & -\Omega [1-\gamma_-^{-1}] \to 0 
    &  \to \pm {\textstyle {1\over2}} \sqrt{\frac{{\cal M}^3}{r^5}} 
                      \\ \hline \Strut
 \zeta_{\rm(sc,cor)}^{\hat z}
    & \Omega \gamma_- [\gamma_--1]  \to 0 
    & 2\Omega [1 - 1/C] \to 0
    & \to \pm \sqrt{\frac{{\cal M}^3}{r^5}}
                      \\ \hline \Strut
 \zeta_{\rm(cor)}^{\hat z}
    & -\Omega
    & \Omega[1-2/\sqrt{1-S^2}] \to -\Omega 
    & \to [\pm {\textstyle {3\over2}}  
             -\frac{a/{\cal M}}{\sqrt{{\cal M}/r}}]
                 \sqrt{\frac{{\cal M}^3}{r^5}}
                      \\ \hline
\end{array}
\end{eqnarray*}	
\label{tab:spinprec}
\end{table}

Table \ref{tab:spinprec} summarizes the corresponding results for the
rotating Minkowski and G\"odel spacetimes, including their small $r$
(low relative velocity) limits. This table also lists the small ${\cal
M}$, small $a$ limits for Kerr for comparison. These latter
expressions agree with the well known post-Newtonian results. but it
is important to note that the geodesic term in the precession formula
and the corrected space curvature term, arising from very different
phenomena in the strong field case, combine to form the single
``geodetic precession" term in this weak field limit. For the
counter-rotating circular geodesics in rotating Minkowski spacetime,
the precession corresponds to the rotation of the inertial coordinate
axes with respect to the threading observer axes. For the corotating
circular geodesics in the G\"odel spacetime, only the gravitomagnetic
precession is nonzero and has the same value as for the
counter-rotating geodesics. 

These results agree with those of Rindler and Perlick \cite{rinper}
and Iyer and Vish\-vesh\-wa\-ra \cite{iyevis88,iyevis93}, 
both of whom use
rotating coordinates $t'=t,\ \phi '=\phi - \omega t$ for which the
circular orbit has a fixed angular coordinate in order to find the
precession angle $\Delta\phi$ of the gyroscope after one loop. This 
precession angle $\Delta \phi$ and the precession angular velocity
$\zeta_{\rm(cor)}(U_\pm,m,E)^{\hat z}$ are related by a scale factor
and an integral over the loop 
\begin{eqnarray}
   \Delta \phi_\pm
   &=& \oint
      \zeta_{\rm(cor)}(U_\pm,m,E)^{\hat z} d\tau_{U_\pm} 
  = \pm \int_0^{2\pi} \zeta_{\rm(cor)}(U_\pm,m,E)^{\hat z}     
            \frac{d\tau_{U_\pm}}{d\phi} d\phi     
                              \nonumber\\
  &=& \pm [2\pi/\nu(U_\pm,m)^\phi] \gamma(U_\pm,m)^{-1}
           \zeta_{\rm(cor)}(U_\pm,m,E)^{\hat z} \ ,  
\end{eqnarray}
using equation (\ref{eq:dphidtau}) and recalling the proper time
conversion factor. Comparing this with equation (\ref{eq:Delta}) and
recalling the additional gamma factor in equation (\ref{eq:spinpre})
shows that the contribution to the total precession angle from the
corrected space curvature precession equals the threading deficit
angle $\Delta(m)$ modulo the sign 
\begin{equation}
  \Delta\phi_\pm{}_{\rm(sc,cor)}  = \pm\Delta(m) \ .
\end{equation}

Of course, with some more formula juggling one can obtain the spin
precession for the case of circular orbits with arbitrary
acceleration. To express this as a function of the relative velocity
and radius, one need only replace $F(U,m)^{\hat r}$ in the radial
component of the $u=m$ version of Eq.~(\ref{eq:zetacfw}) by
$\gamma(U,m)^{-1} f(U,m)^{\hat r}$, where the physical force
$f(U,m)^{\hat r}$ is itself expressed in terms of those variables by
Eq.~(\ref{eq:ffactorized}). For example, the Thomas precession term
has the form 
\begin{eqnarray}\label{eq:zetathom}
   \zeta_{\rm(thom)}(U,m)^{\hat z} 
    &=& \nu(U,m)^{\hat\phi} [ \gamma(U,m) + 1]^{-1} 
                              f(U,m)^{\hat r}
               \nonumber\\ 
    &=& \kappa(\phi,m)^{\hat r} \nu(U,m)^{\hat\phi}
            [ \gamma(U,m) + 1]^{-1} 
               \gamma(U,m)^2
               \nonumber\\ 
    & & \times
               [\nu(U,m)^{\hat\phi} - \nu(U_-,m)^{\hat\phi}]
               [\nu(U,m)^{\hat\phi} - \nu(U_+,m)^{\hat\phi}]\ .
\end{eqnarray}
Gathering the preliminary expressions for the other precession terms
and doing some involved algebra in which Eqs.~(\ref{eq:kgH}) are
helpful leads to the following general result for the total precession
angular velocity 
\begin{eqnarray}\label{eq:zetaacc}
\typeout{*** needs flushleft macro here in this eqnarray or spills into
             right margin}
 & &   \zeta(U,m,E)^{\hat z} 
       = {\textstyle\frac12} \gamma(U,m)^{-2} (d {\cal F}/d\nu)
                (\nu(U,m)^{\hat\phi}; \kappa(\phi,m)^{\hat\phi},
                 \nu(U_-,m)^{\hat\phi}, \nu(U_+,m)^{\hat\phi})
               \ ,\nonumber\\ 
 & &  \zeta_{\rm(cor)}(U,m,E)^{\hat z} 
        =  \zeta(U,m,E)^{\hat z} 
             +\gamma(U,m) \nu(U,m)^{\hat\phi} 
                    / \gamma_{\phi\phi}{}^{1/2} \ ,
\end{eqnarray}
generalizing Eq.~(\ref{eq:zetasph}) to the accelerated case for all
the spacetimes under consideration. 

The precession angular velocity $\zeta(U,m,E)^{\hat z}$ thus vanishes
at the critical points of the radial physical force function,
corresponding to the locking of the spin to the static spherical frame
for the extremal force observers. These observers act as the boundary
between the counter-rotating and corotating spin precession orbits
relative to the radial direction. Table \ref{tab:spinprecsign}
correlates the sign of this precession with these intervals of
relative velocity values in each of the three regions A, B, and C 
assuming $\nu_- + \nu_+ < 0$. In the limiting Schwarzschild case where
region B disappears and $\nu_{\rm(ext)} = 0$, the spin precession 
$\zeta(U,m,E)^{\hat z}$ has the opposite sign compared to the angular
velocity in region A and the same sign in region C. 

\typeout{*** Table 8. (spinprecsign)}
\begin{table}[htbp]\footnotesize
\tcaption
{Signs of the spin precession for circular orbits of a given  
angular relative velocity.
} 
 \typeout{*** eqnarray struts inserted}
 \def\Strut{\relax\hbox{\vrule width0pt height 10.5pt depth 5.5pt}}
\begin{eqnarray*}
\begin{array}{|c|c|c|c|} \hline
 & {\rm Region\ A} \qquad & {\rm Region\ B} \qquad & {\rm Region\ C} 
     \\ \hline \Strut
 \zeta(U,m,E)^{\hat z} > 0 
         & (-1, \nu_{\rm(ext)} ) & \hbox{---} &  (\nu_{\rm(ext)}, 1)
     \\ \hline \Strut
 \zeta(U,m,E)^{\hat z} = 0 
         & \nu_{\rm(ext)}  & \hbox{---} &  \nu_{\rm(ext)} 
     \\ \hline \Strut
 \zeta(U,m,E)^{\hat z} < 0 
         & ( \nu_{\rm(ext)}, 1 ) & (-1,1) &  (-1, \nu_{\rm(ext)} )
     \\ \hline
\end{array}
\end{eqnarray*}	
\label{tab:spinprecsign}
\end{table}

Using the properties of the function ${\cal F}$ introduced in
Eq.~(\ref{eq:calF}) and the relation (\ref{eq:kgH}), the gyro spin
precession angular velocity with respect to the static spherical frame
from its own point of view takes a very simple form analogous to the
physical acceleration it experiences 
\begin{eqnarray} 
     \zeta(U,m,E)^{\hat z} 
        &=&  - {\cal F}(\nu(U,m)^{\hat\phi}; H(m)^{\hat z}/2, 
                      \nu_{\rm(crit)-}(U,m)^{\hat\phi},
                      \nu_{\rm(crit)+}(U,m)^{\hat\phi} )
          \ ,\nonumber\\
     a(U)^{\hat r} 
        &=&  {\cal F}(\nu(U,m)^{\hat\phi}; \kappa(\phi,m)^{\hat r}, 
                      \nu(U_-,m)^{\hat\phi},
                      \nu(U_+,m)^{\hat\phi} )
          \ .
\end{eqnarray}
One can then carry over to the spin precession the entire analysis of
the extrema of the physical force function $a(U)^{\hat r}$. Defining a
relative difference velocity by Eq.~(\ref{eq:nurel}) with $\nu_\pm$
replaced by $\nu_{\rm(crit)\pm}$ leads to $\nu_{\rm(rel,spin)} =
1/\nu_{\rm(rel)}$ which is in the physical range only in region B
where $|\nu_{\rm(rel)}|>1$, $\nu_{\rm(crit)\pm}$ are complex, and the
spin precession is always negative. 
 
Only in this region are the corresponding critical values related to
this new relative velocity by Eq.~(\ref{eq:nucrit}) real 
\begin{equation}
    \nu_{\rm(crit, spin)\pm}
     = {\cal V}_\pm (1/\nu_{\rm(rel)})
     = \nu_{\rm(rel)} \left[1\pm \sqrt{1-1/\nu_{\rm(rel)}^2}\,\right] \ .
\end{equation}
As before only the minus root is in the physical range leading to the
spin extremal velocity 
\begin{equation}
    \nu_{\rm(ext,spin)}
     = \nu_{\rm(rel)} \left[1 - \sqrt{1-1/\nu_{\rm(rel)}^2}\,\right] 
\end{equation}
at which a maximum negative (but minimum absolute value) spin
precession angular velocity occurs with value 
\begin{equation}\label{eq:spinFext}
   -{\cal F}(\nu_{\rm(ext,spin)};
             H/2,\nu_{\rm(crit)-},\nu_{\rm(crit)+})
    =
   - {\textstyle \frac12} H [2 + \gamma_{\rm(rel,spin)}] 
                            / \gamma_{\rm(rel,spin)}\ .
\end{equation}
Figure \ref{fig:rok}$(b)$ illustrates the geometry of the relative
observer plane associated with this analysis. 

\begin{figure}[p!]\footnotesize
\centerline{\epsfbox{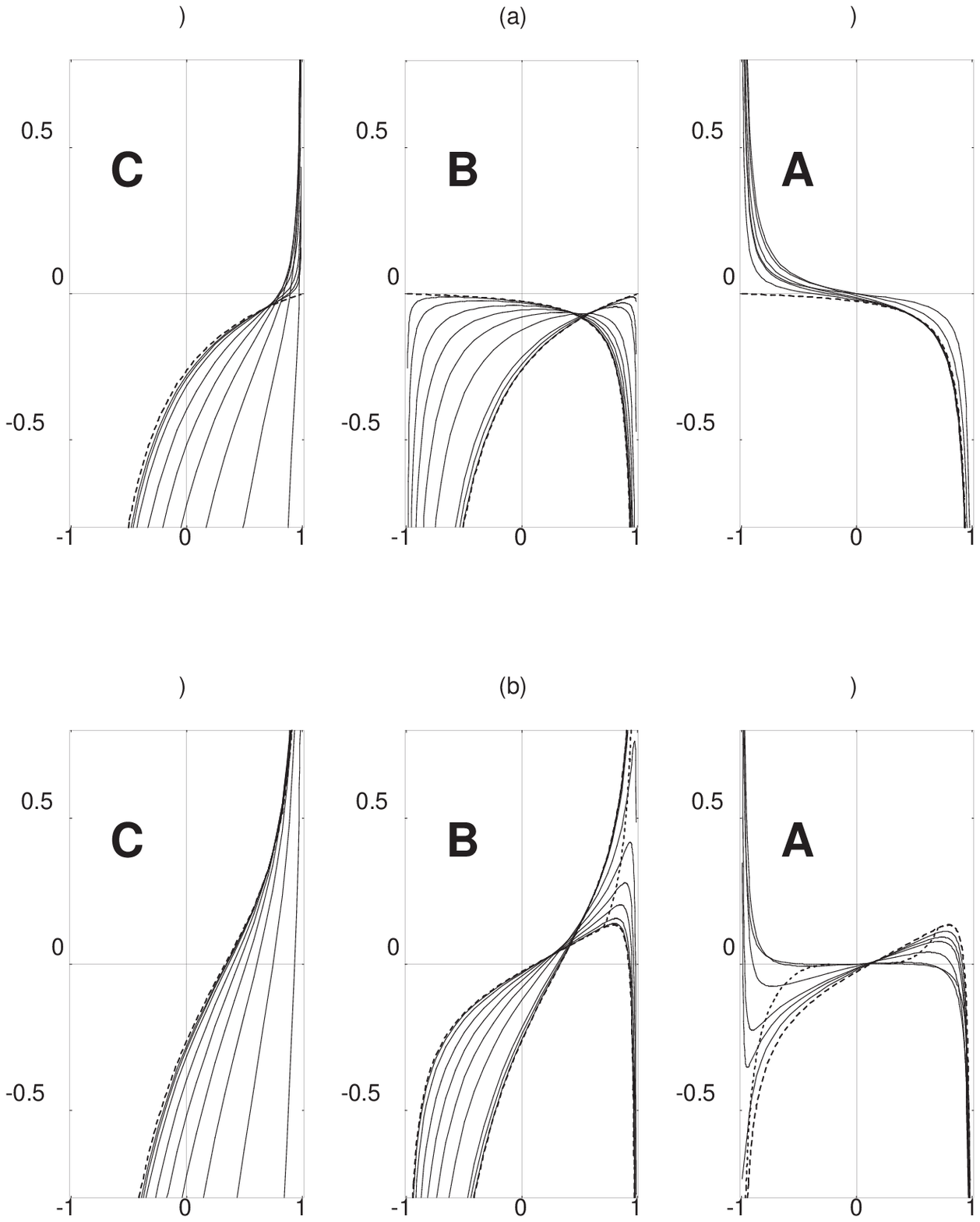}}
\fcaption
{$\zeta$ versus $\nu$: plots of $(a)$ the precession angular velocity
$\zeta(U,m,E)^{\hat z}$ and $(b)$ the corrected precession angular
velocity $\zeta_{\rm(cor)}(U,m,E)^{\hat z}$ of the spin of a gyroscope
versus the orbital physical angular velocity for selected radii for
equatorial circular orbits of a Kerr black hole with $a=0.5$.
}
\label{fig:prec}
\end{figure}

Figure~\ref{fig:prec} shows both $(a)$ the spin precession
$\zeta(U,m,E)^{\hat z}$ and $(b)$ the corrected spin precession
$\zeta_{\rm(cor)}(U,m,E)^{\hat z}$ at the same selected radii as a
function of the threading relative velocity for a Kerr black hole with
$\bar a=0.5$ outside the ergosphere where the spin precession with
respect to the distant stars makes sense. The intersections of these
curves with the short dashed curves in Figure~\ref{fig:prec}$(b)$
occur at the velocities of the geodesic orbits, corresponding to
Figure~\ref{fig:gyropre}. The long dashed curves mark the boundaries
$r_{\rm(AB)}$, $r_{\rm(BC)}$ between the regions A, B, and C. 
One can see three families of observers in region A for which the
corrected precession vanishes. Upon closer inspection one sees that
there is one pair which respectively corotate and counterrotate faster
than the pair of timelike circular geodesics and share the same
respective observer horizons, and another family near the slicing
observers which has the ergosphere as its horizon.

\typeout{*** Figure 9.  (prec)}
\typeout{*** (Full page figure on page 44.)}

Working instead with the coordinate angular velocity, de Felice
\cite{def95} has shown (following from a comparison of \Refs\citen{def95}
and [\citen{defuss91}], see also \Ref\citen{sem96}) 
and implied the fact that the corotating
Fermi-Walker spin precession vanishes exactly at the two critical
points of the radial force function, and is positive between them and
negative outside. In the approach of Iyer and Vishveshwara
\cite{iyevis93} for stationary axially symmetric spacetimes,
specialized to their equatorial plane case $\tau_2 = 0 = {\cal
A}_{(2)}$, the precession angular velocity $ \zeta(U,m,E)^{\hat z}$ is
just the sign-reversal of the first Serret-Frenet torsion $\tau_1$ and
the acceleration $a(U)^{\hat r}$ of the test particle (physical force)
is the Serret-Frenet curvature $\kappa$. For the equatorial plane case
their Eq.~(50) in their notation reduces to 
\begin{eqnarray}
  \tau_1 &=& {\textstyle\frac12} {\cal A} / \sqrt{\Delta_3} 
                   \partial_\omega \kappa
        = -{\textstyle\frac12} 
               [ M\Gamma(U,m)^{2} \gamma_{\phi\phi}{}^{1/2} ]^{-1}
                   \partial_\zeta a(U)^{\hat r}
               \nonumber\\
        &=& -{\textstyle\frac12} \gamma(U,m)^{-2}
                   \partial_\nu {\cal F}
        = -\zeta(U,m,E)^{\hat z} \ .
\end{eqnarray}
This first torsion is called the Fermi drag by de Felice. In the case
of general motion along a Killing trajectory in a stationary axially
symmetric spacetime, one can show that the gyro spin precession
angular velocity (Eq.~(20) of \Ref\citen{iyevis93}) relative to the
Frenet-Serret frame $E_a{}^\alpha$ is just 
\begin{equation}
   -\omega(U)_{\rm(FS)}{}^\alpha 
      = P(U,u)^\alpha{}_\beta \zeta(U,u,E)^\beta \ .
\end{equation}

\section{Sagnac effect and synchronization defect}

The Sagnac effect \cite{pos,ash,van,hen} and its timelike analog, both
of which in turn are connected to the synchronization defect
\cite{baz}, refer to the asymmetry in the arrival times of a pair of
oppositely rotating timelike geodesic or null circular orbits at a
given radius as seen by a given rotating observer.
If $(\zeta_1,\zeta_2)$ is the ordered pair of coordinate angular
velocities of such a pair (either $(\zeta_-,\zeta_+)$ or
$(\dot\phi_-,\dot\phi_+)$), and $\zeta$ is the angular velocity of a
rotating observer with 4-velocity $U$ distinct from this pair, one 
easily finds that
the difference and average of the coordinate arrival times after one
complete revolution with respect to this observer are 
\begin{eqnarray}
   \Delta t 
     &=& {\cal S}(\zeta;\zeta_1,\zeta_2)
       = t_2 - t_1
       =  2\pi\left[ 1/(\zeta_2-\zeta) - 1/(\zeta-\zeta_1) \right]
                \nonumber\\
     &=& -4\pi [ \zeta - (\zeta_1+\zeta_2)/2]
                           / [(\zeta-\zeta_1)(\zeta-\zeta_2)] \ ,
                \nonumber\\
   t_{\rm(avg)} &=&  (t_1+t_2)/2
     = 2\pi\left[ 1/(\zeta_2-\zeta) + 1/(\zeta-\zeta_1) \right]
                \nonumber\\
     &=& \pi [ \zeta_1-\zeta_2]
                     / [(\zeta-\zeta_1)(\zeta-\zeta_2)] \ ,
\end{eqnarray}
the ratio of which gives the relative difference in the coordinate
arrival times 
\begin{equation}
   \Delta t/ t_{\rm(avg)} 
      = 4[ \zeta - (\zeta_1+\zeta_2)/2] / 
                              [ (\zeta_1 - \zeta_2)/2 ] \ .
\end{equation}

For the pair of oppositely rotating timelike geodesics one has 
\begin{equation}
   \Delta t_{\rm(geo)}(U) 
       = {\cal S}(\zeta;\dot\phi_-,\dot\phi_+)
       = -4\pi [\zeta - \zeta_{\rm(gmp)}] /
                    [(\zeta-\dot\phi_-)(\zeta-\dot\phi_+)] \ ,
\end{equation}
while for the pair of oppositely rotating null orbits one has
\begin{equation}
   \Delta t_{\rm(null)}(U) = {\cal S}(\zeta;\zeta_-,\zeta_+)
            = -4\pi [\zeta - \zeta_{\rm(nmp)}] /
                               [(\zeta-\zeta_-)(\zeta-\zeta_+)] \ ,
\end{equation}
recalling that $\zeta_{\rm(nmp)} = \zeta_{\rm(sl)}$. Each of these may
be evaluated for the slicing, threading, and extremal force observers,
when the three orbits are distinct for a given application. Note that
the null arrival time difference is proportional to the angular
momentum (\ref{eq:angmom}) of $U$. 

The Sagnac time difference and its timelike geodesic analog are given
when the observer is taken to be the threading observer 
\begin{eqnarray}
     \Delta t_{\rm(null)}(m) 
            &=& {\cal S}(0;\zeta_-,\zeta_+)
             = 4\pi \zeta_{\rm(nmp)} / (\zeta_- \zeta_+)
                   \nonumber\\
            &=& 4\pi (\zeta_-{}^{-1} + \zeta_+{}^{-1})/2
             = 4\pi M_\phi \ ,
           \nonumber\\
     \Delta t_{\rm(geo)}(m) 
            &=& {\cal S}(0;\dot\phi_-,\dot\phi_+)
             = 4\pi \zeta_{\rm(gmp)} / (\dot\phi_- \dot\phi_+)
                  \nonumber\\
            &=& 4\pi (\dot\phi_-{}^{-1} + \dot\phi_+{}^{-1})/2
             = 4\pi a \quad \hbox{[Kerr]} \ .
\end{eqnarray}

The null difference is positive/negative when the threading observers
corotate/\-coun\-ter-rotate with respect to the slicing observers 
(${\rm sgn}\, \Delta t_{\rm(null)} = {\rm sgn}\, N^\phi = {\rm sgn}\,
M_\phi$).
\typeout{*** margin problem here: 4 pts; enough to worry about??}
The geodesic difference is positive assuming an upward 
gravitomagnetic field 
\hfil\break\typeout{*** hfil break inserted}
(${\rm sgn}\, \Delta t_{\rm(geo)} = {\rm sgn}\,
H^{\hat z}$), with the counterrotating geodesic returning first to the
threading observer ($t_2<t_1$, where 2 labels the corotating
geodesic). The geodesic formula is not valid for G\"odel since the
threading observer is itself geodesic, nor for Minkowski which only
has one geodesic. 

The synchronization defect is just half the Sagnac time difference and
indicates the change in coordinate time which occurs during one
corotating spatial loop ($\phi$ coordinate line in spacetime)
\begin{equation}
    \Delta t_{\rm(SD)}(m) = \int_0^{2\pi} M_\phi\, d\phi
                 = 2\pi M_\phi = \Delta t_{\rm(null)}(m) /2 \ .
\end{equation}
The analogous coordinate time difference for the geodesic case
in the Kerr spacetime
\begin{equation}
   2\pi / \bar\zeta_{\rm(car)} = \Delta t_{\rm(geo)}(m) /2
       = 2\pi a
\end{equation}
corresponds exactly to one threading loop of a circular curve which is
spatial with respect to a Carter observer ($\zeta_{\rm(car)} =
a/(a^2+r^2)$, so $\bar\zeta_{\rm(car)} = 1/a$ by Eq.~(\ref{eq:barU})),
connecting the average-time-of-return point on the threading observer
world line with either return point on the same world line. 

Figure \ref{fig:sagnac} shows the relationship between the arrival
times of oppositely rotating circular geodesic and null orbits and
the Sagnac effect and its timelike geodesic analog in region A. This
is illustrated for the case in which the shift is negative as in the
Kerr spacetime, i.e., the threading observers counter-rotate with
respect to the slicing observers, so the corotating null orbit returns
first to the threading observer from which it originated. Independent
of the sign of the shift but assuming that the gravitomagnetic field
is upward, the geodesic meeting point observer counter-rotates with
respect to both the slicing and threading observers, and the extremal
force observers counter-rotate in turn with respect to the geodesic
meeting point observers. Even though the extremal force observers see
the oppositely rotating pair of geodesics moving with the same speed,
the counter-rotating geodesic arrives after the corotating one as
shown. 

\typeout{*** Figure 10. (sagnac)}

\typeout{*** (Full page figure on page 47
         with caption on bottom of page 46 to left.)}

\begin{figure}[b]\footnotesize
\fcaption
{The conformal diagram of the front half $[-\pi/2,\pi/2]\times R$ of
the flattened $t$-$\phi$ coordinate cylinder for a fixed $r$ in region
A, showing the null and geodesic time differences for corotating and
counter-rotating orbits for the various observer families. The
vertical axis is along the (null meeting point) slicing observers
($n$), the time coordinate axis is along the threading observer
direction ($m$), and the horizontal axis is along the $\phi$
coordinate direction (constant $t$). The diagram illustrates the case
in which the slicing observers corotate with respect to the threading
observers as in Kerr so that the time lines tilt to the left. The
relative observer plane tangent space (region A case) is superimposed
on the cylinder together with the corresponding world lines emanating
from a given single initial point. The slicing, threading, geodesic
meeting point, and extremal force observer 4-velocities and world
lines are shown, together with the null 4-momenta $P_\pm^\alpha =
n^\alpha \pm \hat e(n)^\alpha$ of the two oppositely rotating null
orbits. The synchronization defect is shown at the bottom of the
figure corresponding to one loop of a threading purely spatial curve
with unit tangent $\hat e(m)^\alpha$. 
}
\label{fig:sagnac}
\end{figure}

\begin{figure}[p]\footnotesize

\typeout{*** *** pictex  loading figure 10: 
                    relative observer tube}
$$ \vbox{
\beginpicture

\typeout{*** *** Units set at 0.75cm tickmarks.}
  \setcoordinatesystem units <0.75cm,0.75cm> point at 0 0 


    \putrule from  -4.575 23   to 3.425 23 
    \putrule from  -3.975 -1 to 4.025 -1   
    \putrule from  -4 0   to 5 0           
    \putrule from  0 0   to 0 23           
    \putrule from  0 0   to 0 -1           

\setdashes

    \putrule from 3.425 23 to 4 23 
    \putrule from 4 19.32  to 4 23      
    \putrule from 4 0  to 4 18.33       
   

\setlinear
\setsolid                    

  \plot 0 0 -0.625 24 /          
  \plot 0 0  -1 0.0075   /       

  \plot -4 0 -4.575 23 /      
  \plot  4 0  3.425 23 /      

  \plot -4 0 -3.975 -1 /      
  \plot  4 0 4.025 -1  /      

  \plot 0 0 4 -0.1 /          
  \plot -4 -0.3 0 -0.4  /     
  \plot 0 -0.4 2.0625 -0.4525 / 


\setlinear
\setsolid                    
  \plot 0 0 2 2 /               
  \plot 0 0   -2 2 /            

\setsolid


\setdashes 
\setquadratic

\plot
0 2 -0.4 2.03961 -0.8 2.15407 -1.2 2.33238 -1.6 2.56125 
-2 2.82 -2.4 3.1241 -2.8 3.44093 -3.2 3.77359 -3.4 3.94462
-3.8 4.29418  /

\plot
0 2 0.4 2.03961 0.8 2.15407 1.2 2.33238 1.6 2.56125 
2 2.82 2.4 3.1241 2.8 3.44093 3.2 3.77359 3.4 3.94462
3.8 4.29418 /

\setlinear

\setdashes
  \plot  2 2   3.9 3.9  /       
  \plot  -2 2   -4.1 4.1 /      

  \plot -4.25 11.75 3.5 19.5 /  
  \plot 3.7 12.3 -4.5125 20.5 / 

\setsolid

\plot 0 0 -3.88 4.37  /              
\plot 0 0 -4.1 4.5817 /              
\plot 3.7 13.6 -4.575 22.9 /         

\plot 0 0  2.67 3.33  /              
\plot 0 0  3.9 4.9    /              
\plot -4.3 14.6 2.42 23 /            

\plot 0 0 -1.115 23      /           
\plot 0 0 -3.62 23      /            

\plot 0 0 -1.21 7.70 /               

\setdashes <2pt>


\plot -3.88 4.37 -1.21 7.70 /        
\plot 2.67 3.33  -1.21 7.70 /        
\plot -2 2 2 2 /                     


\put {\mathput{\bullet}}                at  -2.62 16.68

\put {\mathput{\bullet}}                at  0 19.95
\put {\mathput{\bullet}}                at  -0.5 19.32
\put {\mathput{\bullet}}                at  -0.9 18.83
\put {\mathput{\bullet}}                at  -0.45 18.3
\put {\mathput{\bullet}}                at  0 17.8

\put {\mathput{\bullet}}                at  -3.375 21.55

\put {\mathput{\bullet}}                at  -3 19
\put {\mathput{\bullet}}                at  -0.83 16.87
\put {\mathput{\bullet}}                at  -0.4 16.4
\put {\mathput{\bullet}}                at  0 16
\put {\mathput{\bullet}}                at  -2.18 13.82
\put {\mathput{\bullet}}                at  -0.737 15.22
\put {\mathput{\bullet}}                at  -0.3725 15.63

\put {\mathput{\bullet}}                at  0 0
\put {\mathput{\bullet}}                at  0 -0.4


\plot -3.375 21.55 3.45 21.55 / 

\plot 0 19.95 3.4 19.95 /      
\plot -0.5 19.32 3.515 19.32 / 
\plot -0.9 18.83 3.53 18.83 /  
\plot -0.45 18.3 3.55 18.3 /   
\plot  0 17.8 3.57 17.8 /      

\plot -2.6 16.7 3.6 16.7 /     

\setdashes <4pt>


\plot -4 -0.4 0 -0.4 /              


\plot -4.47 19 -3 19 /              

\plot -4.41 16.87 -0.83 16.87 /     
\plot -4.4 16.4 -0.4 16.4 /         
\plot -4.385 16 0 16 /              
\plot -4.3725 15.63 -0.3725 15.63 / 
\plot -4.36 15.22 -0.737 15.22 /     

\plot -4.32 13.82 -2.18 13.82 /     

\setsolid

\put {\mathput{n}}                  [lb]   at  0.2 23.25
\put {\mathput{m}}                  [lb]   at  -0.4 23.25
\put {\mathput{U_{\rm (gmp)}}}      [rb]   at  -0.7 23.25
\put {\mathput{U_{\rm (ext)}}}      [b]   at  -3.5 23.25
\put {\mathput{U_-}}                [b]   at  -4.575 23.25
\put {\mathput{U_+}}                [b]   at  2.5 23.25

\put {\mathput{U_-}}                [lb]   at  -3 4
\put {\mathput{U_+}}                [rt]   at  2 3.5
\put {\mathput{U_+ + U_-}}          [rb]   at  -1.7 7.5

\put {\mathput{P_+}}                [lt]   at  -3.5 12
\put {\mathput{P_-}}                [rt]   at  3 12.5

\put {\mathput{n}}                  [rt]   at  0.4 1.5
\put {\mathput{P_+}}                     [lt]   at  2 1.5
\put {\mathput{P_-}}                     [rt]   at  -2 1.5

\put {\mathput{\hat e(m)}}             [l]   at  2.5 -0.5
\put {\mathput{\hat e(n)}}             [rt]   at   1.5 0.6

\put {\mathput{t}}                     [rb]   at  -1 24
\put {\mathput{\phi}}                  [b]   at  4.5 0.5

\put {\mathput{-\frac{\pi}{2}}}         [lb]   at  -3.85 0.3
\put {\mathput{\frac{\pi}{2}}}          [rb]   at   3.7 0.3

\put {\mathput{\Delta t_{\rm (SD)}}}   [r]   at -4.15 -0.2


\betweenarrows 
  {\mathput{\Delta t_{\rm (geo)}}} [l] from 3.65 17.03 to 3.65 20.63
\betweenarrows 
  {\mathput{\Delta t_{\rm (null)}}} [r] from -4.5 14.5 to -4.5 17.5


\arrow <.5cm> [.1,.4]    from  4.5 0 to  5  0               
\arrow <.5cm> [.1,.4]    from  -0.6250 24 to -0.6275 24.5   

\arrow <.5cm> [.1,.4]    from  1.5 0 to  2  0              
\arrow <.5cm> [.1,.4]    from  1.5 -0.45 to 2.0625 -0.4525 

\arrow <.5cm> [.1,.4]    from  0 1.5 to  0 2           

\arrow <.5cm> [.1,.4]    from   0 0  to    2.67 3.33           
\arrow <.5cm> [.1,.4]    from   0 0  to -3.88 4.37             
\arrow <.5cm> [.1,.4]    from   1.65 1.65 to    2 2            
\arrow <.5cm> [.1,.4]    from  -1.65 1.65 to   -2 2            

\arrow <.5cm> [.1,.4]    from  0 0 to -1.21 7.70               

\endpicture}$$

\end{figure}


The horizontal unit vector $\hat e(n)$ in Figure~\ref{fig:sagnac} is
along the slicing relative velocity direction tangent to the $\phi$
coordinate line (constant $t$). The horizontal short dashed line
connecting $P_\pm$ contains all the (horizontal) slicing relative
velocities. The unit vector $\hat e(m)$ is along the threading
relative velocity direction and indicates the threading local rest
space angular direction. Extending this forward around one loop of the
cylinder leads to the change in the coordinate time equal to the
synchronization defect. 

One can express the threading observer synchronization defect, Sagnac
effect, and timelike geodesic generalization of it using a single
formula involving the physical fields. The 1-form 
\begin{eqnarray}
  \gamma(U,m)^{-1} \bar U_\alpha dx^\alpha
    &=& [ m_\alpha +   1/\nu(U,m)^{\hat\phi} e(m)_\alpha ] 
             \,dx^\alpha
                 \nonumber\\
    &=& - M (dt - M_\phi d\phi)  
        + \gamma_{\phi\phi}{}^{1/2}/\nu(U,m)^{\hat\phi} \, d\phi
\end{eqnarray}
orthogonal to the 4-velocity $U^\alpha$ restricts to zero along the
world line, so the observer proper time arrival time is
\begin{eqnarray}
       \tau(m) &=& M t(m) 
           = [{\rm sgn}\, \nu(U,m)^{\hat\phi}]
              \int_0^{2\pi} [MM_\phi 
        + \gamma_{\phi\phi}{}^{1/2}/\nu(U,m)^{\hat\phi}] \, d\phi
             \nonumber\\
         &=&  2\pi M/|\zeta| \ .
 \end{eqnarray}
Thus difference in arrival times of two oppositely directed world
lines is then 
\begin{equation}
    \Delta \tau(m) = M[t_2(m) - t_1(m)]
     = 2\pi \left[MM_\phi + \gamma_{\phi\phi}{}^{1/2}
               [1/\nu(U_1,m)^{\hat\phi} + 1/\nu(U_2,m)^{\hat\phi}]
            \right]
         \ .
\end{equation}
    
For the oppositely directed null orbits ($\nu(P,m)^\phi = \pm1$), this
just gives 
\begin{equation}
    \Delta \tau_{\rm(null)}(m) = 4\pi M M_\phi = 4\pi p_\phi(m) \ ,
\end{equation}
namely the conserved angular momentum of the threading observers. For
the oppositely directed timelike geodesic orbits this just gives 
\begin{equation}
    \Delta \tau_{\rm(geo)}(m) 
        = 4\pi p_\phi(m)  -  C(m) H(m)^{\hat z}/g(m)^{\hat r} \ ,
\end{equation}
where $C(m)= 2\pi \gamma_{\phi\phi}^{1/2}$ is the threading
circumference of a $\phi$ coordinate circle, using Eq.~(\ref{eq:kgH})
to re-express the quotient of the sum and product of the geodesic
relative velocities. For the limit $\nu(U_i,m)\to\pm\infty$ 
in which the
trajectories become spatial curves with respect to $m$, this gives
twice the synchronization defect (defined to be the change in time
around one such loop, not two) which has the same value as the proper
time Sagnac effect 
\begin{equation}
    \Delta \tau_{\rm(SD)}(m) 
        = {\textstyle \frac12}  \Delta \tau_{\rm(null)}(m) \ .
\end{equation}

These same considerations should apply to any stationary family of
observers, once they are expressed in terms of the threading
potentials and fields associated with the family. For the slicing
observers, for example, both the angular momentum and the threading
gravitomagnetic field vanish so the Sagnac and synchronization defects
vanish. 

Similarly the slicing point of view expression for the proper time
geodesic effect has an analogous formula with the substitutions
$H(m)^{\hat z} \to -2 \theta(n)_{\hat\phi \hat r}$, $C(m) \to C(n) =
2\pi g_{\phi\phi}{}^{1/2}$ 
\begin{equation}
    \Delta \tau_{\rm(geo)}(n) 
      = 0 + 2 C(n) \theta(n)^{\hat\phi \hat r}/g(n)^{\hat r} \ ,
\end{equation}
since the slicing angular momentum vanishes.

\section{Concluding Remarks}

Test particle behavior is one of the most important tools in
interpreting the local spacetime structure of a given spacetime,
whether it is a nice exact solution with symmetry which may be worked
with analytically or an approximate solution with no symmetry derived
with very complicated mathematical machinery. The three familiar
spacetimes analyzed here in terms of natural test observer families
each have a rich and interesting geometry that weaves together in
slightly different ways the various effects that their gravitational
fields have on the motion and spin of test particles. 

By adopting a general approach to the test observer splitting and the
geometry they use to interpret their measurements, many diverse issues
heretofore described in widely differing languages are brought
together in a single discussion based on familiar observer-defined
quantities.  Instead of vague references to centrifugal and Coriolis
forces that motivate the interpretation of fields in linearized
approximations to general relativity or the somewhat obscure formal
representations of such forces in the fully nonlinear theory, this
general approach gives a clear and precise representation of the many
ways these concepts can manifest themselves in the arena of Einstein's
theory. By analyzing these three familiar spacetimes in this approach,
and showing the power that it has in interpreting physical aspects of
their geometry and showing the wider applicability of the analysis to
allow comparisons among them, we hope to encourage others to consider
its use when appropriate. 


\section*{Acknowledgements}

We thank Remo Ruffini of the International Center for Relativistic
Astrophysics at the University of Rome and Francis Everitt of the
Gravity Probe B Relativity Mission group at Stanford University for
their support and encouragement of this work. 


\appendix

\section{Embedding cross-section diagrams}

For the 2-dimensional metrics in polar-like coordinates $\{r,\phi\}$
under consideration 
\begin{equation}
  h_{rr} dr^2+ h_{\phi\phi} d\phi^2 = 
  h_{RR} dR^2+ R^2 d\phi^2  \ ,
\end{equation}
where the components depend only on the radial coordinate, the
embedding into $E_3$ or $M_3$ with cylindrical coordinates
$\{R,\phi,Z\}$ can be accomplished in two steps. First one inverts the
transformation from the radial coordinate $r$ to the radial
circumferential arclength coordinate $R = h_{\phi\phi}{}^{1/2}$, for
which the circumference of a radial coordinate circle has the usual
expression $2\pi R$. Then one performs an integral to get the curve in
the $R$-$Z$ half plane which is the constant $\phi$ cross-section of
the embedding surface. It is convenient for graphing purposes to
introduce the rescaled variables $\bar R = R/{\cal R}$ and $\bar Z =
Z/{\cal R}$ analogous to $\bar r = r/{\cal R}$, where ${\cal R}$ is 
the length scale of Table II.  The case
$h_{RR}=1$ corresponds to the flat case of $E_2$ in polar coordinates.
When $h_{RR}>1$, one next identifies the result with the flat
cylindrical coordinate metric on $E_3$ restricted to the surface
$Z=Z(R)$ 
\begin{equation}
      d R^2 + R^2 d \phi^2 + d Z(R)^2
    = [1 + (dZ(R)/dR)^2] d R^2 + R^2 d \phi^2 \ .
\end{equation}
If instead $h_{RR}<1$, one identifies the result with the flat
cylindrical coordinate metric on $M_3$ restricted to the surface
$Z=Z(R)$, where $Z$ is now a timelike coordinate 
\begin{equation}
      d R^2 + R^2 d \phi^2 - d Z(R)^2
    = [1 - (dZ(R)/dR)^2] d R^2 + R^2 d \phi^2 \ .
\end{equation}
In both cases this gives a first order differential equation for
$Z(R)$ with an integral solution as in the Schwarzschild case
\cite{mtw} 
\begin{equation}
 1 \pm (dZ(R)/dR)^2 = h_{RR} = h_{rr}(dr/dR)^2 \ ,
\end{equation}
or equivalently 
\begin{eqnarray}\label{eq:embeqexp}
	|dZ(R)/dR| &=& \sqrt{ \pm(h_{RR}-1)} \ ,
                 \nonumber\\
	|Z(R)-Z_0| &=& \int_{R_0}^R \sqrt{ \pm(h_{RR}-1)} \,dR \ ,
\end{eqnarray}
where the plus/minus sign corresponds to the embedding in $E_3$/$M_3$.
Dropping the absolute value sign, $Z$ may be chosen initially to
increase with $R$, but a negative derivative may later be required if
the embedding turns back towards the $Z$-axis as $Z$ increases. 

In general it is not possible to express the argument of the square
root explicitly as a function of $R$ but one can make a parametric
plot of $Z(r)$ and $R(r)$ using 
\begin{eqnarray}\label{eq:embeqimp}
  |dZ(r)/dr| &=& \sqrt{ \pm[h_{rr}-(dR(r)/dr)^2] }\ ,
               \nonumber\\
  R(r)     &=& \sqrt{h_{\phi\phi}} \ ,
\end{eqnarray}
which has the integral solution
\begin{equation}\label{eq:embZr}
   |Z(r)-Z_0| 
    = \int_{r_0}^r \sqrt{ \pm[h_{rr}-(dR(r)/dr)^2] } \,dr \ .
\end{equation}
This may be numerically integrated if necessary.

For cases with a regular origin $R\to0$ as $r\to0$, the initial
conditions $Z_0=0$ and $r_0=0$ puts the origin of the $\{r,\phi\}$
coordinate system at the origin of the embedding space. In other
cases, one may start at a physically interesting value of $r$ for
$r_0$, which then determines $R$, and choose $Z_0$ conveniently. 

Once the curve in the  $R$-$Z$ plane is obtained either analytically
or numerically, the full surface is obtained by revolving it around
the $Z$ axis. Introducing the Lie relative radius of curvature from
[BCJ1] for this context 
\begin{equation}
  \rho(\phi,u) = |\kappa(\phi,u)^{\hat r}|^{-1} 
               = |-(\ln R)_{,r}/h_{rr}{}^{1/2}|^{-1}
\end{equation}
leads to the result
$h_{RR}^{1/2} = \rho(\phi,u)/R$ and
\begin{equation}\label{eq:cone}
   |dZ(R)/dR|  = \sqrt{\pm[\rho(\phi,u)^2/R^2 - 1]}
                  = \sqrt{\pm[\rho(\phi,u)^2 - R^2]}/R \ .
\end{equation}
The ratio $\rho(\phi,u)/R$ is identically 1 for a flat space,
greater than 1 for a Euclidean embedding, and less  than 1 for a
Minkowski embedding. In the latter cases, if this ratio passes through
1 at some radius, the tangent to the $R$-$Z$ cross-section is
horizontal and the embedding space signature switches sign. This tangent
line is instead vertical when $|dZ(R)/dR| \to\infty$, which occurs
when $\rho(\phi,u)\to\infty$ at a Lie relatively straight circle. On
either side of the radius $r_{\rm(ss)}$
at which the signature switches sign, the choice
of sign for the derivative $dZ(r)/dr$ is free. This sign will be
always be chosen to be positive to join the two embedding spaces on
either side of a half plane, in which case 
\begin{equation}
  \mathop{\rm sgn}(\kappa(\phi,u)^{\hat{r}})
    = - \mathop{\rm sgn}( dZ/dR ) \ .
\end{equation}
Embedding diagrams for which the signature changes have been used by
Smarr \cite{sma} in studying the 2-surface cross-section of the
horizon of a charged Kerr black hole. 

When the tangent to the embedding cross-section is not vertical, one
can introduce the tangent cone obtained by revolving the tangent line
at a point $(R,Z(R))$ on the cross-section. The relation
(\ref{eq:cone}) then shows that $\rho(\phi,u)$ must be identified with
the distance of the tangent line point of contact from the vertex of
the cone calculated in the embedding space geometry, thus giving a
nice geometrical interpretation to the Lie relative radius of
curvature. For the $E_3$ case, the defect angle of the cone is easily
found to have the positive value 
\begin{equation}
    \Delta = 2\pi [1- R/\rho(\phi,u)] \ ,
\end{equation}
which in the usual case in which $\mathop{\rm
sgn}(\kappa(\phi,m)^{\hat{r}})<0$ determines the space curvature
contribution to the precession angle of the spin of a gyroscope
following the circular orbit after one revolution modulo the sign, as
discussed by Thorne \cite{tho81}. The same formula holds for the
spacelike cones in $M_3$, where the defect angle is instead negative.
Note that in approaching a relatively straight trajectory (vertical
tangent in the $R$-$Z$ cross-section) this defect angle approaches
$2\pi$, while in approaching a horizontal tangent in the $R$-$Z$
cross-section the angle approaches the flat value 0. 

The Gaussian curvature of this surface is easily evaluated from
standard formulas to be 
\begin{equation}\label{gauscur} 
  K = - (R h_{rr}{}^{1/2})^{-1} (R_{,r}/h_{rr}{}^{1/2})_{,r} 
    = - R^{-1} R_{,\hat r \hat r} \ .
\end{equation} 
For the rotating Minkowski slicing and the G\"odel threading cases
where $h_{rr}=1$, this reduces to  $K = -R^{-1} R_{,rr}$. 
This may also be rewritten as 
\begin{equation}
   K = \mathop{\rm sgn}(\kappa(\phi,u)^{\hat r})\, 
                       R^{-1} [R/\rho(\phi,u)]_{,\hat r} \ .
\end{equation}
Thus the sign of the Gaussian curvature depends on the signs of the
radial relative centripetal acceleration (sign of $-R_{,r}$) and of
the radial derivative of the ratio $R/\rho(\phi,u)$. It is also
straightforward to show that 
\begin{equation}
    \mathop{\rm sgn}(K) = \mp \mathop{\rm sgn}(d^2 R/dZ^2)\ ,
\end{equation}
which implies that when the cross-section curve is concave away/toward
the $Z$-axis in $E_3$, the Gaussian curvature is negative/positive,
and vice versa in $M_3$. 

\subsection{Rotating Minkowski spacetime} 

The spatial geometry in the slicing point of view is flat, so the
usual Euclidean geometry applies and the Lie signed relative curvature
of the $\phi$ coordinate lines is just the reciprocal of the radial
coordinate $r$. 

\begin{figure}[p]\footnotesize
\centerline{\epsfbox{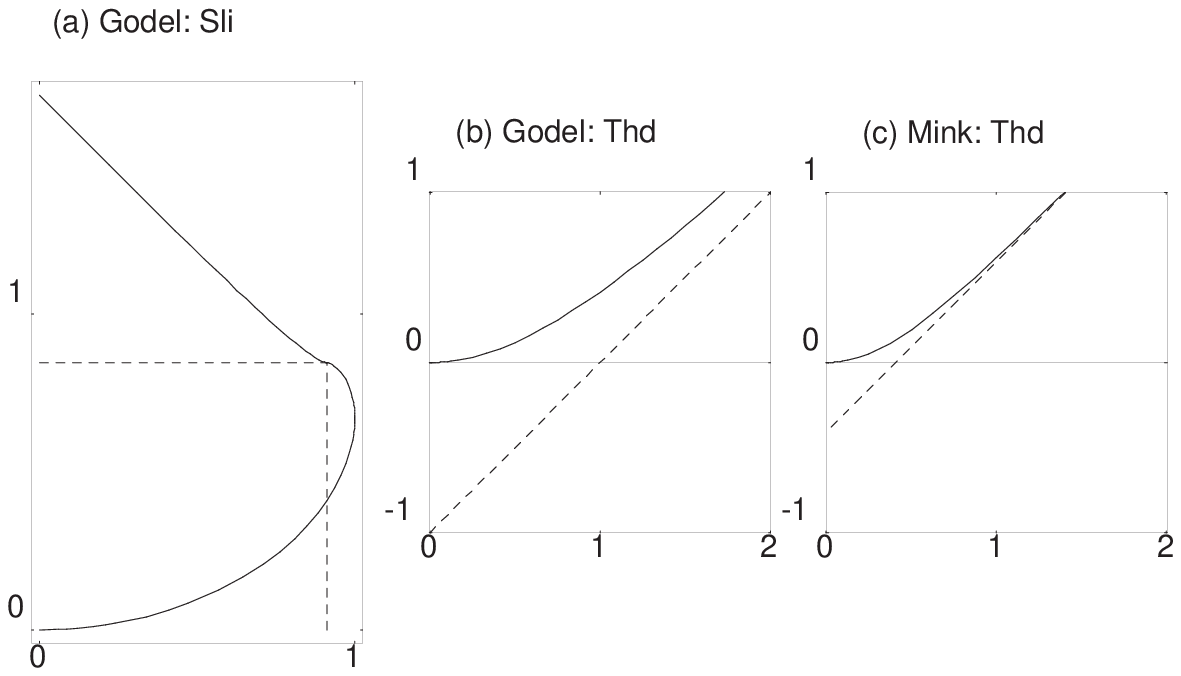}}
\fcaption
{$\bar Z$ versus $\bar R$: embedding cross-section diagram for the
$r$-$\phi$ surfaces in the threading point of view for the G\"odel 
$(a)$ slicing and $(b)$ threading points of view and $(c)$ the
rotating Minkowski threading point of view. In the latter two cases
the asymptotic null lines in the $M_3$ embedding are shown with
respective intercepts $\bar Z\approx -0.567 \bar r_{\rm(hor)}$ and
$\bar Z\approx -0.406 \bar r_{\rm(hor)}$ which are comparable
fractions of the horizon scale, but which appear much different since
units of vorticity rather than curvature are used in the graph. In the
first case starting at the origin with a Euclidean embedding, the
embedding surface moves away from the vertical axis and then turns
back at the radius of the single relatively straight circle, next
switching signature  and finally asymptotically approaching the
vertical axis again at the observer horizon along an asymptotic null
cone. 
}
\label{fig:embdiagramsMG}
\end{figure}

The $r$-$\phi$ surface in the threading point of view is instead an
inhomogeneous surface of revolution of negative Gaussian curvature $
-3\gamma^4 / {{\cal R}^2 }$ which starts at the value  $-3 / {{\cal
R}^2 } $ at the origin and as $r$ increases it decreases to $-\infty$
at the observer horizon at the value $\bar r_{\rm(h)}=1$. The
embedding is in $M_3$ and the singularity at the observer horizon
occurs at $R\to\infty$. Figure \ref{fig:embdiagramsMG}$(a)$ shows the
constant $\phi$ cross-section of this surface, with the horizontal and
vertical axes showing the rescaled variables $\bar{\cal R}$ and $\bar
Z$ respectively. 

\typeout{*** Figure 11.  (embdiagramsMG)}

\typeout{*** (Partial page figure but apparently no room for extra text.)}
\typeout{*** (Made into full page figure on page 54.)}

\subsubsection*{The threading case} 

Since $dR/dr = \gamma^3$ is always positive, then $h_{RR} =
\gamma^{-6} <1$ and the embedding occurs in $M_3$ with a regular
origin and a single sign for $dZ(R)/dR$ which may be taken to be
positive. The relationship $\bar R=\gamma \bar r$ (with $\gamma =
(1-\bar r{}^2)^{-1/2}$) inverts to $\bar r=\gamma^{-1}\bar R$ (with
$\gamma = (1+\bar R{}^2)^{1/2}$). Dropping the absolute value signs in
the embedding equation (\ref{eq:embeqexp}) and choosing $\bar Z_0 = 0$
leads to 
\begin{equation}
    \bar Z(\bar R) 
     = \int_0^{\bar R} \sqrt{ 1- (1+ x{}^2)^{-3} } \,dx \ .
\end{equation}
The change of variable $w= 2\ln\gamma = \ln (1+\bar R{}^2) = - \ln
(1-\bar r{}^2)$ leads to 
\begin{equation} 
  \bar Z(w) 
  =  {\textstyle\frac{1}{2}} \int_0^w\sqrt{1 + 2\cosh y} \,dy  \ ,
\end{equation} 
which in turn suggests the further change of variable
$u = \sinh(w/2) = (\gamma - \gamma^{-1})/2$ and
\begin{eqnarray} 
    \bar Z(u) 
        &=&  2 \int_0^u \sqrt{(3/4 + v^2)/(1 + v^2)} \,dv
                   \nonumber\\
        &=& - {\textstyle\frac{3}{2}} [F(\alpha,1/2) - F(\pi/2,1/2)]
                               + 2 [E(\alpha,1/2) - E(\pi/2,1/2)]
                   \nonumber\\
        & &
                    + 2 u \sqrt{(1+u^2)/(3/4 + u^2)} \ ,
\end{eqnarray} 
where $\alpha = \arctan(1/u)$ by a slightly corrected version of
formula 3.169.2 on p.276 of Gradshteyn and Ryzhik \cite{graryz}, and
$F(\varphi,k)$ and  $E(\varphi,k)$ are elliptic integrals of the first
and second kind in their notation. (When using MAPLE V and Mathematica
respectively, the correspondence is $E(\varphi, k)$ = {\tt
LegendreE}$(\sin\varphi,k)$ = {\tt EllipticE}$(\varphi,k^2)$.) 
\typeout{*** margin problem here: 4.5pts; enough to worry??}

\subsection{G\"odel spacetime} 

Figure
\ref{fig:embdiagramsMG} shows the constant $\phi$ cross-section of the
$r$-$\phi$ surface with the horizontal and vertical axes showing the
rescaled variables $\bar{\cal R}$ and $\bar Z$ respectively in the
$(b)$ threading and $(c)$ slicing points of view. 

In the threading point of view the spatial metric of the $r$-$\phi$
surface is in one of the standard forms for a constant
negative-curvature geometry ($K = - 1/{\cal R}^2$) and is easily
embedded in $M_3$ as a spacelike pseudosphere of radius ${\cal R}$.
The circumference $R$ of the circles of constant radial coordinate
increases without bound but the magnitude of the Lie signed relative
curvature approaches a lower limit corresponding to a maximum Lie
relative radius of curvature of magnitude ${\cal R}$, a consequence of
the negative curvature geometry. 

In the slicing point of view where $r$ is a radial arclength
coordinate this surface has positive Gaussian curvature $K = (2{\cal
R}^{2})^{-1} (1+2s^2-2s^4) / (1-s^2)^2$ which increases from the value
$1/ (2{\cal R}^{2}) $ at the origin to infinity at the observer
horizon. $R$ begins to decrease with increasing $r$ after reaching the
Lie relatively straight circle at the value $\bar r_{\rm(rs)} =
2{\mathop{\rm arcsinh}\nolimits} (2^{-1/2}) \approx 1.317$ ($\bar
R_{\rm(rs)} = 1$) where the Lie signed relative curvature changes sign
and the embedding turns back toward the $Z$-axis. Beyond that the Lie
relative centripetal acceleration points outward, i.e., towards larger
$r$ values (but smaller $R$ values) as occurs for the optical relative
centripetal acceleration close to the horizon in the Schwarzschild
spacetime as shown by Abramowicz \cite{acl88,a90a,a90b}. The embedding
then switches from $E_3$ to $M_3$ at the value $\bar r_{\rm(ss)}  =
2{\mathop{\rm arcsinh}\nolimits} (2^{-1/4}) \approx 1.529$ ($\bar
R_{\rm(ss)}  = \sqrt{2\sqrt2-2} \approx 0.910$, $\bar Z_{\rm(ss)}  =
\approx 0.847$). Finally the embedding surface returns to the $Z$-axis
($R=0$) at the value $\bar r_{\rm(h)} =2{\mathop{\rm
arcsinh}\nolimits} (1) \approx 1.763$ ($\bar Z = 2\bar Z_{\rm(h)}$)
corresponding to the observer horizon, where the embedding is
singular. 

\subsubsection*{The threading case} 

Here the standard embedding with $r=0$ at the origin of $M_3$ is a
pseudosphere of radius ${\cal R}$ 
\begin{equation}
         Z = - {\cal R} + \sqrt{R^2 + {\cal R}^2} \ .
\end{equation}

\subsubsection*{The slicing case} 

The origin $r=0$ is again regular, so $Z_0=0$ is appropriate, and one
may drop the absolute value signs initially in the explicit embedding
equation (\ref{eq:embeqexp}). Since $dR/dr = c(1-2s^2)/ \sqrt{1-s^2}$
changes sign as $r$ increases, this embedding turns back toward the
$Z$-axis, after which $dR/dr$ passes through the value 1 where the
embedding switches from $E_3$ to $M_3$. Attempting to invert the
relationship $\bar R(u) = 2\sqrt{u(1-u)} = 2 s\sqrt{1-s^2}$ by solving
a quadratic equation for $u=s^2$ 
\begin{equation} 
 u =\cases{ 
    \frac 12\left( 1-\sqrt{1-\bar R{}^2} \right) \in [0,1/2] \ ,
           &\cr 
    \frac 12\left( 1+\sqrt{1-\bar R{}^2} \right) \in [1/2,1]\ ,
          & \cr}
\end{equation}
in terms of which ($\bar r = 2 \mathop{\rm arcsinh}(\sqrt{u})$).
However,  $\bar R$ is a doublevalued function of $\bar r$ or $u$, so
the latter variables are more appropriate to use. The integral 
\begin{eqnarray}
   \bar Z &=& \int_0^r 2 s \sqrt{ (1/2-s^4)/(1-s^2) } \,dr
                 \nonumber\\
          &=& \int_0^u 2 \sqrt{ (1/2-v^2)/(1-v^2) } \,dv
\end{eqnarray}
can be integrated in terms of elliptic integrals depending on the
range of $u$. 

\paragraph*{$\bullet$ {\bf Embedding in $E_3$.}}

Starting at the regular origin with $u\in [0,\frac{1}{\sqrt{2}}]$ this
evaluates to 
\begin{equation} 
  \bar Z = 2 E \left(\alpha, 1/\sqrt{2} \right)  
              - F \left(\alpha, 1/\sqrt{2}  \right) \ ,
\end{equation} 
where $\alpha = \arcsin (\sqrt{2} u)$, by formula (3.169.9) on p.276
of Gradshteyn and Ryzhik \cite{graryz}. Note that the endpoint value
$u = 1/\sqrt{2}$ at the signature change corresponds to $\alpha =
\pi/2$, while the relatively straight circle at $u = 1/2$ corresponds
to $\alpha = \pi/4$. Let $\bar Z_{\rm(ss)} = 2 E ( \pi/2 ,1/\sqrt{2}) 
- F ( \pi/2 , 1/\sqrt{2} ) \approx 0.847$. 

\paragraph*{$\bullet$ {\bf Embedding in $M_3$.}}

The new initial condition at $u=1/\sqrt{2}$ for $u \in [1/\sqrt{2},1]$
with the change in sign of the square root argument leads to 
\begin{equation} 
  \bar Z - \bar Z_{\rm(ss)}   
      =  2 E(\beta, 1/\sqrt{2}) - F(\beta, 1/\sqrt{2})
                    - 2 \sqrt{(u^2-1/2)(1-u^2)} / u  \ ,
\end{equation} 
where $\beta = \arcsin( \sqrt{2u^2-1} / u)$, by formula (3.169.11) on
p.277 of Gradshteyn and Ryzhik \cite{graryz}. The observer horizon
then occurs as the limiting point where $u=1$, $\bar R = 0$ and
$\beta=\pi/2$, so that $\bar Z =2 \bar Z_{\rm(ss)}$. 

\subsection{The Kerr spacetime}

The Kerr case can only be handled analytically in the limit of the
Schwarzschild spacetime. The embedding geometry of the $r$-$\phi$
equatorial plane in the Schwarzschild spacetime is well known to be a
parabola of revolution in $E_3$ with its vertex at the observer
horizon which occurs at the relatively straight circle (``throat")
\cite{mtw}. The integral (\ref{eq:embZr})  leads to 
\begin{equation}
   \bar Z(\bar r) = \int_2^{\bar r} \sqrt{2/(u-2)} \,du
                  = 2\sqrt{2} (\bar r-2)^{1/2} \ ,
\end{equation}
with $\bar Z_0=0$ and $\bar r_0 = 2$ while $\bar R = \bar r$,
explicitly confirming the parabolic embedding cross-section. 

The Kerr equatorial plane must be studied numerically in both points
of view since inverting the relationship between $r$  and the
circumferential coordinate $R$ or performing the integral
(\ref{eq:embZr}) cannot be done analytically. The slicing embedding
diagrams have been given by Sharp \cite{sha} while a useful discussion
of the general problem has been given by Romano and Price
\cite{rompri}. 

For the slicing point of view the plus sign is always valid and the
embedding occurs in $E_3$. The initial conditions are taken to be
$\bar Z_0=0$ and $\bar r_0 = \bar r_{\rm(h)} =1+\sqrt{1-\bar a{}^2}
\in[1,2]$, leading to $\bar R(r_{\rm(h)}) = 2$. Note that $\bar R{}^2
= \bar r{}^2 + \bar a{}^2 + 2\bar a{}^2/\bar r$ implies $d\bar R/d\bar
r = (\bar r{}^3-\bar a{}^2)/(\bar r{}^2\bar R)>0$, which is positive
and finite for the physical ranges $0 \leq \bar a\leq 1$ and $\bar r
\geq \bar r_{\rm(h)}$. Since $\Delta=0$ at $r=r_{\rm(h)}$, then
$h_{rr}\to\infty$ as $r\to r_{\rm(h)}{}^+$, so that also $d Z/d R 
= [dZ(r)/dr] \, / \,  [dR(r) \,/\, dr] \to\infty$. Thus all the 
slicing embedding cross-sections start out at $\bar R=2, \bar Z=0$ in
the $R$-$Z$ plane with a vertical tangent as in the Schwarzschild case
where that point is the vertex of a parabola. 

For the threading point of view 
the integrand of (\ref{eq:embZr}) switches sign, while
as $\bar r\to \bar r_{\rm(erg)}{}^+ = 2^+$,  $\bar R \to\infty$,
so the observer horizon is pushed out to infinity in $M_3$ as in
the Minkowski threading case.
Therefore initial conditions are taken to be $\bar Z_0=0$ and 
$\bar r_0$ is chosen to be the single real zero
of the integrand which occurs for the physical ranges
$0\leq \bar a \leq1$, $\bar r\geq2$, equivalent to
\begin{equation}
    2 \bar r{}^2 (\bar r -2)^3 
    + \bar a{}^2 [ 2 \bar r (\bar r-2)^2 - \bar a{}^2 ] = 0 \ .
\end{equation}
As shown in Figure~\ref{fig:rversusa}, these roots lie in the interval
$[2,2.344]$ (approximately) with the endpoint values respectively
corresponding to $\bar a=0$ and $\bar a=1$. The example $\bar a= 0.5$
has $\bar r_{\rm(ss)} \approx 2.157$. 

For $\bar r > \bar r_0$ the plus sign is relevant and the embedding is
in $E_3$, while for $\bar r < \bar r_0$ the minus sign is relevant and
the embedding is in $M_3$. Thus the $R$-axis is taken to be the
signature change line for all values of $a$, with $E_3$ above and
$M_3$ below. One also finds that $d R(r)/dr = 0$ at the single real
solution of $\bar r(\bar r -2)^2 - \bar a{}^2 = 0$ for the allowed
values or $\bar r$ and $\bar a$, leading to a vertical tangent in the
$R$-$Z$ plane at the Lie relatively straight circle, where the
embedding cross-section turns away from the $Z$-axis. 

\begin{figure}[htbp]\footnotesize
\centerline{\epsfbox{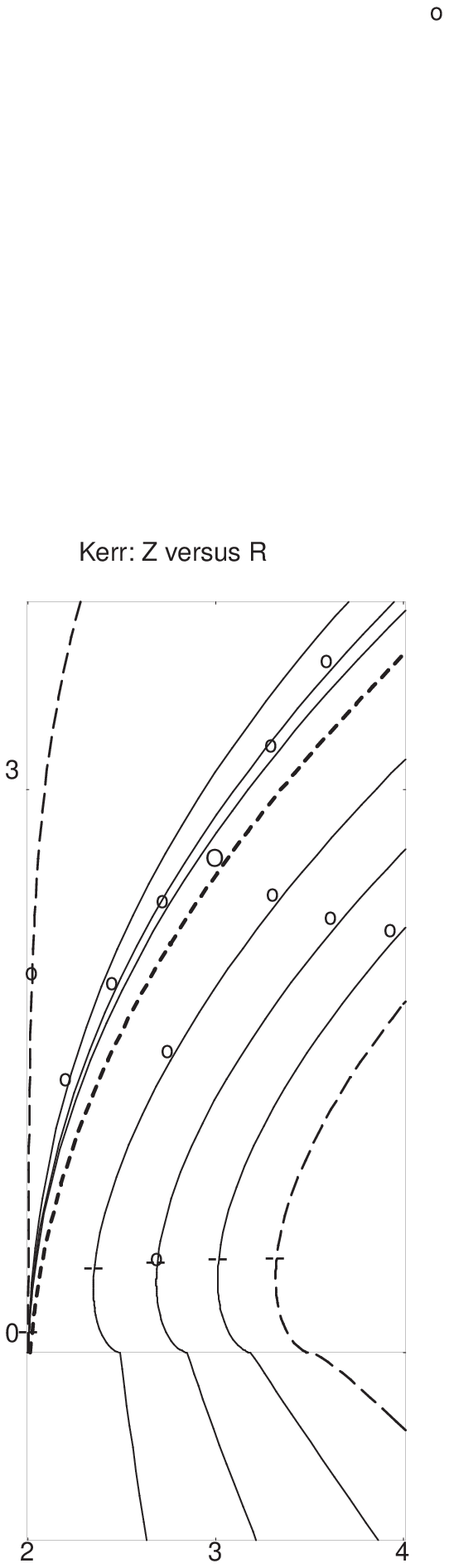}}
\fcaption
{$\bar Z$ versus $\bar R$: embedding cross-section diagrams for the
$r$-$\phi$ surfaces in both the threading and slicing points of view
for the Kerr spacetimes, separated by the thick dashed line at the
Schwarzschild case $\bar a = 0$. Cross-sections for the Euclidean
slicing embedding are shown for the  values $\bar a=0, .25, .5, .75,
0.999$ increasing upward above the thick dashed line, while those for
the Minkowski threading embedding are below, with the same values
increasing downward below, except for the regular final value $\bar a=
1$. At nonzero values of $\bar a$, the threading embedding surface
starts out at the variable ergosphere outer radius at $R\to\infty$,
moves towards the $Z$-axis switching signature along the way, and then
turns away from the axis at the radius of the single relatively
straight circle indicated by a tick mark. The slicing embedding
surface starts with a vertical tangent at the horizon $\bar r = 2$ for
all values of $\bar a$ and moves away from the $Z$-axis, but because
the proper distance near the horizon (fixed at $Z=0$) blows up at
$\bar a\to1$, the interesting part of the embedding diagram is pushed
up to infinity \cite{membrane}. The circles mark the interface between
regions A and B (lower) and then B and C (upper) on each embedding
curve, which merge in a single interface at $\bar a=0$ and expand from
each other with increasing $\bar a$ as in Fig.~\ref{fig:rversusa}.
}
\label{fig:embdiagramsK}
\end{figure}

The Gaussian curvature for both the threading and slicing cases is
negative 
\begin{eqnarray}
    {\cal M}^2 K_{\rm(th)} 
       &=& - \frac{[\bar r(\bar r-2)^2 + 3\bar a{}^2(\bar r-1)]}
                  {\bar r{}^4(\bar r-2)^2}
    \ ,\\
    {\cal M}^2 K_{\rm(sl)} 
       &=& 
    - \frac{(\bar r{}^7 + 5 \bar a{}^2 \bar r{}^5 
              - 11 \bar a{}^2 \bar r{}^4 + 7 \bar a{}^4 \bar r{}^3 
              - 2 \bar a{}^4 \bar r{}^2 
              + \bar a{}^4(3 \bar a{}^2 - 8)\bar r - 7 \bar a{}^6)}
           {\bar r{}^4[\bar r(\bar r{}^2+\bar a{}^2)+2\bar a{}^2]^2}
                  \ ,
    \nonumber
\end{eqnarray}
with the common Schwarzschild limit  ${\cal M}^2 K = -1/\bar r{}^3$.
For Kerr it becomes infinite at the threading observer horizon as in
the rotating Minkowski case, while remaining finite at the slicing
observer horizon (where it vanishes for $\bar a = 1$, representing a
relatively straight circle). Figure \ref{fig:embdiagramsK} shows the
constant $\phi$ cross-section of the embedding surface for different
values of the rotation parameter $a$ in both points of view. 

\typeout{*** Figure 12.  (embdiagramsK)}

\typeout{*** (Final figure full page on last page 59.)}

\section*{Corrections}

This reformatted version contains the following misprint corrections of the original article (to which the page numbers refer) and one reference publication update:

\begin{itemize}
\item
p.~146. Table 1 Line 1: 
G\"odel lapse $N$ change $\sqrt{1-c^2}$ to $\sqrt{1-s^2}$
\item
p.~147. Eq.~(2.3) Line 1: 
change second minus sign to an equal sign
\item
p.~150 Table 2. Row for signed relative curvature, Line 2, G\"odel (last) column: equal sign missing after kappa symbol before minus sign
\item
p.~169. Line following Eq.~(4.25): 
change $\phi$ to $\dot\phi$ in both occurrences
\item
p.~174. Figure 7.(a),(b): The steep downward curve in region B was replaced and the last line of figure caption (a) changed from:
``The steep nearly straight curve in region B is $\ldots$"\\
to\\
``The curve decreasing from the upper geodesic photon point to the lower
geodesic photon point in region B is ``$\ldots$" 

$\zeta_{\rm(crit,spin)}$ where the threading precession is extremal. 
\item
p.~176. Eq.~(5.4): second subscript  ``lie" should be in roman type
\item
p.~182. Eq.~(6.16) Line 1: remove $\gamma^2$ factor
\item
p.~188. Eq.~(7.10): change $4 \pi$ to $2\pi$
\item
p.~189. Eq.~(7.12): change $2 C(m)$ to $C(m)$
\end{itemize}

\end{document}